\documentclass[longauth]{aa}  
\usepackage{graphicx}
\usepackage{pdflscape}
\usepackage{tabulary}
\usepackage{txfonts}

\usepackage{url} 
\usepackage{epstopdf}
\usepackage{color}
\usepackage{hyperref}
\usepackage{url}
\usepackage{amsmath}
\usepackage{bm}
\usepackage{comment}
\usepackage{soul}
\usepackage{mathtools}
\usepackage{tablefootnote}
\usepackage{bold-extra}
\usepackage{xspace}
\usepackage{relsize}
\usepackage[utf8]{inputenc}
 \usepackage[T1]{fontenc}
\usepackage{xcolor}
\usepackage[symbol]{footmisc}
\usepackage[normalem]{ulem}
\usepackage{longtable,array} %
\usepackage[normalem]{ulem} 
\usepackage{pdflscape}
\usepackage{orcidlink}
\usepackage{CJKutf8} 
\newcommand{\cntext}[1]{\begin{CJK*}{UTF8}{gbsn}#1\end{CJK*}}
\newcommand{\HESS}{H.E.S.S.\xspace}

\begin{document} 
\title{Broadband multiwavelength properties of the archetypal blazar 3C\,279 during the 2017 Event Horizon Telescope campaign}

\author{
{G.~Principe}\thanks{First author and project coordinator,  \textcolor{blue}{giacomo.principe@inaf.it}}\orcidlink{0000-0003-0406-7387}\inst{\ref{PALinst31},\ref{PALinst32},\ref{PALinst8}}\and
{J.C.~Algaba}\orcidlink{0000-0001-6993-1696}\inst{\ref{PALinst0}}\and
{E.~Aviano}\thanks{\textit{Fermi}-LAT corresponding author. For questions concerning \textit{Fermi}-LAT results contact \textcolor{blue}{ermes.aviano@phd.units.it}}\orcidlink{0009-0001-2927-8968}\inst{\ref{PALinst31},\ref{PALinst32}}\and
{W.Y.~Cheong}\orcidlink{0009-0002-1871-5824}\inst{\ref{PALinst43},\ref{PALinst44}}\and
{K.~Hada}\orcidlink{0000-0001-6906-772X}\inst{\ref{PALinst7},\ref{PALinst37}}\and 
{D.~Haggard}\orcidlink{0000-0001-6803-2138}\inst{\ref{PALinst10},\ref{PALinst16}}\and 
{A.~Hahn}\orcidlink{0000-0003-0827-5642}\inst{\ref{PALinst38}}\and
{S.G.~Jorstad}\orcidlink{0000-0001-6158-1708}\inst{\ref{PALinst17}}\and
{E.~V.~Kravchenko}\orcidlink{0000-0003-4540-4095}\inst{}
\and
{Y.~Kovalev}\orcidlink{0000-0001-9303-3263}\inst{\ref{GALinst7}}
\and
{S.S.~Lee}\orcidlink{0000-0002-6269-594X}\inst{\ref{PALinst43}}\and 
{M.~Lisakov}\orcidlink{0000-0001-6088-3819}\inst{\ref{PALinst39}}\and
{S.~Markoff}\orcidlink{0000-0001-9564-0876}\inst{\ref{PALinst25},\ref{PALinst26},\ref{PALinst41}}\and
{A.P.~Marscher}\orcidlink{0000-0001-7396-3332}\inst{\ref{PALinst17}}\and
{M.~Sasada}\orcidlink{0000-0001-5946-9960}\inst{\ref{PALinst42},\ref{PALinst7},\ref{PALinst36}}
\and
{P.~Voitsik}\orcidlink{}\inst{}
\and\\
{{(The Event Horizon Telescope Collaboration)}}\and\\
Kazunori Akiyama\orcidlink{0000-0002-9475-4254}\inst{\ref{GALinst1},\ref{GALinst2},\ref{GALinst3},\ref{GALinst4}}\and
Ezequiel Albentosa-Ruíz\orcidlink{0000-0002-7816-6401}\inst{\ref{GALinst5}}\and
Antxon Alberdi\orcidlink{0000-0002-9371-1033}\inst{\ref{GALinst6}}\and
Walter Alef\inst{\ref{GALinst7}}\and
Rohan Ganesh Amanaganti\orcidlink{0009-0002-3431-4625}\inst{\ref{GALinst9}}\and
Richard Anantua\orcidlink{0000-0003-3457-7660}\inst{\ref{GALinst10},\ref{GALinst11},\ref{GALinst12},\ref{GALinst4}}\and
Eleni Antonopoulou\orcidlink{0009-0004-7747-7760}\inst{\ref{GALinst13},\ref{GALinst14}}\and
Keiichi Asada\orcidlink{0000-0001-6988-8763}\inst{\ref{GALinst15}}\and
Rebecca Azulay\orcidlink{0000-0002-2200-5393}\inst{\ref{GALinst5},\ref{GALinst16},\ref{GALinst7}}\and
Anne-Kathrin Baczko\orcidlink{0000-0003-3090-3975}\inst{\ref{GALinst17},\ref{GALinst7}}\and
David Ball\inst{\ref{GALinst18}}\and
Bidisha Bandyopadhyay\orcidlink{0000-0002-2138-8564}\inst{\ref{GALinst19}}\and
John Barrett\orcidlink{0000-0002-9290-0764}\inst{\ref{GALinst2}}\and
Michi Bauböck\orcidlink{0000-0002-5518-2812}\inst{\ref{GALinst20}}\and
Bradford A. Benson\orcidlink{0000-0002-5108-6823}\inst{\ref{GALinst21},\ref{GALinst22}}\and
Dan Bintley\inst{\ref{GALinst23},\ref{GALinst24}}\and
Lindy Blackburn\orcidlink{0000-0002-9030-642X}\inst{\ref{GALinst4},\ref{GALinst12}}\and
Raymond Blundell\orcidlink{0000-0002-5929-5857}\inst{\ref{GALinst4}}\and
Katherine L. Bouman\orcidlink{0000-0003-0077-4367}\inst{\ref{GALinst25}}\and
Geoffrey C. Bower\orcidlink{0000-0003-4056-9982}\inst{\ref{GALinst23},\ref{GALinst24},\ref{GALinst26},\ref{GALinst27}}\and
Michael Bremer\orcidlink{0000-0001-7511-3745}\inst{\ref{GALinst28}}\and
Roger Brissenden\orcidlink{0000-0002-2556-0894}\inst{\ref{GALinst4}}\and
Silke Britzen\orcidlink{0000-0001-9240-6734}\inst{\ref{GALinst7}}\and
Avery E. Broderick\orcidlink{0000-0002-3351-760X}\inst{\ref{GALinst29},\ref{GALinst30},\ref{GALinst31}}\and
Dominique Broguiere\orcidlink{0000-0001-9151-6683}\inst{\ref{GALinst28}}\and
Thomas Bronzwaer\orcidlink{0000-0003-1151-3971}\inst{\ref{GALinst32}}\and
Sandra Bustamante\orcidlink{0000-0001-6169-1894}\inst{\ref{GALinst33}}\and
Douglas F. Carlos\orcidlink{0000-0002-1340-7702}\inst{\ref{GALinst34}}\and
John E. Carlstrom\orcidlink{0000-0002-2044-7665}\inst{\ref{GALinst35},\ref{GALinst22},\ref{GALinst36},\ref{GALinst37}}\and
Andrew Chael\orcidlink{0000-0003-2966-6220}\inst{\ref{GALinst38}}\and
Chi-kwan Chan\orcidlink{0000-0001-6337-6126}\inst{\ref{GALinst18},\ref{GALinst39},\ref{GALinst40}}\and
Chin-Shin Chang\orcidlink{0000-0001-9910-3234}\inst{\ref{GALinst41}}\and
Dominic O. Chang\orcidlink{0000-0001-9939-5257}\inst{\ref{GALinst4},\ref{GALinst12}}\and
Koushik Chatterjee\orcidlink{0000-0002-2825-3590}\inst{\ref{GALinst42},\ref{GALinst12},\ref{GALinst4}}\and
Erandi Chavez\orcidlink{0000-0003-4143-9717}\inst{\ref{GALinst4}}\and
Ming-Tang Chen\orcidlink{0000-0001-6573-3318}\inst{\ref{GALinst26}}\and
Yongjun Chen (\cntext{陈永军})\orcidlink{0000-0001-5650-6770}\inst{\ref{GALinst43},\ref{GALinst44}}\and
Xiaopeng Cheng\orcidlink{0000-0003-4407-9868}\inst{\ref{GALinst45}}\and
Paul Chichura\orcidlink{0000-0002-5397-9035}\inst{\ref{GALinst36},\ref{GALinst35}}\and
Ilje Cho\orcidlink{0000-0001-6083-7521}\inst{\ref{GALinst46},\ref{GALinst45},\ref{GALinst6}}\and
Nicholas S. Conroy\orcidlink{0000-0003-2886-2377}\inst{\ref{GALinst47},\ref{GALinst4}}\and
John E. Conway\orcidlink{0000-0003-2448-9181}\inst{\ref{GALinst17}}\and
Thomas M. Crawford\orcidlink{0000-0001-9000-5013}\inst{\ref{GALinst22},\ref{GALinst35}}\and
Geoffrey B. Crew\orcidlink{0000-0002-2079-3189}\inst{\ref{GALinst2}}\and
Alejandro Cruz-Osorio\orcidlink{0000-0002-3945-6342}\inst{\ref{GALinst48}}\and
Yuzhu Cui (\cntext{崔玉竹})\orcidlink{0000-0001-6311-4345}\inst{\ref{GALinst49}}\and
Brandon Curd\orcidlink{0000-0002-8650-0879}\inst{\ref{GALinst10},\ref{GALinst12},\ref{GALinst4}}\and
Rohan Dahale\orcidlink{0000-0001-6982-9034}\inst{\ref{GALinst6},\ref{GALinst50},\ref{GALinst51}}\and
Jordy Davelaar\orcidlink{0000-0002-2685-2434}\inst{\ref{GALinst52},\ref{GALinst53}}\and
Joost de Kleuver\orcidlink{0009-0001-9624-1751}\inst{\ref{GALinst32}}\and
Mariafelicia De Laurentis\orcidlink{0000-0002-9945-682X}\inst{\ref{GALinst54},\ref{GALinst55}}\and
Roger Deane\orcidlink{0000-0003-1027-5043}\inst{\ref{GALinst56},\ref{GALinst57},\ref{GALinst58}}\and
Jason Dexter\orcidlink{0000-0003-3903-0373}\inst{\ref{GALinst59}}\and
Vedant Dhruv\orcidlink{0000-0001-6765-877X}\inst{\ref{GALinst20}}\and
Indu K. Dihingia\orcidlink{0000-0002-4064-0446}\inst{\ref{GALinst60},\ref{GALinst61}}\and
Sheperd S. Doeleman\orcidlink{0000-0002-9031-0904}\inst{\ref{GALinst4},\ref{GALinst12}}\and
Sergio A. Dzib\orcidlink{0000-0001-6010-6200}\inst{\ref{GALinst7}}\and
Razieh Emami\orcidlink{0000-0002-2791-5011}\inst{\ref{GALinst4}}\and
Heino Falcke\orcidlink{0000-0002-2526-6724}\inst{\ref{GALinst32}}\and
Joseph Farah\orcidlink{0000-0003-4914-5625}\inst{\ref{GALinst62},\ref{GALinst63}}\and
Vincent L. Fish\orcidlink{0000-0002-7128-9345}\inst{\ref{GALinst2}}\and
Edward Fomalont\orcidlink{0000-0002-9036-2747}\inst{\ref{GALinst64}}\and
H. Alyson Ford\orcidlink{0000-0002-9797-0972}\inst{\ref{GALinst18}}\and
Marianna Foschi\orcidlink{0000-0001-8147-4993}\inst{\ref{GALinst6},\ref{GALinst25}}\and
Raquel Fraga-Encinas\orcidlink{0000-0002-5222-1361}\inst{\ref{GALinst32}}\and
William T. Freeman\inst{\ref{GALinst65},\ref{GALinst66}}\and
Per Friberg\orcidlink{0000-0002-8010-8454}\inst{\ref{GALinst23},\ref{GALinst24}}\and
Christian M. Fromm\orcidlink{0000-0002-1827-1656}\inst{\ref{GALinst67},\ref{GALinst68},\ref{GALinst7}}\and
Antonio Fuentes\orcidlink{0000-0002-8773-4933}\inst{\ref{GALinst6}}\and
Peter Galison\orcidlink{0000-0002-6429-3872}\inst{\ref{GALinst12},\ref{GALinst69},\ref{GALinst70}}\and
Charles F. Gammie\orcidlink{0000-0001-7451-8935}\inst{\ref{GALinst20},\ref{GALinst47},\ref{GALinst71}}\and
Roberto García\orcidlink{0000-0002-6584-7443}\inst{\ref{GALinst28}}\and
Olivier Gentaz\orcidlink{0000-0002-0115-4605}\inst{\ref{GALinst28}}\and
Boris Georgiev\orcidlink{0000-0002-3586-6424}\inst{\ref{GALinst18}}\and
Ciriaco Goddi\orcidlink{0000-0002-2542-7743}\inst{\ref{GALinst34},\ref{GALinst72},\ref{GALinst73},\ref{GALinst74}}\and
Roman Gold\orcidlink{0000-0003-2492-1966}\inst{\ref{GALinst75},\ref{GALinst76},\ref{GALinst77}}\and
Arturo I. Gómez-Ruiz\orcidlink{0000-0001-9395-1670}\inst{\ref{GALinst78},\ref{GALinst79}}\and
Brissa Gomez Miller\orcidlink{0009-0006-8345-6805}\inst{\ref{GALinst80}}\and
José L. Gómez\orcidlink{0000-0003-4190-7613}\inst{\ref{GALinst6}}\and
Minfeng Gu (\cntext{顾敏峰})\orcidlink{0000-0002-4455-6946}\inst{\ref{GALinst43},\ref{GALinst81}}\and
Mark Gurwell\orcidlink{0000-0003-0685-3621}\inst{\ref{GALinst4}}\and
Ronald Hesper\orcidlink{0000-0003-1918-6098}\inst{\ref{GALinst85}}\and
Dirk Heumann\orcidlink{0000-0002-7671-0047}\inst{\ref{GALinst18}}\and
Luis C. Ho (\cntext{何子山})\orcidlink{0000-0001-6947-5846}\inst{\ref{GALinst86},\ref{GALinst87}}\and
Paul Ho\orcidlink{0000-0002-3412-4306}\inst{\ref{GALinst15},\ref{GALinst24},\ref{GALinst23}}\and
Mareki Honma\orcidlink{0000-0003-4058-9000}\inst{\ref{GALinst3},\ref{GALinst88},\ref{GALinst89}}\and
Chih-Wei L. Huang\orcidlink{0000-0001-5641-3953}\inst{\ref{GALinst15}}\and
Lei Huang (\cntext{黄磊})\orcidlink{0000-0002-1923-227X}\inst{\ref{GALinst43},\ref{GALinst81}}\and
David H. Hughes\inst{\ref{GALinst78}}\and
Shiro Ikeda\orcidlink{0000-0002-2462-1448}\inst{\ref{GALinst90},\ref{GALinst91},\ref{GALinst92},\ref{GALinst93}}\and
C. M. Violette Impellizzeri\orcidlink{0000-0002-3443-2472}\inst{\ref{GALinst94},\ref{GALinst64}}\and
Makoto Inoue\orcidlink{0000-0001-5037-3989}\inst{\ref{GALinst15}}\and
Sara Issaoun\orcidlink{0000-0002-5297-921X}\inst{\ref{GALinst4},\ref{GALinst53}}\and
Yuhei Iwata\orcidlink{0000-0002-9255-4742}\inst{\ref{GALinst3},\ref{GALinst88}}\and
David J. James\orcidlink{0000-0001-5160-4486}\inst{\ref{GALinst95},\ref{GALinst96}}\and
Buell T. Jannuzi\orcidlink{0000-0002-1578-6582}\inst{\ref{GALinst18}}\and
Michael Janssen\orcidlink{0000-0001-8685-6544}\inst{\ref{GALinst32},\ref{GALinst7}}\and
Britton Jeter\orcidlink{0000-0003-2847-1712}\inst{\ref{GALinst97},\ref{GALinst98}}\and
Wu Jiang (\cntext{江悟})\orcidlink{0000-0001-7369-3539}\inst{\ref{GALinst43}}\and
Alejandra Jiménez-Rosales\orcidlink{0000-0002-2662-3754}\inst{\ref{GALinst32}}\and
Michael D. Johnson\orcidlink{0000-0002-4120-3029}\inst{\ref{GALinst4},\ref{GALinst12}}\and
Adam C. Jones\inst{\ref{GALinst22}}\and
Abhishek V. Joshi\orcidlink{0000-0002-2514-5965}\inst{\ref{GALinst20}}\and
Taehyun Jung\orcidlink{0000-0001-7003-8643}\inst{\ref{GALinst45},\ref{GALinst100}}\and
Tomohisa Kawashima\orcidlink{0000-0001-8527-0496}\inst{\ref{GALinst101}}\and
Garrett K. Keating\orcidlink{0000-0002-3490-146X}\inst{\ref{GALinst4}}\and
Mark Kettenis\orcidlink{0000-0002-6156-5617}\inst{\ref{GALinst102}}\and
Dong-Jin Kim\orcidlink{0000-0002-7038-2118}\inst{\ref{GALinst103}}\and
Jae-Young Kim\orcidlink{0000-0001-8229-7183}\inst{\ref{GALinst104}}\and
Jongsoo Kim\orcidlink{0000-0002-1229-0426}\inst{\ref{GALinst45}}\and
Junhan Kim\orcidlink{0000-0002-4274-9373}\inst{\ref{GALinst105}}\and
Motoki Kino\orcidlink{0000-0002-2709-7338}\inst{\ref{GALinst90},\ref{GALinst106}}\and
Jakob Knollmüller\orcidlink{0000-0002-9906-0040}\inst{\ref{GALinst107},\ref{GALinst32}}\and
Jun Yi Koay\orcidlink{0000-0002-7029-6658}\inst{\ref{GALinst108},\ref{GALinst15}}\and
Prashant Kocherlakota\orcidlink{0000-0001-7386-7439}\inst{\ref{GALinst12},\ref{GALinst4}}\and
Yutaro Kofuji\inst{\ref{GALinst6},\ref{GALinst3}}\and
Patrick M. Koch\orcidlink{0000-0003-2777-5861}\inst{\ref{GALinst15}}\and
Shoko Koyama\orcidlink{0000-0002-3723-3372}\inst{\ref{GALinst108},\ref{GALinst15}}\and
Carsten Kramer\orcidlink{0000-0002-4908-4925}\inst{\ref{GALinst28}}\and
Joana A. Kramer\orcidlink{0009-0003-3011-0454}\inst{\ref{GALinst109},\ref{GALinst7}}\and
Michael Kramer\orcidlink{0000-0002-4175-2271}\inst{\ref{GALinst7}}\and
Thomas P. Krichbaum\orcidlink{0000-0002-4892-9586}\inst{\ref{GALinst7}}\and
Cheng-Yu Kuo\orcidlink{0000-0001-6211-5581}\inst{\ref{GALinst110},\ref{GALinst15}}\and
Noemi La Bella\orcidlink{0000-0002-8116-9427}\inst{\ref{GALinst32}}\and
Deokhyeong Lee\orcidlink{0009-0003-2122-9437}\inst{\ref{GALinst46}}\and
Aviad Levis\orcidlink{0000-0001-7307-632X}\inst{\ref{GALinst50},\ref{GALinst111},\ref{GALinst112}}\and
Shaoliang Li\orcidlink{0009-0005-0338-9490}\inst{\ref{GALinst23},\ref{GALinst24}}\and
Zhiyuan Li (\cntext{李志远})\orcidlink{0000-0003-0355-6437}\inst{\ref{GALinst113},\ref{GALinst114}}\and
Rocco Lico\orcidlink{0000-0001-7361-2460}\inst{\ref{GALinst115},\ref{GALinst6}}\and
Greg Lindahl\orcidlink{0000-0002-6100-4772}\inst{\ref{GALinst116}}\and
Michael Lindqvist\orcidlink{0000-0002-3669-0715}\inst{\ref{GALinst17}}\and
Jun Liu (\cntext{刘俊})\orcidlink{0000-0002-7615-7499}\inst{\ref{GALinst7}}\and
Kuo Liu\orcidlink{0000-0002-2953-7376}\inst{\ref{GALinst118}}\and
Elisabetta Liuzzo\orcidlink{0000-0003-0995-5201}\inst{\ref{GALinst119}}\and
Wen-Ping Lo\orcidlink{0000-0003-1869-2503}\inst{\ref{GALinst15},\ref{GALinst120}}\and
Andrei P. Lobanov\orcidlink{0000-0003-1622-1484}\inst{\ref{GALinst7}}\and
Laurent Loinard\orcidlink{0000-0002-5635-3345}\inst{\ref{GALinst80},\ref{GALinst12}}\and
Colin J. Lonsdale\orcidlink{0000-0003-4062-4654}\inst{\ref{GALinst2}}\and
Amy E. Lowitz\orcidlink{0000-0002-4747-4276}\inst{\ref{GALinst18}}\and
Ru-Sen Lu (\cntext{路如森})\orcidlink{0000-0002-7692-7967}\inst{\ref{GALinst43},\ref{GALinst44},\ref{GALinst7}}\and
Nicholas R. MacDonald\orcidlink{0000-0002-6684-8691}\inst{\ref{GALinst121},\ref{GALinst7}}\and
Jirong Mao (\cntext{毛基荣})\orcidlink{0000-0002-7077-7195}\inst{\ref{GALinst122},\ref{GALinst123},\ref{GALinst124}}\and
Nicola Marchili\orcidlink{0000-0002-5523-7588}\inst{\ref{GALinst119},\ref{GALinst7}}\and
Daniel P. Marrone\orcidlink{0000-0002-2367-1080}\inst{\ref{GALinst18}}\and
Iván Martí-Vidal\orcidlink{0000-0003-3708-9611}\inst{\ref{GALinst5},\ref{GALinst16}}\and
Satoki Matsushita\orcidlink{0000-0002-2127-7880}\inst{\ref{GALinst15}}\and
Lynn D. Matthews\orcidlink{0000-0002-3728-8082}\inst{\ref{GALinst2}}\and
Lia Medeiros\orcidlink{0000-0003-2342-6728}\inst{\ref{GALinst9}}\and
Karl M. Menten\orcidlink{0000-0001-6459-0669}\inst{\ref{GALinst7},\ref{GALinst127}}\and
Hugo Messias\orcidlink{0000-0002-2985-7994}\inst{\ref{GALinst128},\ref{GALinst129}}\and
Izumi Mizuno\orcidlink{0000-0002-7210-6264}\inst{\ref{GALinst23},\ref{GALinst24}}\and
Yosuke Mizuno\orcidlink{0000-0002-8131-6730}\inst{\ref{GALinst61},\ref{GALinst130},\ref{GALinst68}}\and
Joshua Montgomery\orcidlink{0000-0003-0345-8386}\inst{\ref{GALinst84},\ref{GALinst22}}\and
Kotaro Moriyama\orcidlink{0000-0003-1364-3761}\inst{\ref{GALinst6},\ref{GALinst3}}\and
Monika Moscibrodzka\orcidlink{0000-0002-4661-6332}\inst{\ref{GALinst32}}\and
Wanga Mulaudzi\orcidlink{0000-0003-4514-625X}\inst{\ref{GALinst109}}\and
Cornelia Müller\orcidlink{0000-0002-2739-2994}\inst{\ref{GALinst7},\ref{GALinst32}}\and
Hendrik Müller\orcidlink{0000-0002-9250-0197}\inst{\ref{GALinst7}}\and
Alejandro Mus\orcidlink{0000-0003-0329-6874}\inst{\ref{GALinst72},\ref{GALinst115},\ref{GALinst131},\ref{GALinst132},\ref{GALinst133}}\and
Gibwa Musoke\orcidlink{0000-0003-1984-189X}\inst{\ref{GALinst109},\ref{GALinst32}}\and
Ioannis Myserlis\orcidlink{0000-0003-3025-9497}\inst{\ref{GALinst134}}\and
Hiroshi Nagai\orcidlink{0000-0003-0292-3645}\inst{\ref{GALinst90},\ref{GALinst88}}\and
Neil M. Nagar\orcidlink{0000-0001-6920-662X}\inst{\ref{GALinst19}}\and
Dhanya G. Nair\orcidlink{0000-0001-5357-7805}\inst{\ref{GALinst19},\ref{GALinst7}}\and
Masanori Nakamura\orcidlink{0000-0001-6081-2420}\inst{\ref{GALinst135},\ref{GALinst15}}\and
Gopal Narayanan\orcidlink{0000-0002-4723-6569}\inst{\ref{GALinst33}}\and
Iniyan Natarajan\orcidlink{0000-0001-8242-4373}\inst{\ref{GALinst136},\ref{GALinst4},\ref{GALinst12}}\and
Antonios Nathanail\orcidlink{0000-0002-1655-9912}\inst{\ref{GALinst13}}\and
Santiago Navarro Fuentes\inst{\ref{GALinst134}}\and
Joey Neilsen\orcidlink{0000-0002-8247-786X}\inst{\ref{GALinst137}}\and
Chunchong Ni\orcidlink{0000-0003-1361-5699}\inst{\ref{GALinst30},\ref{GALinst31},\ref{GALinst29}}\and
Andy Nilipour\orcidlink{0000-0002-5956-5167}\inst{\ref{GALinst138},\ref{GALinst139}}\and
Michael A. Nowak\orcidlink{0000-0001-6923-1315}\inst{\ref{GALinst140}}\and
Hiroki Okino\orcidlink{0000-0003-3779-2016}\inst{\ref{GALinst3},\ref{GALinst89}}\and
Héctor Raúl Olivares Sánchez\orcidlink{0000-0001-6833-7580}\inst{\ref{GALinst141}}\and
Feryal Özel\orcidlink{0000-0003-4413-1523}\inst{\ref{GALinst142}}\and
Daniel C. M. Palumbo\orcidlink{0000-0002-7179-3816}\inst{\ref{GALinst12},\ref{GALinst4}}\and
Georgios Filippos Paraschos\orcidlink{0000-0001-6757-3098}\inst{\ref{GALinst97},\ref{GALinst98},\ref{GALinst7}}\and
Jongho Park\orcidlink{0000-0001-6558-9053}\inst{\ref{GALinst143},\ref{GALinst144},\ref{GALinst15}}\and
Harriet Parsons\orcidlink{0000-0002-6327-3423}\inst{\ref{GALinst23},\ref{GALinst24}}\and
Nimesh Patel\orcidlink{0000-0002-6021-9421}\inst{\ref{GALinst4}}\and
Ue-Li Pen\orcidlink{0000-0003-2155-9578}\inst{\ref{GALinst15},\ref{GALinst29},\ref{GALinst51},\ref{GALinst112},\ref{GALinst145}}\and
Dominic W. Pesce\orcidlink{0000-0002-5278-9221}\inst{\ref{GALinst4},\ref{GALinst12}}\and
Vincent Piétu\orcidlink{0009-0006-3497-397X}\inst{\ref{GALinst28}}\and
Alexander Plavin\orcidlink{0000-0003-2914-8554}\inst{\ref{GALinst12},\ref{GALinst4},\ref{GALinst7}}\and
Aleksandar PopStefanija\inst{\ref{GALinst33}}\and
Oliver Porth\orcidlink{0000-0002-4584-2557}\inst{\ref{GALinst109},\ref{GALinst68}}\and
Cora Prather\orcidlink{0000-0002-0393-7734}\inst{\ref{GALinst12}}\and
Dimitrios Psaltis\orcidlink{0000-0003-1035-3240}\inst{\ref{GALinst142}}\and
Hung-Yi Pu\orcidlink{0000-0001-9270-8812}\inst{\ref{GALinst148},\ref{GALinst149},\ref{GALinst15}}\and
Alexandra Rahlin\orcidlink{0000-0003-3953-1776}\inst{\ref{GALinst22}}\and
Venkatessh Ramakrishnan\orcidlink{0000-0002-9248-086X}\inst{\ref{GALinst150},\ref{GALinst97},\ref{GALinst98}}\and
Ramprasad Rao\orcidlink{0000-0002-1407-7944}\inst{\ref{GALinst4}}\and
Mark G. Rawlings\orcidlink{0000-0002-6529-202X}\inst{\ref{GALinst64},\ref{GALinst23},\ref{GALinst24}}\and
Angelo Ricarte\orcidlink{0000-0001-5287-0452}\inst{\ref{GALinst12},\ref{GALinst4}}\and
Luca Ricci\orcidlink{0000-0002-4175-3194}\inst{\ref{GALinst153}}\and
Bart Ripperda\orcidlink{0000-0002-7301-3908}\inst{\ref{GALinst51},\ref{GALinst154},\ref{GALinst112},\ref{GALinst29}}\and
Jan Röder\orcidlink{0000-0002-2426-927X}\inst{\ref{GALinst6}}\and
Freek Roelofs\orcidlink{0000-0001-5461-3687}\inst{\ref{GALinst32}}\and
Cristina Romero-Cañizales\orcidlink{0000-0001-6301-9073}\inst{\ref{GALinst15}}\and
Eduardo Ros\orcidlink{0000-0001-9503-4892}\inst{\ref{GALinst7}}\and
Arash Roshanineshat\orcidlink{0000-0002-8280-9238}\inst{\ref{GALinst18}}\and
Helge Rottmann\inst{\ref{GALinst7}}\and
Alan L. Roy\orcidlink{0000-0002-1931-0135}\inst{\ref{GALinst7}}\and
Ignacio Ruiz\orcidlink{0000-0002-0965-5463}\inst{\ref{GALinst134}}\and
Chet Ruszczyk\orcidlink{0000-0001-7278-9707}\inst{\ref{GALinst2}}\and
Kazi L. J. Rygl\orcidlink{0000-0003-4146-9043}\inst{\ref{GALinst119}}\and
León D. S. Salas\orcidlink{0000-0003-1979-6363}\inst{\ref{GALinst109}}\and
Salvador Sánchez\orcidlink{0000-0002-8042-5951}\inst{\ref{GALinst134}}\and
David Sánchez-Argüelles\orcidlink{0000-0002-7344-9920}\inst{\ref{GALinst78},\ref{GALinst79}}\and
Miguel Sánchez-Portal\orcidlink{0000-0003-0981-9664}\inst{\ref{GALinst134}}\and
Ali SaraerToosi\orcidlink{0009-0003-4620-8448}\inst{\ref{GALinst50}}\and
Kaushik Satapathy\orcidlink{0000-0003-0433-3585}\inst{\ref{GALinst18}}\and
Saurabh\orcidlink{0000-0001-7156-4848}\inst{\ref{GALinst7}}\and
Tuomas Savolainen\orcidlink{0000-0001-6214-1085}\inst{\ref{GALinst157},\ref{GALinst98},\ref{GALinst7}}\and
Karl-Friedrich Schuster\orcidlink{0000-0003-2890-9454}\inst{\ref{GALinst158}}\and
Zhiqiang Shen (\cntext{沈志强})\orcidlink{0000-0003-3540-8746}\inst{\ref{GALinst43},\ref{GALinst44}}\and
Sasikumar Silpa\orcidlink{0000-0003-0667-7074}\inst{\ref{GALinst19}}\and
Randall Smith\orcidlink{0000-0003-4284-4167}\inst{\ref{GALinst4}}\and
Bong Won Sohn\orcidlink{0000-0002-4148-8378}\inst{\ref{GALinst45},\ref{GALinst100}}\and
Jason SooHoo\orcidlink{0000-0003-1938-0720}\inst{\ref{GALinst2}}\and
Kamal Souccar\orcidlink{0000-0001-7915-5272}\inst{\ref{GALinst33}}\and
Joshua S. Stanway\orcidlink{0009-0003-7659-4642}\inst{\ref{GALinst159}}\and
He Sun (\cntext{孙赫})\orcidlink{0000-0003-1526-6787}\inst{\ref{GALinst160},\ref{GALinst161}}\and
Alexandra J. Tetarenko\orcidlink{0000-0003-3906-4354}\inst{\ref{GALinst162}}\and
Paul Tiede\orcidlink{0000-0003-3826-5648}\inst{\ref{GALinst4},\ref{GALinst12}}\and
Remo P. J. Tilanus\orcidlink{0000-0002-6514-553X}\inst{\ref{GALinst18}}\and
Michael Titus\orcidlink{0000-0001-9001-3275}\inst{\ref{GALinst2}}\and
Kenji Toma\orcidlink{0000-0002-7114-6010}\inst{\ref{GALinst163},\ref{GALinst164}}\and
Pablo Torne\orcidlink{0000-0001-8700-6058}\inst{\ref{GALinst134},\ref{GALinst7}}\and
Teresa Toscano\orcidlink{0000-0003-3658-7862}\inst{\ref{GALinst6},\ref{GALinst7}}\and
Efthalia Traianou\orcidlink{0000-0002-1209-6500}\inst{\ref{GALinst6},\ref{GALinst7}}\and
Sascha Trippe\orcidlink{0000-0003-0465-1559}\inst{\ref{GALinst165},\ref{GALinst166}}\and
Matthew Turk\orcidlink{0000-0002-5294-0198}\inst{\ref{GALinst47}}\and
Akhil Uniyal\orcidlink{0000-0001-8213-646X}\inst{\ref{GALinst61}}\and
Ilse van Bemmel\orcidlink{0000-0001-5473-2950}\inst{\ref{GALinst167}}\and
Bram van den Berg\orcidlink{0009-0000-9340-4204}\inst{\ref{GALinst32}}\and
Huib Jan van Langevelde\orcidlink{0000-0002-0230-5946}\inst{\ref{GALinst102},\ref{GALinst94}}\and
Daniel R. van Rossum\orcidlink{0000-0001-7772-6131}\inst{\ref{GALinst32}}\and
Sebastiano D. von Fellenberg\orcidlink{0000-0002-9156-2249}\inst{\ref{GALinst51},\ref{GALinst7}}\and
Jesse Vos\orcidlink{0000-0003-3349-7394}\inst{\ref{GALinst168}}\and
Jan Wagner\orcidlink{0000-0003-1105-6109}\inst{\ref{GALinst7}}\and
Zhiren Wang\orcidlink{0009-0004-9417-2213}\inst{\ref{GALinst29},\ref{GALinst30},\ref{GALinst31}}\and
Derek Ward-Thompson\orcidlink{0000-0003-1140-2761}\inst{\ref{GALinst159}}\and
John Wardle\orcidlink{0000-0002-8960-2942}\inst{\ref{GALinst169}}\and
Jasmin E. Washington\orcidlink{0000-0002-7046-0470}\inst{\ref{GALinst18}}\and
Jonathan Weintroub\orcidlink{0000-0002-4603-5204}\inst{\ref{GALinst4},\ref{GALinst12}}\and
Maciek Wielgus\orcidlink{0000-0002-8635-4242}\inst{\ref{GALinst6}}\and
Kaj Wiik\orcidlink{0000-0002-0862-3398}\inst{\ref{GALinst170},\ref{GALinst97},\ref{GALinst98}}\and
Michael F. Wondrak\orcidlink{0000-0002-6894-1072}\inst{\ref{GALinst109},\ref{GALinst125},\ref{GALinst32},\ref{GALinst171}}\and
George N. Wong\orcidlink{0000-0001-6952-2147}\inst{\ref{GALinst172},\ref{GALinst38}}\and
Jompoj Wongphexhauxsorn\orcidlink{0000-0002-7730-4956}\inst{\ref{GALinst153},\ref{GALinst7}}\and
Qingwen Wu (\cntext{吴庆文})\orcidlink{0000-0003-4773-4987}\inst{\ref{GALinst173}}\and
Paul Yamaguchi\orcidlink{0000-0002-6017-8199}\inst{\ref{GALinst4}}\and
Aristomenis Yfantis\orcidlink{0000-0002-3244-7072}\inst{\ref{GALinst6}}\and
Doosoo Yoon\orcidlink{0000-0001-8694-8166}\inst{\ref{GALinst109}}\and
André Young\orcidlink{0000-0003-0000-2682}\inst{\ref{GALinst32}}\and
Ziri Younsi\orcidlink{0000-0001-9283-1191}\inst{\ref{GALinst174},\ref{GALinst68}}\and
Wei Yu (\cntext{于威})\orcidlink{0000-0002-5168-6052}\inst{\ref{GALinst4}}\and
Feng Yuan (\cntext{袁峰})\orcidlink{0000-0003-3564-6437}\inst{\ref{GALinst175}}\and
Ye-Fei Yuan (\cntext{袁业飞})\orcidlink{0000-0002-7330-4756}\inst{\ref{GALinst176}}\and
Ai-Ling Zeng (\cntext{曾艾玲})\orcidlink{0009-0000-9427-4608}\inst{\ref{GALinst6}}\and
J. Anton Zensus\orcidlink{0000-0001-7470-3321}\inst{\ref{GALinst7}}\and
Shuo Zhang\orcidlink{0000-0002-2967-790X}\inst{\ref{GALinst177}}\and
Brandon Zhao\orcidlink{0009-0005-3991-9879}\inst{\ref{GALinst25}}\and
Guang-Yao Zhao\orcidlink{0000-0002-4417-1659}\inst{\ref{GALinst7},\ref{GALinst6}}\and\\
{{(The MAGIC collaboration)}}\and\\
J.~Abhir\inst{\ref{MAGIC1}}\and
A.~Abhishek\inst{\ref{MAGIC2}}\and
V.~A.~Acciari\inst{\ref{MAGIC3}}\and
A.~Aguasca-Cabot\inst{\ref{MAGIC4}}\and
I.~Agudo\inst{\ref{MAGIC5}}\and
I.~Albanese\inst{\ref{MAGIC6}}\and
T.~Aniello\inst{\ref{MAGIC7}}\and
S.~Ansoldi\inst{\ref{MAGIC8},\ref{MAGIC36}}\and
L.~A.~Antonelli\inst{\ref{MAGIC7}}\and
A.~Arbet Engels\inst{\ref{PALinst38}}\and
C.~Arcaro\inst{\ref{MAGIC6}}\and
T.~T.~H.~Arnesen\inst{\ref{MAGIC10}}\and
A.~Babi\'c\inst{\ref{MAGIC11}}\and
C.~Bakshi\inst{\ref{MAGIC12}}\and
U.~Barres de Almeida\inst{\ref{MAGIC13}}\and
J.~A.~Barrio\inst{\ref{MAGIC14}}\and
L.~Barrios-Jim\'enez\inst{\ref{MAGIC10}}\and
I.~Batkovi\'c\inst{\ref{MAGIC6}}\and
J.~Becerra Gonz\'alez\thanks{MAGIC corresponding author, For questions concerning MAGIC results contact \href{mailto:contact.magic@mpp.mpg.de}{contact.magic@mpp.mpg.de}}\orcidlink{0000-0002-6729-9022}\inst{\ref{MAGIC10}}\and
W.~Bednarek\inst{\ref{MAGIC15}}\and
E.~Bernardini\inst{\ref{MAGIC6}}\and
J.~Bernete\inst{\ref{MAGIC16}}\and
A.~Berti\inst{\ref{PALinst38}}\and
J.~Besenrieder\inst{\ref{PALinst38}}\and
C.~Bigongiari\inst{\ref{MAGIC7}}\and
A.~Biland\inst{\ref{MAGIC1}}\and
O.~Blanch\inst{\ref{MAGIC3}}\and
G.~Bonnoli\inst{\ref{MAGIC7}}\and
\v{Z}.~Bo\v{s}njak\inst{\ref{MAGIC11}}\and
E.~Bronzini\inst{\ref{MAGIC7}}\and
I.~Burelli\inst{\ref{MAGIC3}}\and
A.~Campoy-Ordaz\inst{\ref{MAGIC17}}\and
A.~Carosi\inst{\ref{MAGIC7}}\and
R.~Carosi\inst{\ref{MAGIC18}}\and
M.~Carretero-Castrillo\inst{\ref{MAGIC4}}\and
A.~J.~Castro-Tirado\inst{\ref{MAGIC5}}\and
D.~Cerasole\inst{\ref{MAGIC19}}\and
G.~Ceribella\inst{\ref{PALinst38}}\and
A.~Cervi\~no\inst{\ref{MAGIC14}}\and
Y.~Chai\inst{\ref{MAGIC20}}\and
G.~Chon\inst{\ref{PALinst38}}\and
J.~L.~Contreras\inst{\ref{MAGIC14}}\and
J.~Cortina\inst{\ref{MAGIC16}}\and
S.~Covino\inst{\ref{MAGIC7},\ref{MAGIC37}}\and
P.~Da Vela\inst{\ref{MAGIC7}}\and
F.~Dazzi\inst{\ref{MAGIC7}}\and
A.~De Angelis\inst{\ref{MAGIC6}}\and
B.~De Lotto\inst{\ref{MAGIC8}}\and
M.~Delfino\inst{\ref{MAGIC3},\ref{MAGIC38}}\and
J.~Delgado\inst{\ref{MAGIC3},\ref{MAGIC38}}\and
C.~Delgado Mendez\inst{\ref{MAGIC16}}\and
F.~Di Pierro\inst{\ref{MAGIC21}}\and
R.~Di Tria\inst{\ref{MAGIC19}}\and
L.~Di Venere\inst{\ref{MAGIC19}}\and
A.~Dinesh\inst{\ref{MAGIC14}}\and
D.~Dominis Prester\inst{\ref{MAGIC22}}\and
A.~Donini\inst{\ref{MAGIC7}}\and
D.~Dorner\inst{\ref{MAGIC23}}\and
M.~Doro\inst{\ref{MAGIC6}}\and
L.~Eisenberger\inst{\ref{MAGIC23}}\and
D.~Elsaesser\inst{\ref{MAGIC24}}\and
L.~Foffano\inst{\ref{MAGIC7}}\and
L.~Font\inst{\ref{MAGIC17}}\and
F.~Fr\'ias Garc\'ia-Lago\inst{\ref{MAGIC10}}\and
S.~Fr\"ose\inst{\ref{MAGIC24}}\and
Y.~Fukazawa\inst{\ref{MAGIC25}}\and
S.~Gasparyan\inst{\ref{MAGIC26}}\and
M.~Gaug\inst{\ref{MAGIC17}}\and
J.~G.~Giesbrecht Paiva\inst{\ref{MAGIC13}}\and
N.~Giglietto\inst{\ref{MAGIC19}}\and
F.~Giordano\inst{\ref{MAGIC19}}\and
P.~Gliwny\inst{\ref{MAGIC15}}\and
T.~Gradetzke\inst{\ref{MAGIC24}}\and
R.~Grau\inst{\ref{MAGIC20}}\and
J.~G.~Green\inst{\ref{PALinst38}}\and
P.~G\"unther\inst{\ref{MAGIC23}}\and
D.~Hadasch\inst{\ref{MAGIC3}}\and
G.~Harutyunyan\inst{\ref{MAGIC26}}\and
J.~Herrera Llorente\inst{\ref{MAGIC10}}\and
D.~Hrupec\inst{\ref{MAGIC27}}\and
D.~Israyelyan\inst{\ref{MAGIC26}}\and
J.~Jahanvi\inst{\ref{MAGIC8}}\and
I.~Jim\'enez Mart\'inez\inst{\ref{PALinst38}}\and
J.~Jim\'enez Quiles\inst{\ref{MAGIC3}}\and
S.~Kankkunen\inst{\ref{MAGIC28}}\and
T.~Kayanoki\inst{\ref{MAGIC25}}\and
J.~Konrad\inst{\ref{MAGIC24}}\and
P.~M.~Kouch\inst{\ref{MAGIC28}}\and
H.~Kubo\inst{\ref{MAGIC20}}\and
J.~Kushida\inst{\ref{MAGIC29}}\and
M.~L\'ainez\inst{\ref{MAGIC14}}\and
A.~Lamastra\inst{\ref{MAGIC7}}\and
E.~Lindfors\inst{\ref{MAGIC28}}\and
S.~Lombardi\inst{\ref{MAGIC7}}\and
F.~Longo\inst{\ref{MAGIC8},\ref{MAGIC39}}\and
R.~L\'opez-Coto\inst{\ref{MAGIC5}}\and
M.~L\'opez-Moya\inst{\ref{MAGIC14}}\and
A.~L\'opez-Oramas\inst{\ref{MAGIC10}}\and
S.~Loporchio\inst{\ref{MAGIC19}}\and
L.~Luli\'c\inst{\ref{MAGIC22}}\and
P.~Majumdar\inst{\ref{MAGIC12}}\and
M.~Makariev\inst{\ref{MAGIC30}}\and
M.~Mallamaci\inst{\ref{MAGIC31}}\and
G.~Maneva\inst{\ref{MAGIC30}}\and
M.~Manganaro\inst{\ref{MAGIC22}}\and
S.~Mangano\inst{\ref{MAGIC16}}\and
K.~Mannheim\inst{\ref{MAGIC23}}\and
S.~Marchesi\inst{\ref{MAGIC7}}\and
M.~Mariotti\inst{\ref{MAGIC6}}\and
M.~Mart\'inez\inst{\ref{MAGIC3}}\and
P.~Maru\v{s}evec\inst{\ref{MAGIC11}}\and
D.~Mazin\inst{\ref{MAGIC20},\ref{PALinst38}}\and
S.~Menchiari\inst{\ref{MAGIC5}}\and
J.~M\'endez Gallego\inst{\ref{MAGIC5}}\and
S.~Menon\inst{\ref{MAGIC7},\ref{MAGIC40}}\and
D.~Miceli\inst{\ref{MAGIC6}}\and
J.~M.~Miranda\inst{\ref{MAGIC2}}\and
R.~Mirzoyan\inst{\ref{PALinst38}}\and
M.~Molero Gonz\'alez\thanks{MAGIC corresponding author, For questions concerning MAGIC results contact \href{mailto:contact.magic@mpp.mpg.de}{contact.magic@mpp.mpg.de}}\orcidlink{0000-0003-0967-715X}\inst{\ref{MAGIC16}}\and
E.~Molina\inst{\ref{MAGIC10}}\and
H.~A.~Mondal\inst{\ref{MAGIC20}}\and
A.~Moralejo\inst{\ref{MAGIC3}}\and
C.~Nanci\inst{\ref{MAGIC7}}\and
A.~Negro\inst{\ref{MAGIC21}}\and
V.~Neustroev\inst{\ref{MAGIC32}}\and
C.~Nigro\inst{\ref{MAGIC3}}\and
L.~Nikoli\'c\inst{\ref{MAGIC2}}\and
K.~Noda\inst{\ref{MAGIC20}}\and
S.~Nozaki\inst{\ref{MAGIC20}}\and
A.~Okumura\inst{\ref{MAGIC33}}\and
J.~Otero-Santos\inst{\ref{MAGIC6}}\and
S.~Paiano\inst{\ref{MAGIC7}}\and
D.~Paneque\inst{\ref{PALinst38}}\and
R.~Paoletti\inst{\ref{MAGIC2}}\and
J.~M.~Paredes\inst{\ref{MAGIC4}}\and
M.~Peresano\inst{\ref{PALinst38}}\and
M.~Persic\inst{\ref{MAGIC8},\ref{MAGIC41}}\and
M.~Pihet\inst{\ref{MAGIC5}}\and
G.~Pirola\inst{\ref{PALinst38}}\and
F.~Podobnik\inst{\ref{MAGIC2}}\and
P.~G.~Prada Moroni\inst{\ref{MAGIC18}}\and
E.~Prandini\inst{\ref{MAGIC6}}\and
M.~Rib\'o\inst{\ref{MAGIC4}}\and
J.~Rico\inst{\ref{MAGIC3}}\and
A.~Roy\inst{\ref{MAGIC25}}\and
N.~Sahakyan\inst{\ref{MAGIC26}}\and
F.~G.~Saturni\inst{\ref{MAGIC7}}\and
F.~Schiavone\inst{\ref{MAGIC19}}\and
K.~Schmitz\inst{\ref{MAGIC24}}\and
A.~Sciaccaluga\inst{\ref{MAGIC7}}\and
G.~Silvestri\inst{\ref{MAGIC6}}\and
A.~Simongini\inst{\ref{MAGIC7}}\and
J.~Sitarek\inst{\ref{MAGIC15}}\and
V.~Sliusar\inst{\ref{MAGIC34}}\and
D.~Sobczynska\inst{\ref{MAGIC15}}\and
A.~Stamerra\inst{\ref{MAGIC7}}\and
J.~Stri\v{s}kovi\'c\inst{\ref{MAGIC27}}\and
D.~Strom\inst{\ref{PALinst38}}\and
Y.~Suda\inst{\ref{MAGIC25}}\and
M.~Takahashi\inst{\ref{MAGIC33}}\and
R.~Takeishi\inst{\ref{MAGIC20}}\and
J.~Tartera Barber\`a\inst{\ref{MAGIC3}}\and
P.~Temnikov\inst{\ref{MAGIC30}}\and
T.~Terzi\'c\inst{\ref{MAGIC22}}\and
M.~Teshima\inst{\ref{PALinst38},\ref{MAGIC20}}\and
A.~Tutone\inst{\ref{MAGIC7}}\and
S.~Ubach\inst{\ref{MAGIC17}}\and
S.~Ventura\inst{\ref{MAGIC2}}\and
G.~Verna\inst{\ref{MAGIC2}}\and
I.~Viale\inst{\ref{MAGIC21}}\and
A.~Vigliano\inst{\ref{MAGIC8}}\and
C.~F.~Vigorito\inst{\ref{MAGIC21}}\and
E.~Visentin\inst{\ref{MAGIC21}}\and
V.~Vitale\inst{\ref{MAGIC35}}\and
M.~Vorbrugg\inst{\ref{MAGIC23}}\and
I.~Vovk\inst{\ref{MAGIC20}}\and
R.~Walter\inst{\ref{MAGIC34}}\and
C.~Walther\inst{\ref{MAGIC24}}\and
F.~Wersig\inst{\ref{MAGIC24}}\and
P.~K.~H.~Yeung\inst{\ref{MAGIC20}}\and
{{(Other Authors)}}\and\\
A.~Acharyya\orcidlink{0000-0002-2028-9230}\inst{\ref{HESSUSD}}\and
F.~Aharonian\orcidlink{0000-0003-1157-3915}\inst{\ref{HESSYSU},\ref{HESSDIAS}}\and
H.~Ashkar\orcidlink{0000-0002-2153-1818}\inst{\ref{HESSLLR}}\and
M.~Backes\orcidlink{0000-0002-9326-6400}\inst{\ref{HESSUNAM},\ref{HESSNWU}}\and
M.~Barnard\orcidlink{0000-0003-1720-7959}\inst{\ref{HESSNWU}}\and
Y.~Becherini\orcidlink{0000-0002-2115-2930}\inst{\ref{HESSAPC}}\and
M.~B\"ottcher\orcidlink{0000-0002-8434-5692}\inst{\ref{HESSNWU}}\and
J.~Bolmont\orcidlink{0000-0003-4739-8389}\inst{\ref{HESSLPNHE}}\and
B.~Bruno\inst{\ref{HESSECAP}}\and
T.~Bylund\orcidlink{0000-0003-2946-1313}\inst{\ref{HESSLUX}}\and
S.~Casanova\orcidlink{0000-0002-6144-9122}\inst{\ref{HESSIFJPAN}}\and
M.~Chernyakova\orcidlink{0000-0002-9735-3608}\inst{\ref{HESSDCU},\ref{HESSDIAS}}\and
J. O.~Chibueze\orcidlink{0000-0002-9875-7436}\inst{\ref{HESSNWU},\ref{HESSUNAM}}\and
O.~Chibueze\orcidlink{0000-0001-8601-2675}\inst{\ref{HESSNWU}}\and
B.~Cornejo\orcidlink{0009-0003-0039-0483}\inst{\ref{HESSIRFU}}\and
J.~Damascene~Mbarubucyeye\orcidlink{0000-0002-4991-6576}\inst{\ref{HESSDESY}}\and
I.D.~Davids\orcidlink{0000-0002-6476-964X}\inst{\ref{HESSUNAM}}\and
J.~de~Assis~Scarpin\orcidlink{0009-0004-4411-236X}\inst{\ref{HESSLLR}}\and
M.~de~Naurois\orcidlink{0000-0002-7245-201X}\inst{\ref{HESSLLR}}\and
A.~Dmytriiev\orcidlink{0000-0003-0102-5579}\inst{\ref{HESSWits}}\and
K.~Egberts\orcidlink{0009-0000-5511-7060}\inst{\ref{HESSUP}}\and
S.~Fegan\orcidlink{0000-0002-9978-2510}\inst{\ref{HESSLLR}}\and
K.~Feijen\orcidlink{0000-0003-1476-3714}\inst{\ref{HESSAPC}}\and
M.~D.~Filipovic\orcidlink{0000-0002-4990-9288}\inst{\ref{HESSSydney}}\and
G.~Fontaine\orcidlink{0000-0002-6443-5025}\inst{\ref{HESSLLR}}\and
S.~Gabici\inst{\ref{HESSAPC}}\and
J.F.~Glicenstein\orcidlink{0000-0003-2581-1742}\inst{\ref{HESSIRFU}}\and
P.~Goswami\orcidlink{0000-0001-5430-4374}\inst{\ref{HESSLSW}}\and
L.~Heckmann\orcidlink{0000-0002-6653-8407}\inst{\ref{HESSAPC}}\and
B.~He\ss{}\orcidlink{0009-0004-9999-171X}\inst{\ref{HESSIAAT}}\and
M.~Holler\orcidlink{0000-0002-0107-8657}\inst{\ref{HESSInnsbruck}}\and
D.~Horns\orcidlink{0000-0003-1945-0119}\inst{\ref{HESSUHAM}}\and
E.~Kasai\orcidlink{0000-0001-9696-7221}\inst{\ref{HESSUNAM}}\and
K.~Katarzy{\'n}ski\orcidlink{0000-0002-8806-4863}\inst{\ref{HESSNCUT}}\and
D.~Kerszberg\orcidlink{0000-0002-5289-1509}\inst{\ref{HESSLPNHE}}\and
B. Kh\'{e}lifi\orcidlink{0000-0001-6876-5577}\inst{\ref{HESSAPC}}\and
K.~Kosack\orcidlink{0000-0001-8424-3621}\inst{\ref{HESSIRFU}}\and
R.G.~Lang\orcidlink{0000-0003-0492-5628}\inst{\ref{HESSECAP}}\and
S.~Lazarevi\'c\orcidlink{0000-0001-6109-8548}\inst{\ref{HESSSydney}}\and
P.~Liniewicz\orcidlink{0009-0008-3575-3965}\inst{\ref{HESSOAUJ}}\and
A.~Luashvili\thanks{For questions concerning H.E.S.S.\ results}\orcidlink{0000-0003-4384-1638}\inst{\ref{HESSNWU}}\and
D.~Malyshev\orcidlink{0000-0001-9689-2194}\inst{\ref{HESSIAAT}}\and
D.~Malyshev\orcidlink{0000-0002-9102-4854}\inst{\ref{HESSECAP}}\and
M.~G.~F.~Mayer\orcidlink{0000-0002-9771-9841}\inst{\ref{HESSECAP}}\and
M.~Meyer\orcidlink{0000-0002-0738-7581}\inst{\ref{HESSUSD}}\and
A.~Mikhno\orcidlink{0000-0002-9996-914X}\inst{\ref{HESSLPNHE}}\and
E.~Moulin\orcidlink{0000-0003-4007-0145}\inst{\ref{HESSIRFU}}\and
H.~Ndiyavala\orcidlink{0000-0001-9279-1775}\inst{\ref{HESSUNAM},\ref{HESSNWU}}\and
J.~Niemiec\orcidlink{0000-0001-6036-8569}\inst{\ref{HESSIFJPAN}}\and
P.~Pichard\orcidlink{0009-0005-9803-0762}\inst{\ref{HESSAPC}}\and
T.~Preis\orcidlink{0009-0001-7110-6764}\inst{\ref{HESSInnsbruck}}\and
G.~P\"uhlhofer\orcidlink{0000-0003-4632-4644}\inst{\ref{HESSIAAT}}\and
A.~Quirrenbach\inst{\ref{HESSLSW}}\and
A.~Reimer\orcidlink{0000-0001-8604-7077}\inst{\ref{HESSInnsbruck}}\and
O.~Reimer\orcidlink{0000-0001-6953-1385}\inst{\ref{HESSInnsbruck}}\and
I.~Reis\orcidlink{0000-0002-0771-3332}\inst{\ref{HESSIRFU}}\and
B.~Rudak\orcidlink{0000-0003-0452-3805}\inst{\ref{HESSNCAC}}\and
K.~Sabri\inst{\ref{HESSMNTPL}}\and
V.~Sahakian\orcidlink{0000-0003-1198-0043}\inst{\ref{HESSYPI}}\and
A.~Santangelo\orcidlink{0000-0003-4187-9560}\inst{\ref{HESSIAAT}}\and
M.~Sasaki\orcidlink{0000-0001-5302-1866}\inst{\ref{HESSECAP}}\and
I.~Shebalkova\inst{\ref{HESSDCU}}\and
W.~Si~Said\orcidlink{0009-0007-6555-6893}\inst{\ref{HESSLLR}}\and
{\L.}~Stawarz\orcidlink{0000-0002-7263-7540}\inst{\ref{HESSOAUJ}}\and
R.~Steenkamp\orcidlink{0009-0009-4130-977X}\inst{\ref{HESSUNAM}}\and
T.~Tanaka\orcidlink{0000-0002-4383-0368}\inst{\ref{HESSKonan}}\and
G.~L.~Taylor\orcidlink{0009-0001-8062-036X}\inst{\ref{HESSLSW}}\and
R.~Terrier\orcidlink{0000-0002-8219-4667}\inst{\ref{HESSAPC}}\and
Y.~Tian\orcidlink{0009-0005-7165-3791}\inst{\ref{HESSDESY}}\and
M.~Tluczykont\inst{\ref{HESSUHAM}}\and
C.~Venter\orcidlink{0000-0002-2666-4812}\inst{\ref{HESSNWU}}\and
J.~Vink\orcidlink{0000-0002-4708-4219}\inst{\ref{HESSGRAPPA}}\and
V.~Voitsekhovskyi\orcidlink{0000-0002-3906-4840}\inst{\ref{HESSGRAPPA}}\and
S.J.~Wagner\thanks{For questions concerning H.E.S.S.\ results}\orcidlink{0000-0002-7474-6062}\inst{\ref{HESSLSW}}\and
A.~Wierzcholska\orcidlink{0000-0003-4472-7204}\inst{\ref{HESSIFJPAN},\ref{HESSLSW}}\and
M.~Zacharias\orcidlink{0000-0001-5801-3945}\inst{\ref{HESSLSW},\ref{HESSNWU}}\and
A.A.~Zdziarski\orcidlink{0000-0002-0333-2452}\inst{\ref{HESSNCAC}}
}
\institute{
Dipartimento di Fisica, Universit\'a di Trieste, I-34127 Trieste, Italy\label{PALinst31}\and
Istituto Nazionale di Fisica Nucleare, Sezione di Trieste, I-34127 Trieste, Italy\label{PALinst32}\and
INAF Istituto di Radioastronomia, Via P. Gobetti, 101, I-40129 Bologna, Italy\label{PALinst8}\and
Department of Physics, Faculty of Science, University of Malaya, 50603 Kuala Lumpur, Malaysia\label{PALinst0}\and
Korea Astronomy and Space Science Institute, 776 Daedeok-daero, Daejeon 34055, Korea \label{PALinst43}\and
University of Science and Technology, 217 Gajeong-ro, Yuseong-gu, Daejeon 34113, Korea \label{PALinst44}\and
Mizusawa VLBI Observatory, National Astronomical Observatory of Japan, 2-12 Hoshigaoka, Mizusawa, Oshu, Iwate 023-0861, Japan\label{PALinst7}\and
Graduate School of Science, Nagoya City University, Yamanohata 1, Mizuho-cho, Mizuho-ku, Nagoya, 467-8501, Aichi, Japan \label{PALinst37}\and
Department of Physics, McGill University, 3600 University Street, Montreal, QC H3A 2T8, Canada\label{PALinst10}\and
Trottier Space Institute at McGill, 3550 rue University, Montréal,  QC H3A 2A7, Canada\label{PALinst16}\and
Max-Planck-Institut f\"ur Physik, D-85748 Garching, Germany\label{PALinst38}\and
Institute for Astrophysical Research, Boston University, 725 Commonwealth Ave., Boston 02215, MA, USA\label{PALinst17}\and
Instituto de F\'isica, Pontificia Universidad Cat\'olica de Valpara\'iso, Casilla 4059, Valpara\'iso, Chile\label{PALinst39}\and
API--Anton Pannekoek Institute for Astronomy, University of Amsterdam, Science Park 904, 1098 XH Amsterdam, The Netherlands\label{PALinst25}\and
GRAPPA--Gravitation and AstroParticle Physics Amsterdam, University of Amsterdam, Science Park 904, 1098 XH Amsterdam, The Netherlands\label{PALinst26}\and
Hiroshima Astrophysical Science Center, Hiroshima University, 1-3-1 Kagamiyama, Higashi-Hiroshima, Hiroshima 739-8526, Japan\label{PALinst36}\and
Department of Physics, Tokyo Institute of Technology, 2-12-1 Ookayama, Meguro-ku, Tokyo 152-8551, Japan\label{PALinst42}\and
Institute of Astronomy, University of Cambridge, Madingley Road, Cambridge CB3 0HA, United Kingdom\label{PALinst41}\and
Institute of Sensors, Signals and Systems, Heriot-Watt University, Edinburgh EH14 4AS, United Kingdom\label{GALinst1}\and
Massachusetts Institute of Technology Haystack Observatory, 99 Millstone Road, Westford, MA 01886, USA\label{GALinst2}\and
Mizusawa VLBI Observatory, National Astronomical Observatory of Japan, 2-12 Hoshigaoka, Mizusawa, Oshu, Iwate 023-0861, Japan\label{GALinst3}\and
Center for Astrophysics $|$ Harvard \& Smithsonian, 60 Garden Street, Cambridge, MA 02138, USA\label{GALinst4}\and
Departament d'Astronomia i Astrofísica, Universitat de València, C. Dr. Moliner 50, E-46100 Burjassot, València, Spain\label{GALinst5}\and
Instituto de Astrofísica de Andalucía-CSIC, Glorieta de la Astronomía s/n, E-18008 Granada, Spain\label{GALinst6}\and
Max-Planck-Institut für Radioastronomie, Auf dem Hügel 69, D-53121 Bonn, Germany\label{GALinst7}\and
Centre for Astronomy and Astrophysics Research, Department of Physics, Faculty of Science, Universiti Malaya, 50603 Kuala Lumpur, Malaysia\label{GALinst8}\and
Center for Gravitation, Cosmology and Astrophysics, Department of Physics, University of Wisconsin–Milwaukee, P.O. Box 413, Milwaukee, WI 53201, USA\label{GALinst9}\and
Department of Physics \& Astronomy, The University of Texas at San Antonio, One UTSA Circle, San Antonio, TX 78249, USA\label{GALinst10}\and
Physics \& Astronomy Department, Rice University, Houston, TX 77005-1827, USA\label{GALinst11}\and
Black Hole Initiative at Harvard University, 20 Garden Street, Cambridge, MA 02138, USA\label{GALinst12}\and
Research Center for Astronomy, Academy of Athens, Soranou Efessiou 4, 115 27 Athens, Greece\label{GALinst13}\and
Department of Physics, National and Kapodistrian University of Athens, Panepistimiopolis, GR 15783 Zografos, Greece\label{GALinst14}\and
Institute of Astronomy and Astrophysics, Academia Sinica, 11F of Astronomy-Mathematics Building, AS/NTU No. 1, Sec. 4, Roosevelt Rd., Taipei 106216, Taiwan, R.O.C.\label{GALinst15}\and
Observatori Astronòmic, Universitat de València, C. Catedrático José Beltrán 2, E-46980 Paterna, València, Spain\label{GALinst16}\and
Department of Physics and Astronomy, Chalmers University of Technology, Onsala Space Observatory, SE-439 92 Onsala, Sweden\label{GALinst17}\and
Steward Observatory and Department of Astronomy, University of Arizona, 933 N. Cherry Ave., Tucson, AZ 85721, USA\label{GALinst18}\and
Astronomy Department, Universidad de Concepción, Casilla 160-C, Concepción, Chile\label{GALinst19}\and
Department of Physics, University of Illinois, 1110 West Green Street, Urbana, IL 61801, USA\label{GALinst20}\and
Fermi National Accelerator Laboratory, MS209, P.O. Box 500, Batavia, IL 60510, USA\label{GALinst21}\and
Department of Astronomy and Astrophysics, University of Chicago, 5640 South Ellis Avenue, Chicago, IL 60637, USA\label{GALinst22}\and
East Asian Observatory, 660 N. A'ohoku Place, Hilo, HI 96720, USA\label{GALinst23}\and
James Clerk Maxwell Telescope (JCMT), 660 N. A'ohoku Place, Hilo, HI 96720, USA\label{GALinst24}\and
California Institute of Technology, 1200 East California Boulevard, Pasadena, CA 91125, USA\label{GALinst25}\and
Institute of Astronomy and Astrophysics, Academia Sinica, 645 N. A'ohoku Place, Hilo, HI 96720, USA\label{GALinst26}\and
Department of Physics and Astronomy, University of Hawaii at Manoa, 2505 Correa Road, Honolulu, HI 96822, USA\label{GALinst27}\and
Institut de Radioastronomie Millimétrique (IRAM), 300 rue de la Piscine, F-38400 Saint-Martin-d'Hères, France\label{GALinst28}\and
Perimeter Institute for Theoretical Physics, 31 Caroline Street North, Waterloo, ON N2L 2Y5, Canada\label{GALinst29}\and
Department of Physics and Astronomy, University of Waterloo, 200 University Avenue West, Waterloo, ON N2L 3G1, Canada\label{GALinst30}\and
Waterloo Centre for Astrophysics, University of Waterloo, Waterloo, ON N2L 3G1, Canada\label{GALinst31}\and
Department of Astrophysics, Institute for Mathematics, Astrophysics and Particle Physics (IMAPP), Radboud University, P.O. Box 9010, 6500 GL Nijmegen, The Netherlands\label{GALinst32}\and
Department of Astronomy, University of Massachusetts, Amherst, MA 01003, USA\label{GALinst33}\and
Instituto de Astronomia, Geofísica e Ciências Atmosféricas, Universidade de São Paulo, R. do Matão, 1226, São Paulo, SP 05508-090, Brazil\label{GALinst34}\and
Kavli Institute for Cosmological Physics, University of Chicago, 5640 South Ellis Avenue, Chicago, IL 60637, USA\label{GALinst35}\and
Department of Physics, University of Chicago, 5720 South Ellis Avenue, Chicago, IL 60637, USA\label{GALinst36}\and
Enrico Fermi Institute, University of Chicago, 5640 South Ellis Avenue, Chicago, IL 60637, USA\label{GALinst37}\and
Princeton Gravity Initiative, Jadwin Hall, Princeton University, Princeton, NJ 08544, USA\label{GALinst38}\and
Data Science Institute, University of Arizona, 1230 N. Cherry Ave., Tucson, AZ 85721, USA\label{GALinst39}\and
Program in Applied Mathematics, University of Arizona, 617 N. Santa Rita, Tucson, AZ 85721, USA\label{GALinst40}\and
Department of Astronomy, University of Geneva, Chemin Pegasi 51, 1290 Versoix, Switzerland\label{GALinst41}\and
Department of Physics, University of Maryland, 7901 Regents Drive, College Park, MD 20742, USA\label{GALinst42}\and
Shanghai Astronomical Observatory, Chinese Academy of Sciences, 80 Nandan Road, Shanghai 200030, People's Republic of China\label{GALinst43}\and
Key Laboratory of Radio Astronomy and Technology, Chinese Academy of Sciences, A20 Datun Road, Chaoyang District, Beijing, 100101, People’s Republic of China\label{GALinst44}\and
Korea Astronomy and Space Science Institute, Daedeok-daero 776, Yuseong-gu, Daejeon 34055, Republic of Korea\label{GALinst45}\and
Department of Astronomy, Kyungpook National University, 80 Daehak-ro, Buk-gu, Daegu 41566, Republic of Korea\label{GALinst46}\and
Department of Astronomy, University of Illinois at Urbana-Champaign, 1002 West Green Street, Urbana, IL 61801, USA\label{GALinst47}\and
Instituto de Astronomía, Universidad Nacional Autónoma de México (UNAM), Apdo Postal 70-264, Ciudad de México, México\label{GALinst48}\and
Institute of Astrophysics, Central China Normal University, Wuhan 430079, People's Republic of China\label{GALinst49}\and
Department of Computer Science, University of Toronto, 40 St. George St., Toronto, ON, M5S 2E4, Canada\label{GALinst50}\and
Canadian Institute for Theoretical Astrophysics, University of Toronto, 60 St. George Street, Toronto, ON M5S 3H8, Canada\label{GALinst51}\and
Department of Astrophysical Sciences, Peyton Hall, Princeton University, Princeton, NJ 08544, USA\label{GALinst52}\and
NASA Hubble Fellowship Program, Einstein Fellow\label{GALinst53}\and
Dipartimento di Fisica ``E. Pancini'', Università di Napoli ``Federico II'', Compl. Univ. di Monte S. Angelo, Edificio G, Via Cinthia, I-80126, Napoli, Italy\label{GALinst54}\and
INFN Sez. di Napoli, Compl. Univ. di Monte S. Angelo, Edificio G, Via Cinthia, I-80126, Napoli, Italy\label{GALinst55}\and
Wits Centre for Astrophysics, University of the Witwatersrand, 1 Jan Smuts Avenue, Braamfontein, Johannesburg 2050, South Africa\label{GALinst56}\and
Department of Physics, University of Pretoria, Hatfield, Pretoria 0028, South Africa\label{GALinst57}\and
Centre for Radio Astronomy Techniques and Technologies, Department of Physics and Electronics, Rhodes University, Makhanda 6140, South Africa\label{GALinst58}\and
JILA and Department of Astrophysical and Planetary Sciences, University of Colorado, Boulder, CO 80309, USA\label{GALinst59}\and
Institute of Fundamental Physics and Quantum Technology, \& School of Physical Science and Technology, Ningbo University, Ningbo, Zhejiang 315211, People’s Republic of China\label{GALinst60}\and
Tsung-Dao Lee Institute, Shanghai Jiao Tong University, Shengrong Road 520, Shanghai, 201210, People’s Republic of China\label{GALinst61}\and
Las Cumbres Observatory, 6740 Cortona Drive, Suite 102, Goleta, CA 93117-5575, USA\label{GALinst62}\and
Department of Physics, University of California, Santa Barbara, CA 93106-9530, USA\label{GALinst63}\and
National Radio Astronomy Observatory, 520 Edgemont Road, Charlottesville, VA 22903, USA\label{GALinst64}\and
Department of Electrical Engineering and Computer Science, Massachusetts Institute of Technology, 32-D476, 77 Massachusetts Ave., Cambridge, MA 02142, USA\label{GALinst65}\and
Google Research, 355 Main St., Cambridge, MA 02142, USA\label{GALinst66}\and
Institut für Theoretische Physik und Astrophysik, Universität Würzburg, Emil-Fischer-Str. 31, D-97074 Würzburg, Germany\label{GALinst67}\and
Institut für Theoretische Physik, Goethe-Universität Frankfurt, Max-von-Laue-Straße 1, D-60438 Frankfurt am Main, Germany\label{GALinst68}\and
Department of History of Science, Harvard University, Cambridge, MA 02138, USA\label{GALinst69}\and
Department of Physics, Harvard University, Cambridge, MA 02138, USA\label{GALinst70}\and
NCSA, University of Illinois, 1205 W. Clark St., Urbana, IL 61801, USA\label{GALinst71}\and
Dipartimento di Fisica, Università degli Studi di Cagliari, SP Monserrato-Sestu km 0.7, I-09042 Monserrato (CA), Italy\label{GALinst72}\and
INAF - Osservatorio Astronomico di Cagliari, via della Scienza 5, I-09047 Selargius (CA), Italy\label{GALinst73}\and
INFN, sezione di Cagliari, I-09042 Monserrato (CA), Italy\label{GALinst74}\and
Institute for Mathematics and Interdisciplinary Center for Scientific Computing, Heidelberg University, Im Neuenheimer Feld 205, Heidelberg 69120, Germany\label{GALinst75}\and
Institut f\"ur Theoretische Physik, Universit\"at Heidelberg, Philosophenweg 16, 69120 Heidelberg, Germany\label{GALinst76}\and
CP3-Origins, University of Southern Denmark, Campusvej 55, DK-5230 Odense, Denmark\label{GALinst77}\and
Instituto Nacional de Astrofísica, Óptica y Electrónica. Apartado Postal 51 y 216, 72000. Puebla Pue., México\label{GALinst78}\and
Consejo Nacional de Humanidades, Ciencia y Tecnología, Av. Insurgentes Sur 1582, 03940, Ciudad de México, México\label{GALinst79}\and
Instituto de Radioastronomía y Astrofísica, Universidad Nacional Autónoma de México, Morelia 58089, México\label{GALinst80}\and
Key Laboratory for Research in Galaxies and Cosmology, Chinese Academy of Sciences, Shanghai 200030, People's Republic of China\label{GALinst81}\and
Graduate School of Science, Nagoya City University, Yamanohata 1, Mizuho-cho, Mizuho-ku, Nagoya, 467-8501, Aichi, Japan\label{GALinst82}\and
Department of Physics, McGill University, 3600 rue University, Montréal, QC H3A 2T8, Canada\label{GALinst83}\and
Trottier Space Institute at McGill, 3550 rue University, Montréal,  QC H3A 2A7, Canada\label{GALinst84}\and
NOVA Sub-mm Instrumentation Group, Kapteyn Astronomical Institute, University of Groningen, Landleven 12, 9747 AD Groningen, The Netherlands\label{GALinst85}\and
Department of Astronomy, School of Physics, Peking University, Beijing 100871, People's Republic of China\label{GALinst86}\and
Kavli Institute for Astronomy and Astrophysics, Peking University, Beijing 100871, People's Republic of China\label{GALinst87}\and
Department of Astronomical Science, The Graduate University for Advanced Studies (SOKENDAI), 2-21-1 Osawa, Mitaka, Tokyo 181-8588, Japan\label{GALinst88}\and
Department of Astronomy, Graduate School of Science, The University of Tokyo, 7-3-1 Hongo, Bunkyo-ku, Tokyo 113-0033, Japan\label{GALinst89}\and
National Astronomical Observatory of Japan, 2-21-1 Osawa, Mitaka, Tokyo 181-8588, Japan\label{GALinst90}\and
The Institute of Statistical Mathematics, 10-3 Midori-cho, Tachikawa, Tokyo, 190-8562, Japan\label{GALinst91}\and
Department of Statistical Science, The Graduate University for Advanced Studies (SOKENDAI), 10-3 Midori-cho, Tachikawa, Tokyo 190-8562, Japan\label{GALinst92}\and
Kavli Institute for the Physics and Mathematics of the Universe, The University of Tokyo, 5-1-5 Kashiwanoha, Kashiwa, 277-8583, Japan\label{GALinst93}\and
Leiden Observatory, Leiden University, Postbus 2300, 9513 RA Leiden, The Netherlands\label{GALinst94}\and
ASTRAVEO LLC, PO Box 1668, Gloucester, MA 01931, USA\label{GALinst95}\and
Applied Materials Inc., 35 Dory Road, Gloucester, MA 01930, USA\label{GALinst96}\and
Finnish Centre for Astronomy with ESO, University of Turku, FI-20014 Turun Yliopisto, Finland\label{GALinst97}\and
Aalto University Metsähovi Radio Observatory, Metsähovintie 114, FI-02540 Kylmälä, Finland\label{GALinst98}\and
Institute for Astrophysical Research, Boston University, 725 Commonwealth Ave., Boston, MA 02215, USA\label{GALinst99}\and
Korea National University of Science and Technology, Gajeong-ro 217, Yuseong-gu, Daejeon 34113, Republic of Korea\label{GALinst100}\and
National Institute of Technology, Ichinoseki College, Takanashi, Hagisho, Ichinoseki, Iwate, 021-8511, Japan\label{GALinst101}\and
Joint Institute for VLBI ERIC (JIVE), Oude Hoogeveensedijk 4, 7991 PD Dwingeloo, The Netherlands\label{GALinst102}\and
CSIRO, Space and Astronomy, PO Box 76, Epping, NSW 1710, Australia\label{GALinst103}\and
Department of Physics, Ulsan National Institute of Science and Technology (UNIST), Ulsan 44919, Republic of Korea\label{GALinst104}\and
Department of Physics, Korea Advanced Institute of Science and Technology (KAIST), 291 Daehak-ro, Yuseong-gu, Daejeon 34141, Republic of Korea\label{GALinst105}\and
Kogakuin University of Technology \& Engineering, Academic Support Center, 2665-1 Nakano, Hachioji, Tokyo 192-0015, Japan\label{GALinst106}\and
Max-Planck-Institut für Astrophysik, Karl-Schwarzschild-Str. 1, 85748 Garching, Germany\label{GALinst107}\and
Graduate School of Science and Technology, Niigata University, 8050 Ikarashi 2-no-cho, Nishi-ku, Niigata 950-2181, Japan\label{GALinst108}\and
Anton Pannekoek Institute for Astronomy, University of Amsterdam, Science Park 904, 1098 XH, Amsterdam, The Netherlands\label{GALinst109}\and
Physics Department, National Sun Yat-Sen University, No. 70, Lien-Hai Road, Kaosiung City 80424, Taiwan, R.O.C.\label{GALinst110}\and
David A. Dunlap Department of Astronomy \& Astrophysics, University of Toronto, 50 St. George St, M5S 3H4, ON, Canada\label{GALinst111}\and
Dunlap Institute for Astronomy and Astrophysics, University of Toronto, 50 St. George Street, Toronto, ON M5S 3H4, Canada\label{GALinst112}\and
School of Astronomy and Space Science, Nanjing University, Nanjing 210023, People's Republic of China\label{GALinst113}\and
Key Laboratory of Modern Astronomy and Astrophysics, Nanjing University, Nanjing 210023, People's Republic of China\label{GALinst114}\and
INAF-Istituto di Radioastronomia, Via P. Gobetti 101, I-40129 Bologna, Italy\label{GALinst115}\and
Common Crawl Foundation, 9663 Santa Monica Blvd. 425, Beverly Hills, CA 90210 USA\label{GALinst116}\and
Instituto de Física, Pontificia Universidad Católica de Valparaíso, Casilla 4059, Valparaíso, Chile\label{GALinst117}\and
Key Laboratory of Radio Astronomy and Technology,  Shanghai Astronomical Observatory, CAS, 80 Nandan Road, Shanghai 200030, People’s Republic of China\label{GALinst118}\and
INAF-Istituto di Radioastronomia \& Italian ALMA Regional Centre, Via P. Gobetti 101, I-40129 Bologna, Italy\label{GALinst119}\and
Department of Physics, National Taiwan University, No. 1, Sec. 4, Roosevelt Rd., Taipei 106216, Taiwan, R.O.C\label{GALinst120}\and
Department of Physics and Astronomy, University of Mississippi, Mississippi 38677, USA\label{GALinst121}\and
Yunnan Observatories, Chinese Academy of Sciences, 650011 Kunming, Yunnan Province, People's Republic of China\label{GALinst122}\and
Center for Astronomical Mega-Science, Chinese Academy of Sciences, 20A Datun Road, Chaoyang District, Beijing, 100012, People's Republic of China\label{GALinst123}\and
Key Laboratory for the Structure and Evolution of Celestial Objects, Chinese Academy of Sciences, 650011 Kunming, People's Republic of China\label{GALinst124}\and
Gravitation and Astroparticle Physics Amsterdam (GRAPPA) Institute, University of Amsterdam, Science Park 904, 1098 XH Amsterdam, The Netherlands\label{GALinst125}\and
Deceased\label{GALinst127}\and
Joint ALMA Observatory, Alonso de C\'ordova 3107, Vitacura 763-0355, Santiago, Chile\label{GALinst128}\and
European Southern Observatory, Alonso de C\'ordova 3107, Vitacura, Casilla 19001, Santiago, Chile\label{GALinst129}\and
School of Physics and Astronomy, Shanghai Jiao Tong University, 800 Dongchuan Road, Shanghai, 200240, People’s Republic of China\label{GALinst130}\and
Soft Computing, Image Processing and Aggregation Research Group (SCOPIA) \& Modelling and Imaging Radio Astronomical Data (MIRADA), Dept. of Mathematics and Computer Science, University of the Balearic Islands, Ctra. Valldemossa, Km 7.5, Palma 07122, Spain\label{GALinst131}\and
Artificial Intelligence Research Institute of the Balearic Islands (IAIB), Palma 07122, Spain\label{GALinst132}\and
Health Research Institute of the Balearic Islands (IdISBa), 07010 Palma de Mallorca, Spain\label{GALinst133}\and
Institut de Radioastronomie Millimétrique (IRAM), Avenida Divina Pastora 7, Local 20, E-18012, Granada, Spain\label{GALinst134}\and
National Institute of Technology, Hachinohe College, 16-1 Uwanotai, Tamonoki, Hachinohe City, Aomori 039-1192, Japan\label{GALinst135}\and
SKA Observatory, Jodrell Bank, Lower Withington, Macclesfield, SK11 9FT, UK\label{GALinst136}\and
Department of Physics, Villanova University, 800 Lancaster Avenue, Villanova, PA 19085, USA\label{GALinst137}\and
Cavendish Astrophysics, University of Cambridge, Madingley Road, Cambridge CB3 0HA, UK\label{GALinst138}\and
Kavli Institute for Cosmology, University of Cambridge, Madingley Road, Cambridge CB3 0HA, UK\label{GALinst139}\and
Physics Department, Washington University, CB 1105, St. Louis, MO 63130, USA\label{GALinst140}\and
Departamento de Matemática da Universidade de Aveiro and Centre for Research and Development in Mathematics and Applications (CIDMA), Campus de Santiago, 3810-193 Aveiro, Portugal\label{GALinst141}\and
School of Physics, Georgia Institute of Technology, 837 State St NW, Atlanta, GA 30332, USA\label{GALinst142}\and
School of Space Research, Kyung Hee University, 1732, Deogyeong-daero, Giheung-gu, Yongin-si, Gyeonggi-do 17104, Republic of Korea\label{GALinst143}\and
G-LAMP NEXUS Institute, Kyung Hee University, Yongin, 17104, Republic of Korea\label{GALinst144}\and
Canadian Institute for Advanced Research, 180 Dundas St West, Toronto, ON M5G 1Z8, Canada\label{GALinst145}\and
Dipartimento di Fisica, Università di Trieste, I-34127 Trieste, Italy\label{GALinst146}\and
INFN Sez. di Trieste, I-34127 Trieste, Italy\label{GALinst147}\and
Department of Physics, National Taiwan Normal University, No. 88, Sec. 4, Tingzhou Rd., Taipei 116, Taiwan, R.O.C.\label{GALinst148}\and
Center of Astronomy and Gravitation, National Taiwan Normal University, No. 88, Sec. 4, Tingzhou Road, Taipei 116, Taiwan, R.O.C.\label{GALinst149}\and
Signal Processing Research Centre, Tampere University, FI-33720 Tampere, Finland\label{GALinst150}\and
Department of Mathematics, New Uzbekistan University, Tashkent 100007, Uzbekistan\label{GALinst151}\and
Julius-Maximilians-Universität Würzburg, Fakultät für Physik und Astronomie, Institut für Theoretische Physik und Astrophysik, Lehrstuhl für Astronomie, Emil-Fischer-Str. 31, D-97074 Würzburg, Germany\label{GALinst153}\and
Department of Physics, University of Toronto, 60 St. George Street, Toronto, ON M5S 1A7, Canada\label{GALinst154}\and
Department of Physics, Tokyo Institute of Technology, 2-12-1 Ookayama, Meguro-ku, Tokyo 152-8551, Japan\label{GALinst155}\and
Hiroshima Astrophysical Science Center, Hiroshima University, 1-3-1 Kagamiyama, Higashi-Hiroshima, Hiroshima 739-8526, Japan\label{GALinst156}\and
Aalto University Department of Electronics and Nanoengineering, PL 15500, FI-00076 Aalto, Finland\label{GALinst157}\and
Institut de Radioastronomie Millimétrique (IRAM), 300 rue de la Piscine, F-38406 Saint Martin d'Hères, France\label{GALinst158}\and
Jeremiah Horrocks Institute, University of Lancashire, Preston PR1 2HE, UK\label{GALinst159}\and
National Biomedical Imaging Center, Peking University, Beijing 100871, People’s Republic of China\label{GALinst160}\and
College of Future Technology, Peking University, Beijing 100871, People’s Republic of China\label{GALinst161}\and
Department of Physics and Astronomy, University of Lethbridge, Lethbridge, Alberta T1K 3M4, Canada\label{GALinst162}\and
Frontier Research Institute for Interdisciplinary Sciences, Tohoku University, Sendai 980-8578, Japan\label{GALinst163}\and
Astronomical Institute, Tohoku University, Sendai 980-8578, Japan\label{GALinst164}\and
Department of Physics and Astronomy, Seoul National University, Gwanak-gu, Seoul 08826, Republic of Korea\label{GALinst165}\and
SNU Astronomy Research Center, Seoul National University, Gwanak-gu, Seoul 08826, Republic of Korea\label{GALinst166}\and
ASTRON, Oude Hoogeveensedijk 4, 7991 PD Dwingeloo, The Netherlands\label{GALinst167}\and
Centre for Mathematical Plasma Astrophysics, Department of Mathematics, KU Leuven, Celestijnenlaan 200B, B-3001 Leuven, Belgium\label{GALinst168}\and
Physics Department, Brandeis University, 415 South Street, Waltham, MA 02453, USA\label{GALinst169}\and
Tuorla Observatory, Department of Physics and Astronomy, University of Turku, FI-20014 Turun Yliopisto, Finland\label{GALinst170}\and
Excellence Fellow at Radboud University, Nijmegen, The Netherlands\label{GALinst171}\and
School of Natural Sciences, Institute for Advanced Study, 1 Einstein Drive, Princeton, NJ 08540, USA\label{GALinst172}\and
School of Physics, Huazhong University of Science and Technology, Wuhan, Hubei, 430074, People's Republic of China\label{GALinst173}\and
Mullard Space Science Laboratory, University College London, Holmbury St. Mary, Dorking, Surrey, RH5 6NT, UK\label{GALinst174}\and
Center for Astronomy and Astrophysics and Department of Physics, Fudan University, Shanghai 200438, People's Republic of China\label{GALinst175}\and
Astronomy Department, University of Science and Technology of China, Hefei 230026, People's Republic of China\label{GALinst176}\and
Department of Physics and Astronomy, Michigan State University, 567 Wilson Rd, East Lansing, MI 48824, USA\label{GALinst177}\and
  ETH Z\"urich, CH-8093 Z\"urich, Switzerland\label{MAGIC1}\and
 Universit\`a di Siena and INFN Pisa, I-53100 Siena, Italy\label{MAGIC2}\and
 Institut de F\'isica d'Altes Energies (IFAE), The Barcelona Institute of Science and Technology (BIST), E-08193 Bellaterra (Barcelona), Spain\label{MAGIC3}\and
 Universitat de Barcelona, ICCUB, IEEC-UB, E-08028 Barcelona, Spain\label{MAGIC4}\and
 Instituto de Astrof\'isica de Andaluc\'ia-CSIC, Glorieta de la Astronom\'ia s/n, 18008, Granada, Spain\label{MAGIC5}\and
 Universit\`a di Padova and INFN, I-35131 Padova, Italy\label{MAGIC6}\and
 National Institute for Astrophysics (INAF), I-00136 Rome, Italy\label{MAGIC7}\and
 Universit\`a di Udine and INFN Trieste, I-33100 Udine, Italy\label{MAGIC8}\and
 Instituto de Astrof\'isica de Canarias and Dpto. de  Astrof\'isica, Universidad de La Laguna, E-38200, La Laguna, Tenerife, Spain\label{MAGIC10}\and
 Croatian MAGIC Group: University of Zagreb, Faculty of Electrical Engineering and Computing (FER), 10000 Zagreb, Croatia\label{MAGIC11}\and
 Saha Institute of Nuclear Physics, A CI of Homi Bhabha National Institute, Kolkata 700064, West Bengal, India\label{MAGIC12}\and
 Centro Brasileiro de Pesquisas F\'isicas (CBPF), 22290-180 URCA, Rio de Janeiro (RJ), Brazil\label{MAGIC13}\and
 IPARCOS Institute and EMFTEL Department, Universidad Complutense de Madrid, E-28040 Madrid, Spain\label{MAGIC14}\and
 University of Lodz, Faculty of Physics and Applied Informatics, Department of Astrophysics, 90-236 Lodz, Poland\label{MAGIC15}\and
 Centro de Investigaciones Energ\'eticas, Medioambientales y Tecnol\'ogicas, E-28040 Madrid, Spain\label{MAGIC16}\and
 Departament de F\'isica, and CERES-IEEC, Universitat Aut\`onoma de Barcelona, E-08193 Bellaterra, Spain\label{MAGIC17}\and
 Universit\`a di Pisa and INFN Pisa, I-56126 Pisa, Italy\label{MAGIC18}\and
 INFN MAGIC Group: INFN Sezione di Bari and Dipartimento Interateneo di Fisica dell'Universit\`a e del Politecnico di Bari, I-70125 Bari, Italy\label{MAGIC19}\and
 Japanese MAGIC Group: Institute for Cosmic Ray Research (ICRR), The University of Tokyo, Kashiwa, 277-8582 Chiba, Japan\label{MAGIC20}\and
 INFN MAGIC Group: INFN Sezione di Torino and Universit\`a degli Studi di Torino, I-10125 Torino, Italy\label{MAGIC21}\and
 Croatian MAGIC Group: University of Rijeka, Faculty of Physics, 51000 Rijeka, Croatia\label{MAGIC22}\and
 Universit\"at W\"urzburg, D-97074 W\"urzburg, Germany\label{MAGIC23}\and
 Technische Universit\"at Dortmund, D-44221 Dortmund, Germany\label{MAGIC24}\and
 Japanese MAGIC Group: Physics Program, Graduate School of Advanced Science and Engineering, Hiroshima University, 739-8526 Hiroshima, Japan\label{MAGIC25}\and
 Armenian MAGIC Group: ICRANet-Armenia, 0019 Yerevan, Armenia\label{MAGIC26}\and
 Croatian MAGIC Group: Josip Juraj Strossmayer University of Osijek, Department of Physics, 31000 Osijek, Croatia\label{MAGIC27}\and
 Finnish MAGIC Group: Finnish Centre for Astronomy with ESO, Department of Physics and Astronomy, University of Turku, FI-20014 Turku, Finland\label{MAGIC28}\and
 Japanese MAGIC Group: Department of Physics, Tokai University, Hiratsuka, 259-1292 Kanagawa, Japan\label{MAGIC29}\and
 Inst. for Nucl. Research and Nucl. Energy, Bulgarian Academy of Sciences, BG-1784 Sofia, Bulgaria\label{MAGIC30}\and
 INFN MAGIC Group: INFN Sezione di Catania and Dipartimento di Fisica e Astronomia, University of Catania, I-95123 Catania, Italy\label{MAGIC31}\and
 Finnish MAGIC Group: Space Physics and Astronomy Research Unit, University of Oulu, FI-90014 Oulu, Finland\label{MAGIC32}\and
 Japanese MAGIC Group: Institute for Space-Earth Environmental Research and Kobayashi-Maskawa Institute for the Origin of Particles and the Universe, Nagoya University, 464-6801 Nagoya, Japan\label{MAGIC33}\and
 University of Geneva, Chemin d'Ecogia 16, CH-1290 Versoix, Switzerland\label{MAGIC34}\and
 INFN MAGIC Group: INFN Roma Tor Vergata, I-00133 Roma, Italy\label{MAGIC35}\and
 also at International Center for Relativistic Astrophysics (ICRA), Rome, Italy\label{MAGIC36}\and
 also at Como Lake centre for AstroPhysics (CLAP), DiSAT, Universit\`a dell'Insubria, via Valleggio 11, 22100 Como, Italy.\label{MAGIC37}\and
 also at Port d'Informaci\'o Cient\'ifica (PIC), E-08193 Bellaterra (Barcelona), Spain\label{MAGIC38}\and
 also at Dipartimento di Fisica, Universit\`a di Trieste, I-34127 Trieste, Italy\label{MAGIC39}\and
 Dipartimento di Fisica, Universit\`a di Roma Tor Vergata, Via della Ricerca Scientifica, 1, Roma I-00133, Italy\label{MAGIC40}\and
 also at INAF Padova\label{MAGIC41}\and
 University of Southern Denmark, Odense, Denmark \label{HESSUSD}\and
 Yerevan State University, 1 Alek Manukyan St, Yerevan 0025, Armenia \label{HESSYSU}\and
 Astronomy \& Astrophysics Section, School of Cosmic Physics, Dublin Institute for Advanced Studies, DIAS Dunsink Observatory, Dublin D15 XR2R, Ireland \label{HESSDIAS}\and
 Laboratoire Leprince-Ringuet, \'Ecole Polytechnique, CNRS, Institut Polytechnique de Paris, F-91128 Palaiseau, France \label{HESSLLR}\and
 University of Namibia, Department of Physics, Private Bag 13301, Windhoek 10005, Namibia \label{HESSUNAM}\and
 Centre for Space Research, North-West University, Potchefstroom 2520, South Africa \label{HESSNWU}\and
 Universit\'e Paris Cit\'e, CNRS, Astroparticule et Cosmologie, F-75013 Paris, France \label{HESSAPC}\and
 LUX, Observatoire de Paris, Universit\'e PSL, CNRS, Sorbonne Universit\'e, 5 Pl. Jules Janssen, 92190 Meudon, France \label{HESSLUX}\and
 Sorbonne Universit\'e, CNRS/IN2P3, Laboratoire de Physique Nucl\'eaire, et de Hautes Energies, LPNHE, 4 place Jussieu, 75005 Paris, France \label{HESSLPNHE}\and
 Friedrich-Alexander-Universit\"at Erlangen-N\"urnberg, Erlangen Centre for Astroparticle Physics,  Nikolaus-Fiebiger-Str. 2, 91058 Erlangen, Germany \label{HESSECAP}\and
 Instytut Fizyki Jadrowej PAN, ul. Radzikowskiego 152, ul. Radzikowskiego 152, 31-342 Kraków, Poland \label{HESSIFJPAN}\and
 School of Physical Sciences and Centre for Astrophysics \& Relativity, Dublin City University, Glasnevin, Dublin D09 W6Y4, Ireland \label{HESSDCU}\and
 IRFU, CEA, Universit\'e Paris-Saclay, F-91191 Gif-sur-Yvette, France \label{HESSIRFU}\and
 Deutsches Elektronen-Synchrotron DESY, Platanenallee 6, 15738 Zeuthen, Germany \label{HESSDESY}\and
 School of Physics, University of the Witwatersrand, 1 Jan Smuts Avenue, Braamfontein, Johannesburg, 2050, South Africa \label{HESSWits}\and
 Institut f\"ur Physik und Astronomie, Universit\"at Potsdam, Karl-Liebknecht-Strasse 24/25, D 14476 Potsdam, Germany \label{HESSUP}\and
 School of Science, Western Sydney University, Locked Bag 1797, Penrith South DC, NSW 2751, Australia \label{HESSSydney}\and
Landessternwarte, Universit\"at Heidelberg, K\"onigstuhl, D 69117 Heidelberg, Germany \label{HESSLSW}\and
 Institut f\"ur Astronomie und Astrophysik, Universit\"at T\"ubingen, Sand 1, D 72076 T\"ubingen, Germany \label{HESSIAAT}\and
 Universit\"at Innsbruck, Institut f\"ur Astro- und Teilchenphysik, Technikerstraße 25, 6020 Innsbruck, Austria \label{HESSInnsbruck}\and
 Universit\"at Hamburg, Institut f\"ur Experimentalphysik, Luruper Chaussee 149, D 22761 Hamburg, Germany \label{HESSUHAM}\and
 Institute of Astronomy, Faculty of Physics, Astronomy and Informatics, Nicolaus Copernicus University, Grudziadzka 5, 87-100 Torun, Poland \label{HESSNCUT}\and
 Laboratoire Univers et Particules de Montpellier, Universit\'{e} Montpellier, CNRS/IN2P3, CC 72, Place Eug\`{e}ne Bataillon, F-34095 Montpellier Cedex 5, France \label{HESSMNTPL}\and
 Obserwatorium Astronomiczne, Uniwersytet Jagielloński, ul. Orla 171, 30-244 Krak\'ow, Poland \label{HESSOAUJ}\and
 Nicolaus Copernicus Astronomical Center, Polish Academy of Sciences, ul. Bartycka 18, 00-716 Warsaw, Poland \label{HESSNCAC}\and
 Yerevan Physics Institute, 2 Alikhanian Brothers St., 0036 Yerevan, Armenia \label{HESSYPI}\and
 Department of Physics, Konan University, 8-9-1 Okamoto, Higashinada, Kobe, Hyogo 658-8501, Japan \label{HESSKonan}\and
 GRAPPA, Anton Pannekoek Institute for Astronomy, University of Amsterdam, Science Park 904, 1098 XH Amsterdam, The Netherlands \label{HESSGRAPPA}
}

   \date{Received February 27, 2026; accepted June 18, 2026}

  \abstract
   {The archetypal blazar 3C\,279 has a prominent relativistic jet and strong broadband variability across the electromagnetic spectrum. In April 2017, the Event Horizon Telescope (EHT) observed 3C\,279 with an unprecedented angular resolution of about 20\,$\mu$as, accompanied by one of the most extensive quasi-simultaneous multiwavelength (MWL) campaigns ever conducted, spanning from radio to TeV $\gamma$-ray energies.}
   {Taking advantage of this comprehensive MWL dataset, we investigated the physical processes governing 3C\,279, with a particular focus on the formation, collimation, and acceleration of its relativistic jet and on the origin of its high-energy emission, including the underlying particle-acceleration mechanisms.}
   {We analyzed individual observations and multiband light curves. We also constructed a new quasi-simultaneous spectral energy distribution covering frequencies from the radio band to very high-energy (VHE) $\gamma$ rays. We further performed a phenomenological modeling using the turbulent extreme multi-zone (TEMZ) model to constrain the fundamental physical properties of the source.}
   {The EHT observations reveal a clear flux increase in the innermost core between April 5 and 11, 2017. Over a broader time span, radio observations at longer wavelengths reveal concurrent enhancements in the core flux and polarization around mid-April, coinciding with the ejection of a superluminal knot moving at $(25 \pm 2)c$. Record UV–optical flares with strong polarization variability occurred in late March, followed by high-energy $\gamma$-ray activity that declined before the end of the EHT observing period. During this time, the source remained in a low X-ray state and exhibited no detectable VHE emission.}
   {The results of the TEMZ modeling indicate that the broadband spectrum and variability of 3C\,279 might be explained with 
   a jet scenario in which turbulent plasma cells are compressed by a stationary conical shock. Nonetheless, alternative interpretations, such as magnetic reconnection or a moving shock-in-jet event, remain possible. This coordinated MWL campaign advances our understanding of the origin of the jet and $\gamma$-ray emission in the blazar 3C\,279, and it also provides a comprehensive publicly available dataset that will serve as a valuable reference for future studies.}
\authorrunning{G. Principe et al.}
\titlerunning{MWL properties of 3C\,279 during the EHT 2017 campaign}
   \maketitle
   \nolinenumbers

\section{Introduction}
Flat-spectrum radio quasars (FSRQs) form a subclass of blazars and represent the most powerful active galactic nuclei (AGNs). Their relativistic jets are oriented close to the observer’s line of sight, resulting in strongly beamed emission. It remains a central challenge in modern astrophysics to understand the mechanisms of jet launching, collimation, evolution, and how particles are accelerated to GeV and TeV energies in such extreme environments (see, e.g., \citet{2019ARA&A..57..467B} for a review).

Blazars produce broadband nonthermal radiation that extends across the entire electromagnetic spectrum, from radio wavelengths to very high-energy (VHE; E$ > $100\,GeV) $\gamma$ rays. Their radio and optical emission is typically highly linearly polarized (a few up to a few dozen percent; see, e.g., \citealt{2016MNRAS.463.3365A}) and varies strongly on timescales ranging from minutes to years \citep{2016ApJ...824L..20A}. Moreover, blazars and jetted AGN constitute the most numerous class of extragalactic $\gamma$-ray emitters \citep{2020ApJS..247...33A}, presenting high-energy emission across a wide range of evolutionary stages and spatial scales, from compact subparsec regions \citep{2016ApJ...821L..31M,2020A&A...635A.185P,2021MNRAS.507.4564P} to more extended kiloparsec-scale structures \citep{2020ApJ...892..105A}. Among them, 3C\,279 is a prominent FSRQ and one of the most distant VHE $\gamma$-ray sources, located at a redshift of z$=$0.5362 \citep{1996ApJS..104...37M}. At the center of its host galaxy resides a supermassive black hole (SMBH) with an estimated mass of 8$\times 10^8 M_{\odot}$ \citep{2009A&A...505..601N}.

The source was first targeted in April 2017 during a four-day campaign while also serving as one of the primary calibrators for the Event Horizon Telescope (EHT) observations of M87$^\ast$ and Sgr\,A$^\ast$ at 230\,GHz \citep{PaperI,EHTC2022a}. The EHT ultra-high angular resolution of 20\,$\mu$as achieved in these observations corresponds to a spatial scale of 0.13\,pc ($\sim$1700 Schwarzschild radii). In 2017, the EHT observed 3C\,279 in four epochs: April 5, 6, 10, and 11 \citep{2020A&A...640A..69K}. On all days, the data revealed a complex multicomponent morphology in the innermost jet region, with the northernmost feature, identified as the VLBI core, elongated roughly perpendicular to the large-scale jet axis seen at longer wavelengths. The overall morphology might correspond to either a broadened resolved jet base or to a spatially bent jet. 
The source morphology also varied significantly on timescales of days, indicating systematic structural evolution in the inner jet. Two compact components are observed to move non-radially at apparent superluminal speeds of $\sim$15$c$ and $\sim$20$c$. Together with inferred Doppler factors of 10–20, these motions support a scenario of traveling shocks or instabilities within a bent and possibly rotating jet. The observed apparent speeds are consistent with measurements at longer (centimeter) wavelengths, suggesting that no substantial jet acceleration occurs between the 1.3\,mm VLBI core and the outer jet regions.
Several mechanisms have been proposed to explain the jet bending at these scales. In thin-disk systems such as 3C\,279, jet precession might arise from Lense–Thirring torques in a misaligned accretion flow \citep{liska2018,liska2021}. Alternatively, jet wobbling can be driven by magnetic eruptions associated with reconnection near the event horizon \citep[e.g.,][]{ripperda2022,ressler2025,lalakos2025}, or by magnetohydrodynamic (MHD) instabilities, such as kink modes induced by interactions of the jet with its surroundings \citep[e.g.,][]{2016MNRAS.461L..46T,bromberg2016,lalakos2024} and Kelvin–Helmholtz instabilities within the flow \citep[e.g.,][]{chow2023}. Moreover, recent 3D MHD simulations showed that recollimation regions can trigger non-axisymmetric instabilities (e.g., kink modes), leading to jet distortion, enhanced dissipation, and quasi-periodic variability, with properties depending on the jet magnetization and external pressure profile \citep[e.g.,][]{boula2025,hu2025}.
Finally, a binary supermassive black hole system has been proposed for 3C\,279 \citep{quian2019}, which might induce quasi-periodic jet orientation changes via orbital motion, spin–orbit coupling, or accretion-flow perturbations \citep[e.g.,][]{1980Natur.287..307B,ressler2025}.

The source 3C\,279 exhibits broadband emission extending from the radio band to VHE $\gamma$ rays \citep{2011A&A...530A...4A,2019A&A...627A.159H}, and the luminosity of the high-energy component generally dominates that of the synchrotron emission \citep{1994ApJ...435L..91M}.
High angular resolution (microarcsecond-scale) observations with RadioAstron resolved the 3C\,279 jet, uncovering several filaments that were interpreted as the result of plasma instabilities in a kinetically dominated flow \citep{2023NatAs...7.1359F}.
The source varies strongly across the entire electromagnetic spectrum (e.g., \citealt{2012ApJ...754..114H}), and the $\gamma$-ray fluxes vary by more than two orders of magnitude, from $\sim 10^{-7}$ to $10^{-5}$ cm$^{-2}$ s$^{-1}$ above 100\,MeV (see, e.g., \citealt{2016ApJ...824L..20A}). Multiwavelength (MWL) monitoring during a $\gamma$-ray flare in 2009 revealed a correlated change in optical polarization, providing evidence of a direct link between high-energy activity and the dynamics of the jet magnetic field \citep{2010Natur.463..919A}. 

The source is known for intense and rapid $\gamma$-ray flaring activity. On June 16, 2015, the Large Area Telescope (LAT) on board the {\em Fermi Gamma-ray Space Telescope} detected a giant outburst with minute-scale variability, showing flux-doubling times shorter than 5\,min \citep{2016ApJ...824L..20A}. Such a rapid variability implies a very compact emission region, located at distances of a few hundred Schwarzschild radii from the central engine in conical jet models. During the same flare, VHE $\gamma$ rays were detected with \HESS, indicating an emission site at $r > 1.7\times10^{17}$\,cm, that is, beyond the expected extent of the broad-line region \citep{2019A&A...627A.159H}. Likewise, {\em Fermi}-LAT observations in April 2018 revealed a distinctive double-peaked $\gamma$-ray variability pattern on minute timescales, consistent with particle acceleration via relativistic magnetic reconnection. The absence of significant $\gamma\gamma$ absorption further constrains the emission region to $\sim 10^4$ gravitational radii, where turbulence is likely sustained by fluid-dynamical kink instabilities \citep{2020NatCo..11.4176S}.

In analogy with the MWL campaigns accompanying the EHT observations of M87$^\ast$ \citep{2021ApJ...911L..11E,2024A&A...692A.140A}, coordinated quasi-simultaneous MWL observations of 3C\,279 were conducted during the 2017 EHT observing run. This dataset provides a unique opportunity to investigate jet physics and particle acceleration processes, supports the interpretation and modeling of the EHT results, and constitutes a valuable legacy resource for the wider astronomical community.

The paper is structured as follows. In Sect.\,\ref{sec:obs} we briefly describe the MWL observational campaign on 3C\,279 accompanying the 2017 EHT observations, while further details on data processing and individual results are provided in Appendix\,\ref{appendix:observations}. In Sect.\,\ref{sec:mwl_results} we present the radio multifrequency jet structure, the MWL variability and quasi-simultaneous spectrum. In Sect.\,\ref{sec:modelling} we describe the phenomenological modeling using the turbulent extreme multi-zone (TEMZ) model. Sect.\,\ref{sec:discussion} discusses the jet structure and the origin of the $\gamma$-ray flaring activity. Finally, Sect.\,\ref{sec:summary} summarizes our main results and conclusions.

\section{Observations}
\label{sec:obs}
\begin{figure*}
\centering
   \includegraphics[width=17cm]{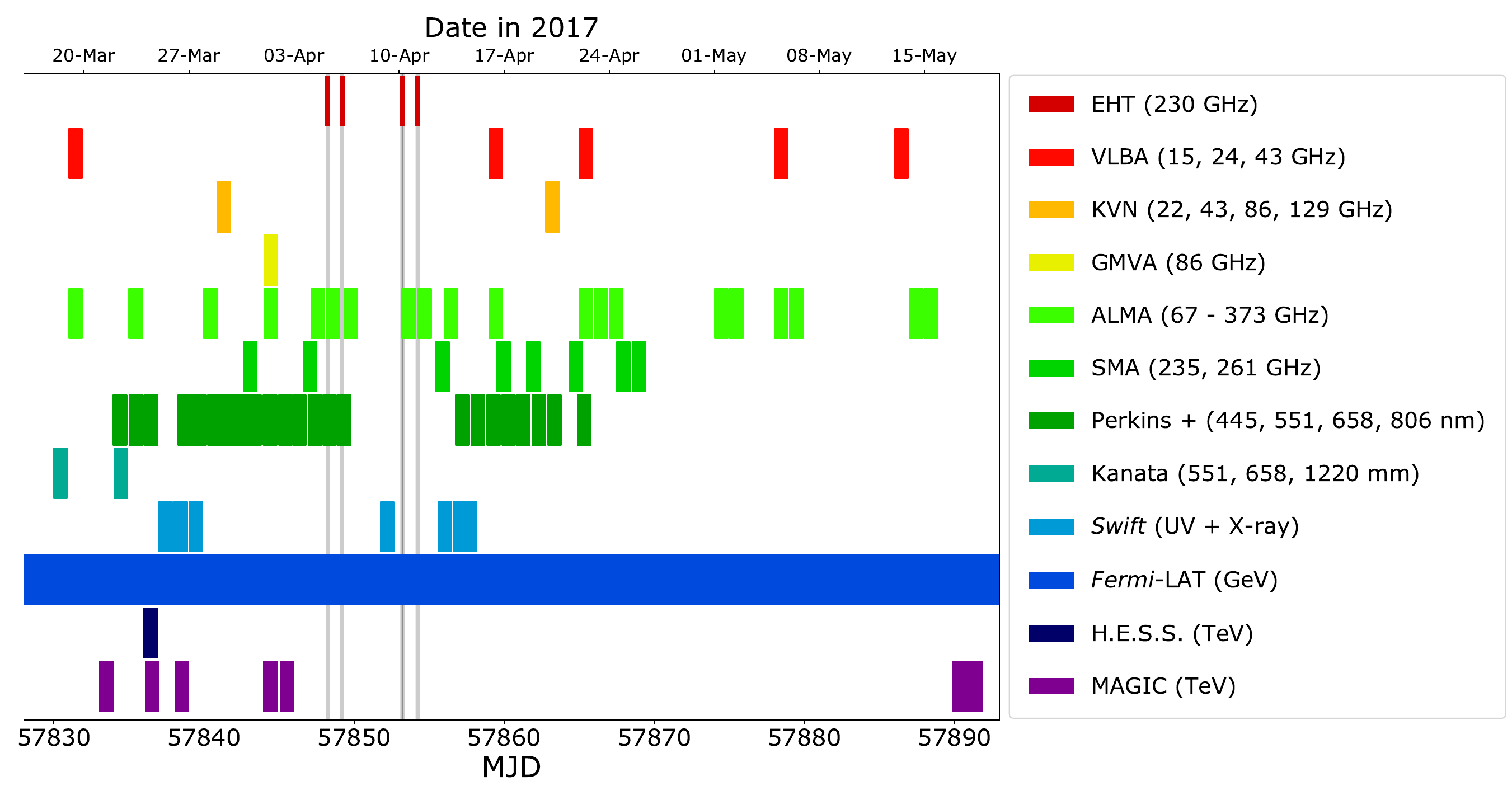}
     \caption{Summary of the instrument coverage during the 2017 MWL campaign on the blazar 3C\,279. The vertical gray bands refer to the EHT observations in the first row. The vertical dark gray line marks the date of the EHT observation we used for the spectral analysis.}
     \label{fig:2017_instrument_coverage}
\end{figure*}

As a main calibrator for EHT primary targets, 3C 279 was monitored during the first EHT observing campaign, conducted between April 5 and 11, 2017. Complementary MWL observations were performed in conjunction with the EHT runs, covering a broad frequency range, from 1.7\,GHz with RadioAstron to VHE $\gamma$ rays with the tHigh Energy Stereoscopic System (\HESS) and the The Major Atmospheric Gamma Imaging Cherenkov (MAGIC). Figure~\ref{fig:2017_instrument_coverage} summarizes the observational coverage between March and April 2017 during the first EHT–MWL campaign. In addition to the instruments shown in the figure, our analysis also incorporates cumulative flux estimates from RadioAstron at 1.7 and 5\,GHz, spanning the period 2016–2018.

The following subsections provide brief descriptions of the individual observations carried out during the 2017 MWL campaign on 3C\,279. Comprehensive details, including data-processing procedures and band-specific analyses and results, are presented in Appendix~\ref{appendix:observations}.

\subsection{Radio}
The blazar 3C\,279 was observed with several radio facilities in the centimeter (cm) and millimeter (mm) wavelength bands in VLBI and single-dish modes; details of the observations and data reduction are provided in Appendix~\ref{sec:radio_observations}.
High-resolution (milliarcsecond-scale) core monitoring of the source was carried out with the Very Long Baseline Array (VLBA) at 43 GHz and with the Korean VLBI Network (KVN) through multifrequency observations at 22, 43, 86, and 129 GHz. These data reveal variability in the core flux density, with the emission increasing during April–May (see Sect.~\ref{sec:mwl_lc}). At millimeter wavelengths, the Submillimeter Array (SMA) and ALMA performed approximately weekly observations and also observed during EHT observations \citep{2021ApJ...910L..14G}, providing information on the more extended (arcsecond-scale) emission.

\subsection{IR, optical, and UV}
The source was monitored in the IR, optical, and UV range from 1220\,nm with the Kanata telescope (see Appendix~\ref{sec:optical-kanata}) to 193\,nm with \textit{Swift}-UVOT (see Appendix~\ref{sec:swift_uvot}).
Moreover, well-sampled optical polarimetric observations of 3C\,279 in the \emph{R} band, together with photometric measurements, were obtained with four telescopes: the 1.8\,m Perkins Telescope, the 40\,cm LX-200 telescope, the 70\,cm AZT-8 telescope, and the 1.54\,m Kuiper Telescope (see Appendix~\ref{sec:optical-perkins} for details).

\subsection{X-rays and $\gamma$ rays}
In the X-ray band, 3C\,279 was observed with \textit{Swift}-XRT in the 0.3–10\,keV energy range (see Appendix~\ref{sec:xray-swiftxrt}). During the 2017 campaign, the source was monitored on March 26 and 27 and on April  8, 12, and 13, and all observations were performed in photon-counting mode.

Above $100$\,MeV, the \textit{Fermi} Large Area Telescope (LAT) detected significant emission from 3C\,279 on three-hour timescales (see also Appendix~\ref{sec:gamma-fermi}). The source entered a high-activity state at the end of March, and the flaring episode ceased around April 9, shortly before the conclusion of the EHT observing window. 

At VHE $\gamma$ rays, the source was observed by \HESS and MAGIC (see Appendices~\ref{sec:gamma-hess} and~\ref{sec:gamma-magic} for details).
\HESS conducted observations totaling 4.2\,h during the night of 24-25 March 2017, following a bright-moon break and recorded optical flaring episodes of the source. No significant VHE emission was detected.
Similarly, MAGIC accumulated 8.51\,h of data between March and April~2017. 3C\,279 was not detected above 5\,$\sigma$ for individual nightly observations or during the stacked analysis for the entire observation campaign. The only data point shown in the MWL light curves (LC) (Fig. \ref{fig:MWL_LC_2017}) together with the U.L.s corresponds to a 2.5\,$\sigma$ fluctuation (the LC shows U.L.s below 2\,$\sigma$ and data points otherwise).

\section{Multiwavelength results}
\label{sec:mwl_results}
\subsection{Radio structure}
\label{sec:radio-structure}

\begin{figure}[th]
\includegraphics[trim={0 0.7cm 0 0.7cm}, clip,width=\columnwidth]{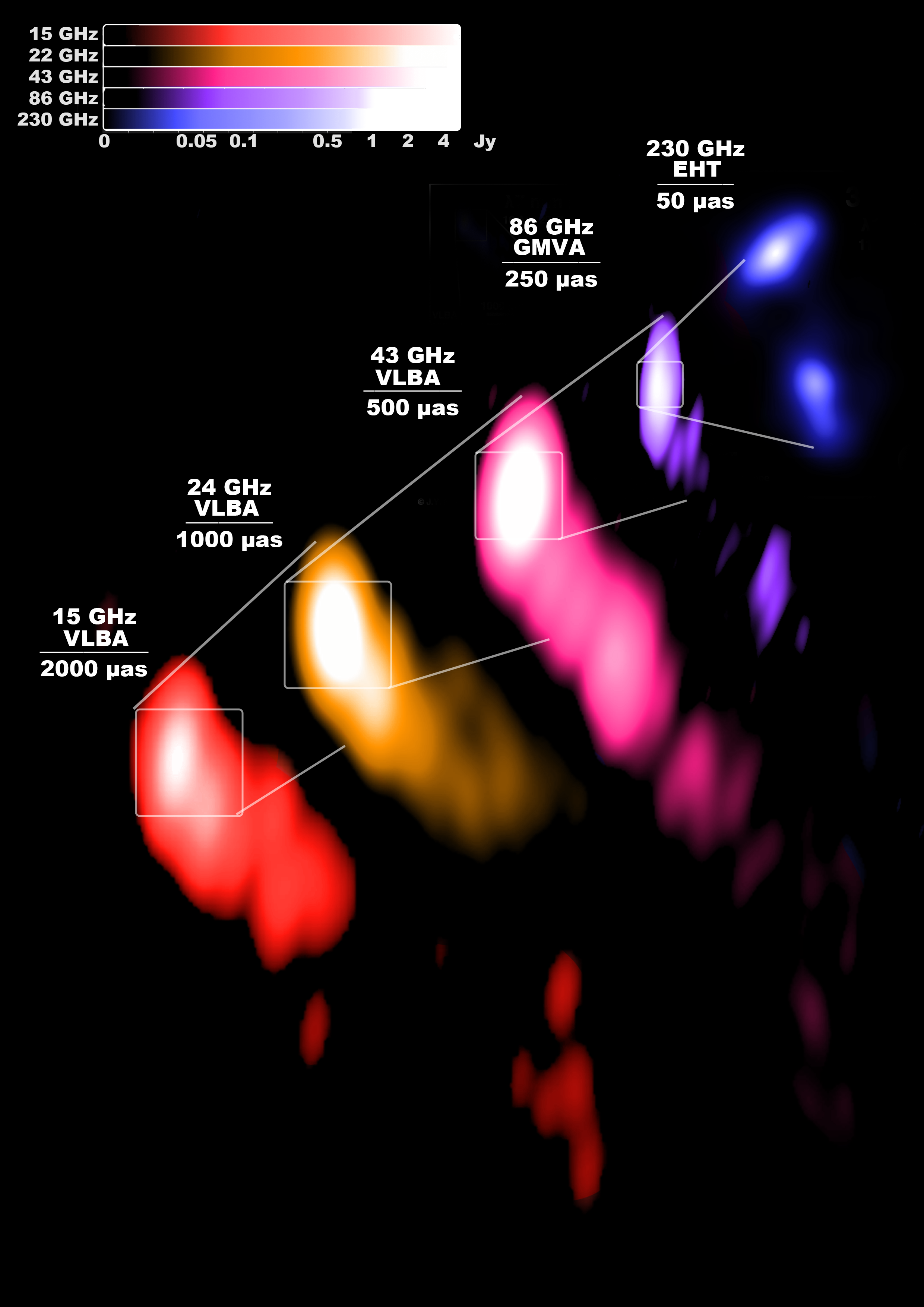}
\caption{Multifrequency (15, 24, 43, 86, and 230 GHz) composite radio image of 3C\,279 obtained in April 2017.}
\label{fig:composite_radio}
\end{figure}

The radio morphology of the jet of 3C\,279 (Fig.~\ref{fig:composite_radio}) consists of a compact core (the brightest region located in the northeast corner of the image) and an extended, nearly straight jet, visible up to deprojected scales of about 700\,pc at 15\,GHz, assuming a viewing angle of $1.5\degr$ \citep{Larionov2020}. 
Figure \ref{fig:composite_radio} presents a composite radio image of the jet structure of 3C\,279 obtained in April 2017 from single-epoch VLBI observations at 15, 24, 43, 86, and 230 GHz, using the Very Long Baseline Array (VLBA), the Global Millimeter VLBI Array (GMVA), and the EHT. The accompanying VLBA 43\,GHz observations were taken with a monthly cadence between February and June, 2017, enabling studies of the total and polarized intensity evolution. 

\begin{figure*}
\centering
\includegraphics[width=17cm]{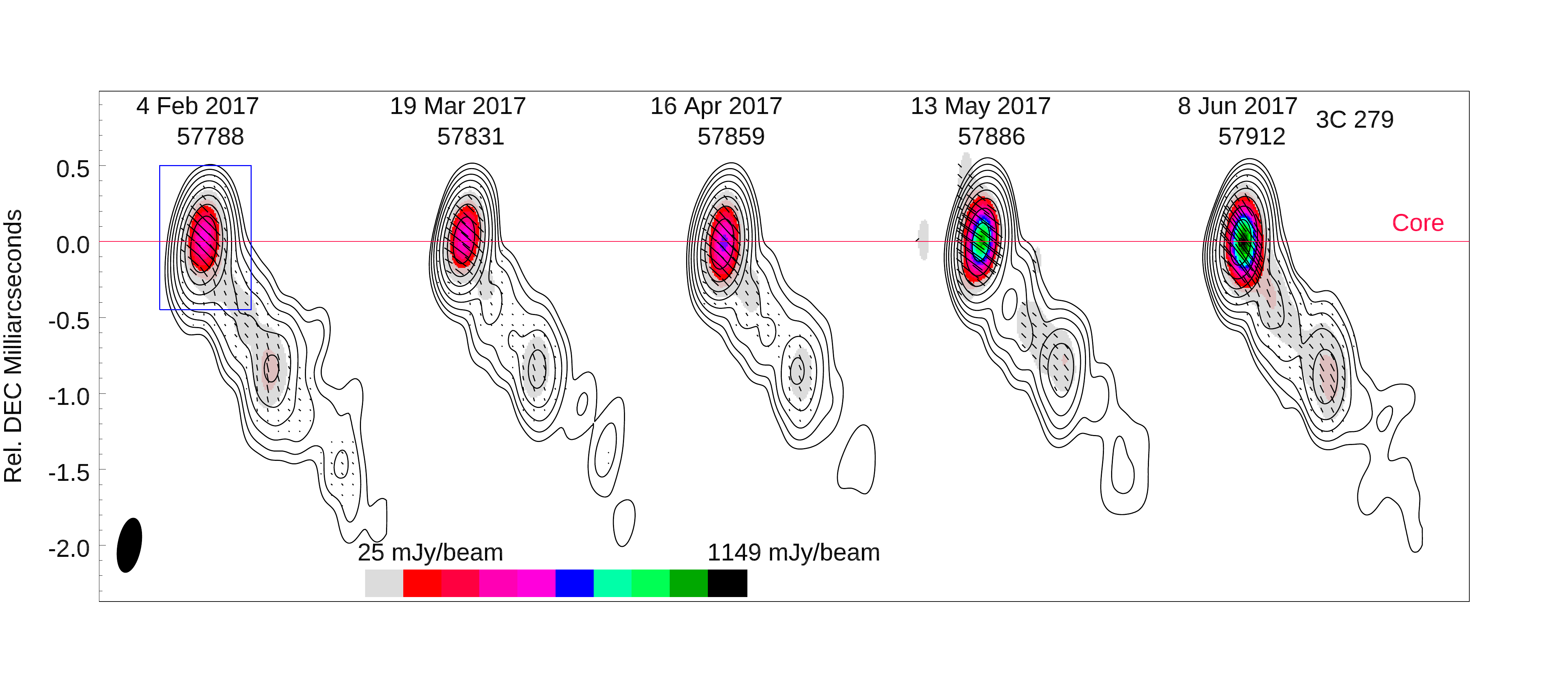}
\caption{Sequence of VLBA total (contours) and polarized (color scale) intensity images of 3C\,279 at 43\,GHz (magenta image in Fig. \ref{fig:composite_radio}).  The global intensity peak is 10.31\,Jy/beam; the contour levels start at 0.25\% of the peak and increase by factors of 2. All image parameters are listed in Table~\ref{tab:MapParm_7mm}. The images are convolved with the same beam of 0.36$\times$0.15 mas$^2$ at PA=$-$10$^\circ$(the ellipse in the bottom left corner);  linear segments within the images indicate the direction of the electric vector position angle (direction of polarization). The blue rectangle in the first image shows the core region used to calculate the polarization parameters presented in Table~\ref{tab:JetPol_7mm}.}
\label{fig:vlba_bu}
\end{figure*}

At 230\,GHz, the nuclear region is elongated nearly transversely to the direction of the larger-scale jet, separated from the other jet component by $\thicksim100\,\mu$as ($\thicksim$24\,pc, deprojected). 
The apparent position angle of the inner jet smoothly changes from about $-138$\degr{} at 15\,GHz to about $-174$\degr{} at 230\,GHz. 
Figure \ref{fig:vlba_bu} presents a sequence of multi-epoch 43\,GHz VLBA images showing the evolution of the total and polarized intensity between February and June 2017. The parameters of these images are presented in Table~\ref{tab:JetPol_7mm}. The images were obtained within the VLBA-BU-BLAZAR program. We used fully calibrated data posted at the program website \footnote{https://www.bu.edu/blazars/BEAM-ME.html} to construct the total and polarized intensity images shown in Figure \ref{fig:vlba_bu} and to calculate the parameters listed in Table~\ref{tab:JetPol_7mm}.

For the kinematics of 3C\,279 at 43\,GHz during the 2017 EHT campaign, we employed the results provided by \citet{2022ApJS..260...12W}. According to their analysis, a superluminal knot (component C37), detected for the first time in the jet around the end of October 2017, was ejected on April 2, 2017 ($\pm$13 days), from the center of the 43\,GHz VLBI core with an apparent speed of (25$\pm$2)\,c, one of the highest apparent speeds of known components seen for this source.
In the image of June 8, 2017, the knot should be $\sim$0.14\,mas from the center of the core, which is well inside the core region designated in Figure~\ref{fig:vlba_bu}. 
The ejection appears to have triggered high total and polarized flux in the core region: the average flux density of the core region during the period from June 2007 to December 2018 is ($8.5\pm2.4$)\,Jy \citet{2022ApJS..260...12W}, while the core flux density on June 8, 2017, is about 1.5 times higher than this value. This is accompanied by an elevated linear polarization degree of about 11\%, which is a few times higher than the quiescent value \citep{Larionov2020}.

\begin{table}
\caption{Polarization parameters of the jet with the VLBA at 43\,GHz.}
\label{tab:JetPol_7mm}

\small
\centering
\scriptsize
\setlength{\tabcolsep}{3pt}
\begin{tabular}{rlrrrrrr}
\hline\hline
Date\tablefootmark{a} &
MJD &
$S_{\rm int}$\tablefootmark{b} &
$P_{\rm int}$\tablefootmark{c} &
$\chi_{\rm int}$\tablefootmark{d} &
$S_{\rm core}$\tablefootmark{e} &
$P_{\rm core}$\tablefootmark{f} &
$\chi_{\rm core}$\tablefootmark{g} \\
\hline
 &  &  [Jy] & [\%] & [$^\circ$] & [Jy] & [\%] & [$^\circ$]\\   
\hline
  Feb. 4 &57788&11.77$\pm$0.59&9.7$\pm$1.0&35$\pm$6&8.66$\pm$0.44&8.7$\pm$0.9&40$\pm$6 \\
Mar. 19 &57831&7.70$\pm$0.39&10.7$\pm$1.1&42$\pm$6&6.14$\pm$0.31&10.1$\pm$1.0&47$\pm$6 \\
 Apr. 16 &57859&8.47$\pm$0.43&11.1$\pm$1.1&40$\pm$5&7.05$\pm$0.36&10.4$\pm$1.1&43$\pm$5 \\
May 13&57886&12.69$\pm$0.64&11.5$\pm$1.2&50$\pm$6&11.03$\pm$0.55&11.0$\pm$1.2&50$\pm$7 \\
  Jun. 8&57912&14.76$\pm$0.74&13.1$\pm$1.3&35$\pm$6&12.87$\pm$0.65&11.3$\pm$1.1&34$\pm$6 \\
\hline
\end{tabular}
\tablefoot{
    \tablefoottext{a}{Epoch of observation}
    \tablefoottext{b}{Total intensity integrated over the image}
    \tablefoottext{c}{Degree of polarization integrated over the image}
    \tablefoottext{d}{Position angle of the polarization integrated over the image}
    \tablefoottext{e}{Total intensity of the core region designated in Figure~\ref{fig:vlba_bu}}
    \tablefoottext{f}{Degree of polarization of the core region; uncertainties are given at the 1$\sigma$ level}
    \tablefoottext{g}{Position angle of the polarization in the core region; uncertainties are given at the 1$\sigma$ level.}
    }
\end{table}

\subsection{Multiwavelength results and variability} 
\label{sec:mwl_lc}

\begin{figure*}
\centering
\includegraphics[width=17cm]{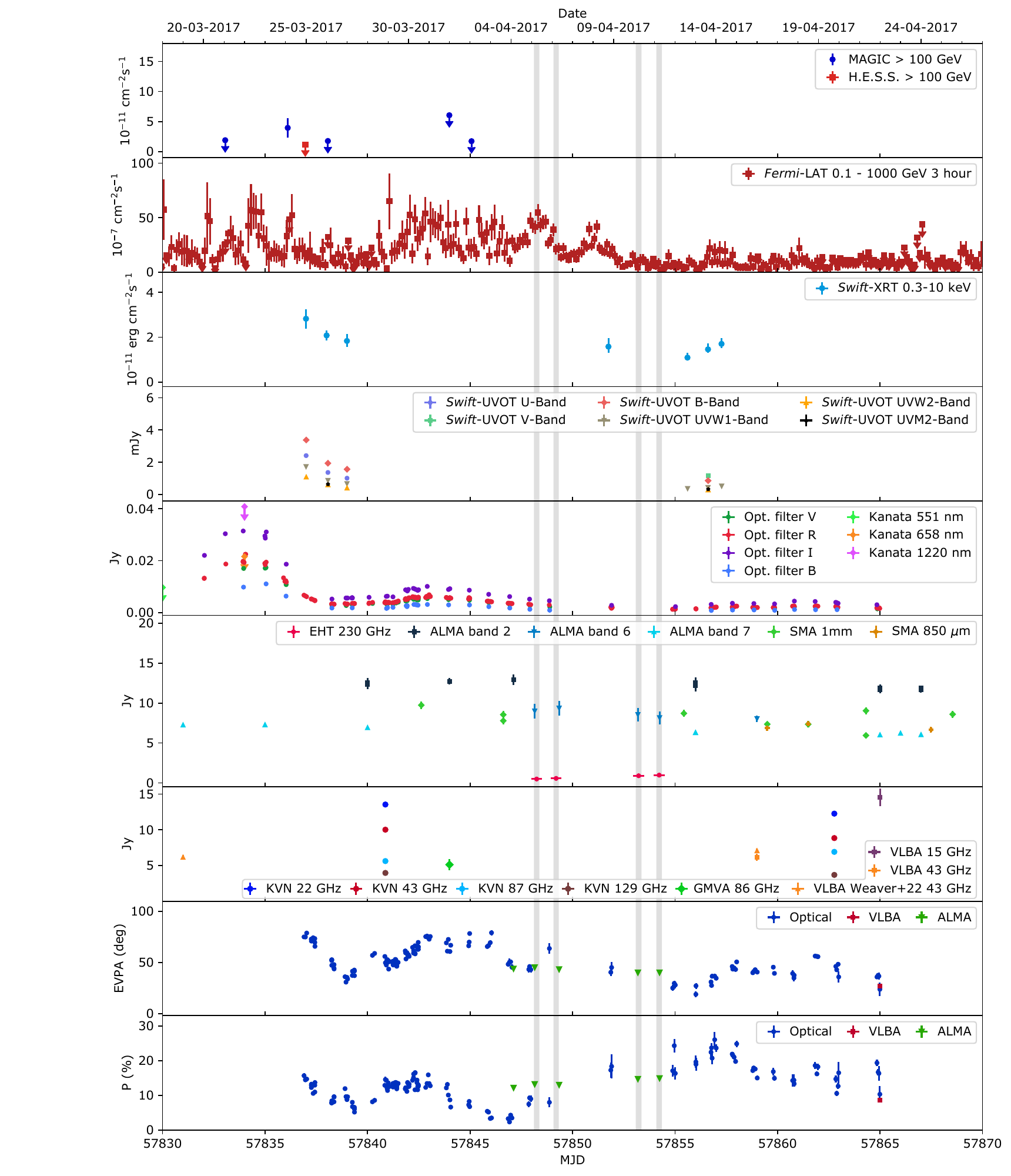}
\caption{MWL light curves of 3C\,279 during MJD 57830--57870 (March 18 -- April 27, 2017). The gray bands indicate the periods of EHT observations. The top seven panels display the source flux, and the bottom two panels show the electric vector position angle and polarization intensity, respectively.}
\label{fig:MWL_LC_2017}
\end{figure*}

Figure~\ref{fig:MWL_LC_2017} presents the short-term (March–April 2017) MWL LC, including the polarization intensity (P) and electric vector position angle (EVPA) of 3C\,279, obtained with the instruments participating in the coordinated EHT–MWL campaign.
The EHT observations of the innermost core region (component C0–0 in \citealt{2020A&A...640A..69K}) reveal a pronounced flux enhancement between April 5–6 and 10–11, with the core flux nearly doubling from $\sim$0.5\,Jy to $\sim$0.95\,Jy. During this period, the corresponding brightness temperature changed from approximately $6\times10^{10}$\,K to $4.5\times10^{10}$\,K. In contrast, the immediately adjacent components C0–1 and C0–2, located at projected separations of 15–30\,$\mu$as (corresponding to $\sim$0.1–0.2\,pc from the core) and exhibiting superluminal motion, show opposite trends: C0–1 displays a slight dimming, while C0–2 brightens.
An increase in the core flux density $F_\nu$ is expected to be accompanied by a shift of the synchrotron self-absorption turnover frequency $\nu_m$ toward higher values. The scaling of $\nu_m$ with $F_\nu$ depends on whether the variability is driven by changes in the relativistic electron density normalization $N_0$, magnetic field strength $B$, or Doppler factor $\delta$. Using Eqs.~(1) and (2) of \citet{marsher1985}, 
\begin{equation}
F_\nu \propto N_0 B^{(s+1)/2} \delta^{(s+3)/2}
\label{eq1}
\end{equation}
and
\begin{equation}
\nu_m \propto N_0^{2/(s+4)} B^{(s+2)/(s+4)} \delta^{(s+2)/(s+4)},
\label{eq2}
\end{equation}
where $-s$ is the slope of the electron energy distribution and the dimensions of the source are assumed to be constant, we can determine how $\nu_m$ is expected to vary when either $N_0$, $B$, or $\delta$ changes while the other parameters remain constant. In each case, we used relation \ref{eq1} to solve for the variable parameter in terms of $F_\nu$ and the two constant parameters, and we then substituted that expression into expression \ref{eq2} to obtain
\begin{equation}
\begin{aligned}
\nu_m &\propto F_\nu^{2/(s+4)}
&& ({\rm variation~in~} N_0), \\
\nu_m &\propto F_\nu^{2(s+2)/[(s+1)(s+4)]}
&& ({\rm variation~in~} B), \\
\nu_m &\propto F_\nu^{2(s+2)/[(s+3)(s+4)]}
&& ({\rm variation~in~} \delta).
\end{aligned}
\label{eq}
\end{equation}
These scalings are valid provided that the observing frequency is not already close to the synchrotron self-absorption turnover frequency, where the spectral shape becomes strongly affected by opacity effects. For the value derived from our modeling, $s=2.4$, the corresponding exponents are 0.31, 0.40, and 0.25, implying an increase in $\nu_m$ by factors of approximately 1.2, 1.3, and 1.2, respectively, when the 230 GHz flux density increases by a factor of 1.9, as we observed.

At larger scales, and at centimeter and millimeter radio frequencies, no significant flux variability on daily timescales is observed.
In contrast, the source exhibits unprecedented flaring activity in the UV–optical bands, peaking on March 21–22 and reaching the maximum in the I-band at 22:26 UTC on March 21, with a magnitude of $12.309 \pm 0.002$. The synchrotron flux increases by a factor of $\sim$20 relative to historical low states, reaching its highest level ever reported \citep{Larionov2020}, while over the one-month timescale considered here, the optical emission varied by approximately an order of magnitude. Moreover, this record optical flare occurred during the rising phase of a strong $\gamma$-ray flaring episode (see also Sect.~\ref{sec:gamma-fermi}).
The optical polarization also varies strongly, with the polarization degree ranging from 5\% to 25\% and the polarization angle varying between 20$^\circ$ and 80$^\circ$.
During the 2017 EHT campaign, the source was in a low X-ray emission state, with a 0.3--10 keV flux of $F_{0.3-10\,\mathrm{keV}}=(1.5\pm0.3)\times10^{-11}$ erg cm$^{-2}$ s$^{-1}$ and a photon index of $\Gamma_X = 1.5 \pm 0.1$. This flux is slightly lower than the long-term Swift-XRT average measured over the 2006–2022 period, $F_{0.3-10\,\mathrm{keV}}=1.93\times10^{-11}$ erg cm$^{-2}$ s$^{-1}$ \citep{Thekkoth2023broadband}.

In the $\gamma$-ray band, 3C\,279 underwent intense flaring episodes, with \textit{Fermi}-LAT three-hour binned fluxes reaching up to ten times the long-term (16-year) average ($\phi_{E>100\,\mathrm{MeV}} = (6.32\pm0.03) \times 10^{-7} \mathrm{ph\ cm^{-2}\ s^{-1}}$). For reference, the integrated flux reported in the 4FGL-DR4 catalog is  $\phi_{E>100\,\mathrm{MeV}} \sim 5 \times 10^{-7} \mathrm{ph\ cm^{-2}\ s^{-1}}$ \citep{2022ApJS..260...53A}.  
During the EHT observing window (April 5–11), the source exhibited an average $\gamma$-ray flux of $\phi_{E>100,\mathrm{MeV}} = (2.07 \pm 0.07)\times10^{-6}\ \mathrm{ph\ cm^{-2}\ s^{-1}}$. The emission was even enhanced on April 5–6, reaching $\phi_{E>100,\mathrm{MeV}} = (2.90 \pm 0.12)\times10^{-6}\ \mathrm{ph\ cm^{-2}\ s^{-1}}$, and it subsequently declined toward its typical level by around April 9, shortly before the end of the EHT observing campaign.
At VHEs, no significant emission was detected by \HESS or MAGIC during this period.

During the first EHT observing window (April 5-6), we measured a fractional variability amplitude (following \citealt{Vaughan_2003}) of $F_{\rm var} = (40 \pm 5)\%$ in the HE $\gamma$-ray light curve. The shortest flux--halving time derived from the Fermi-LAT data (using the method of \citealt{Zhang_1999}) is $t_{1/2} \simeq (9.0 \pm 2.5)\,\mathrm{h}$. The largest variability amplitudes are observed in the optical bands, with $F_{\rm var} \sim 82\%$ in $R$ and $\sim 86\%$ in $I$. The minimum optical variability timescale, $T_{2}^{\rm min} = (0.59 \pm 0.03)\,\mathrm{d}$ ($\sim (14 \pm 1)$\,h), is comparable to previous constraints.

While the \textit{Fermi}-LAT light curve exhibits numerous short-timescale outbursts, the optical light curves display a broader peak and no clear counterpart to the rapid $\gamma$-ray activity. High optical fluxes occasionally coincide with high LAT fluxes, but the converse is not always true: comparable HE $\gamma$-ray fluxes may be observed during lower optical states, and a pronounced LAT flare occurred during an optical low state. A prolonged period of low emission is then observed in both bands. Overall, the Pearson correlation coefficient during the full period considered (see Fig. \ref{fig:MWL_LC_2017}) is $r \sim 0$, and it remained modest ($r = 0.42 \pm 0.19$) up to MJD 57840, corresponding to the period of a simultaneous high state in the optical and $\gamma$-ray bands.
\citet{Emery_2019} reported a slight correlation between the HE $\gamma$-ray and optical fluxes ($r = 0.61 \pm 0.05$) when they restricted the analysis to MJD 57805--57840, immediately preceding our observing window, during which optical flares and increased HE $\gamma$-ray activity were detected. The same study found a much stronger correlation ($r \sim 0.9$) during the June 2018 optical and $\gamma$-ray flaring episode.

\subsection{Multiwavelength spectrum}
\begin{figure}[ht!]
\begin{center}
\includegraphics[width=\columnwidth]
{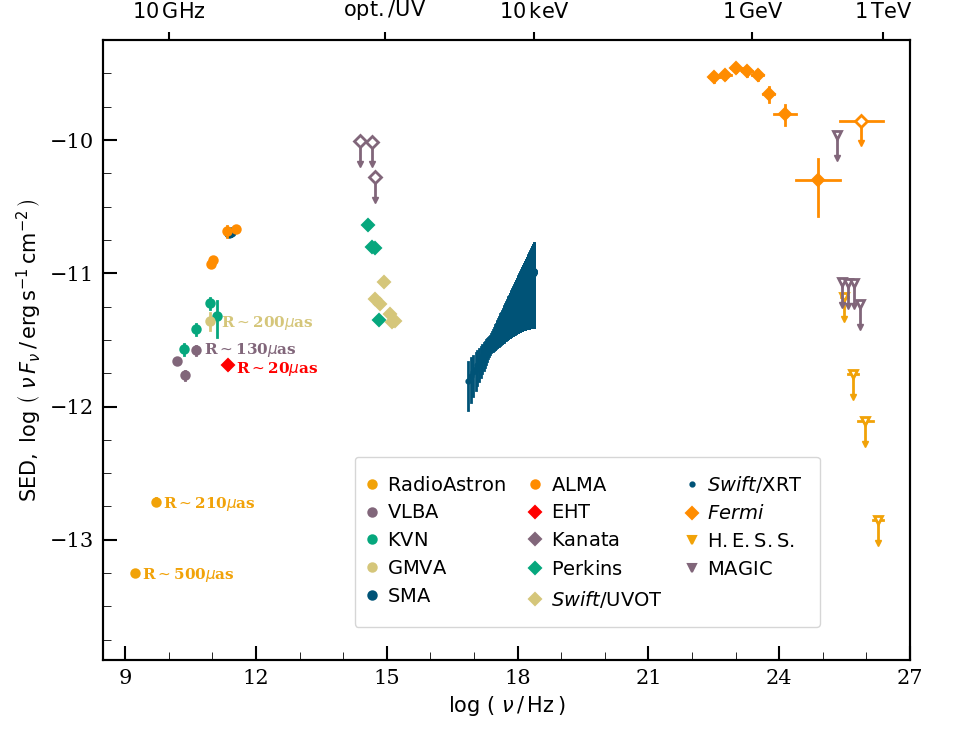}
    \caption{MWL broadband SED of 3C\,279 around the EHT observational campaign (MJD 57836–57861; see Table \ref{tab:sed2017} in Appendix \ref{appendix:mwl_sed}). For RadioAstron, integrated fluxes from cumulative observations between 2016 and 2018 are shown. The Swift-XRT spectrum is shown as a bow tie. The fluxes are indicated with different colors and markers according to the instrument. For radio VLBI and interferometric data, upper limits on the emission size (radius) at selected frequencies are labeled. }
\label{fig:mwl_sed}
\end{center}
\end{figure}

Figure~\ref{fig:mwl_sed} shows the quasi-simultaneous 2017 MWL broadband spectral energy distribution (SED) of 3C\,279 obtained around the time of the EHT observations. To select the closest-in-time MWL data, the EHT observation on April 10 was used as a reference, since the only X-ray observation within the EHT window occurred on April 8. For the radio data included in the MWL SED, we considered the innermost region accessible at each observing frequency, corresponding to either the compact core alone (C0–0 component for the EHT data; see \citealt{2020A&A...640A..69K}) or, for non-EHT observations, to the combined emission from the core and the inner jet (see Table \ref{tab:sed2017}). During the modeling phase (see Sect. \ref{sec:modelling}), our first goal was to test whether the broadband emission (from radio up to $\gamma$ rays) might originate predominantly from the central regions resolved by the EHT. 

The SED exhibits the characteristic two-hump structure typical of blazars. The low-energy hump, produced by synchrotron emission, peaks in the IR–optical band, while the high-energy hump, attributed to inverse-Compton processes in most models (for a review, see \citet{2016ARA&A..54..725M}), peaks in the $\gamma$-ray band at GeV energies. At this epoch, the source shows high Compton-dominance, with the high-energy component clearly dominating the low-energy synchrotron emission.

High-resolution VLBI observations spatially resolve the source structure, whereas at higher frequencies, the source appears to be point-like because the angular resolution is limited (see Sect.~\ref{sec:radio_observations}). The SED also indicates the spatial scales corresponding to the total emission for the highest-resolution observations (for more information, see also Table \ref{tab:sed2017}). All flux points correspond to statistically significant detections ($>3\sigma$) with 1$\sigma$ uncertainties, except for upper limits, which are plotted at the 2$\sigma$ confidence level.

\section{Modeling the source emission}
\label{sec:modelling}

Visual inspection of Fig. \ref{fig:MWL_LC_2017} reveals a wide range of variability in the flux and linear polarization (degree and position angle) of 3C\,279 at all observed wavelengths. There are no statistically significant trends in the individual light curves that persist on scales of months or years. A dedicated analysis will be presented in a forthcoming publication. This implies that stochastic processes play a major role in the emission. Furthermore, the position angles of optical and millimeter-wave polarization fluctuate by roughly $\pm20\degr$ about the jet direction. This requires substantial reconfiguration of the magnetic field pattern.
This coexistence of ordered and turbulent magnetic field components can be produced by the superposition of a turbulent plasma embedded in a partially ordered magnetic field. The ordering can result from a helical magnetic field or from a chaotic field that is partially compressed when it crosses a shock. 

\subsection{The TEMZ model}

Based on these considerations, we employed the Turbulent Extreme Multi-Zone\footnote{TEMZ is implemented via a Fortran-95 code and available at \url{www.bu.edu/blazars/temz.html}.} (TEMZ; \citealt{Marscher2014,Marscher2017}; see also \citealt{Zhang2023} and \citealt{Liodakis2024}) numerical model in an attempt to reproduce the time-variable broadband behavior and polarization of 3C\,279. In this model, magnetized plasma flows at near light speed across a standing conical shock in a jet, as can occur where there is a mismatch between the internal and external pressure. We assumed that the upstream plasma consists of ultra-relativistic electrons and positrons,
presumably energized via many minor magnetic reconnections
and second-order Fermi acceleration, and colder protons (whose radiation was neglected here). 
The plasma was turbulent, although the magnetic field also included a helical component with opening angle $\psi$ (whose extremes correspond to a toroidal field if $\psi=90\degr$ and a longitudinal field if $\psi=0\degr$). The shock compressed the component of the magnetic field that is perpendicular to the shock normal (as measured in the plasma frame), following the relativistic jump conditions \citep[e.g.,][]{Konigl1980,Cawthorne1990}. Diffusive shock acceleration increased the energies of the particles, forming a power-law energy distribution with a slope of $-s=-2.3$ up to a maximum energy $\gamma_{\rm max}mc^2$. At this energy, the acceleration and cooling times are equal for magnetic field directions that allow multiple back-and-forth passages of particles across the shock front \citep[the ``subluminal'' regime; see][]{Baring2017}. To determine the acceleration time, we used the \citet{Baring2017} formulation, within which the mean-free path of pitch-angle scattering is proportional to the energy of the particle.

\subsection{TEMZ implementation for 3C\,279}

\begin{figure}
\includegraphics[width=\columnwidth]{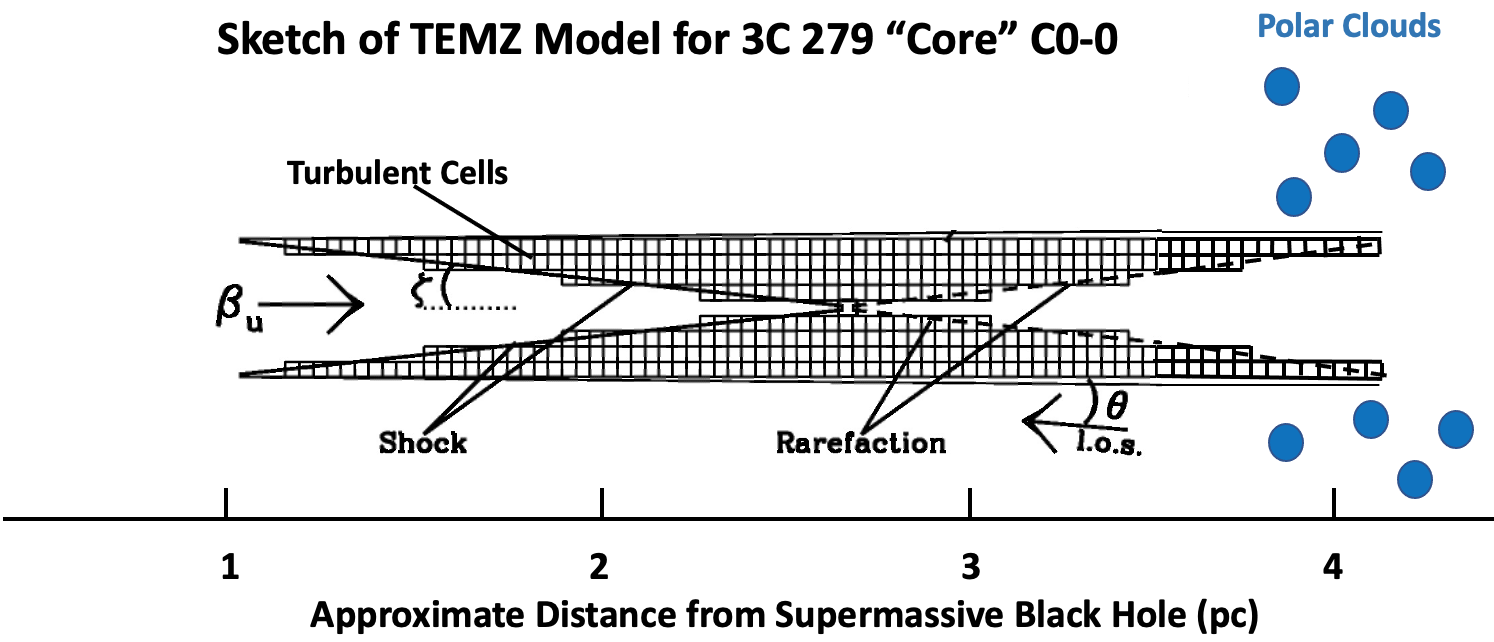}
    \caption{Sketch of the TEMZ model for 3C\,279 adapted from \citet{Marscher2014}. The emission region, ranging from the shock to the rarefaction, comprises 2880 turbulent cells. The line of sight (l.o.s) is inclined by angle $\theta$ to the jet axis. }
    \label{fig:TEMZ_sketch}
\end{figure}

We applied the TEMZ model by interpreting component C0-0 (EHT inner core) as a standing conical shock, located $\sim1$\,pc from the supermassive black hole. The emission region extended from the shock front to a downstream rarefaction, as sketched in Figure \ref{fig:TEMZ_sketch}. In order to reproduce the apparent superluminal motion of knots in 3C\,279, observed to be as high as $\sim40c$ in 2017 \citep{Larionov2020}, we set the bulk Lorentz factor of the flow to be 40\footnote{The bulk Lorentz factor needs to be high enough to attain the highest apparent speed observed during this period. While the Lorentz factor might increase downstream, the most rapid variability of 3C\,279 (in the $\gamma$-ray band) also requires $\Gamma \sim$ 40. A slightly wider or narrower viewing angle in C0 would cause the apparent speed there to be lower.} (the minimum value given the apparent speed) upstream of the shock, and the angle between the line of sight and jet axis to be $1.5\degr$. This assumes no change in the flow speed between the high-energy emission region and the knots observed at 43 GHz. This requires that all of the bulk acceleration of the jet occurs
upstream of component C0-0. We adopted an angle $\zeta=6\degr$ between the shock front and the jet axis, which is inclined enough to allow substantial compression without decelerating the plasma flow too much \citep[see][]{Cawthorne1990}.

The emission region contained a total of 2880 cylindrical cells, each with its own electron density, electron energy distribution, and unidirectional magnetic field. This number of cells reproduces the observed mean degree of optical polarization, which varies on a timescale of years over a range of $\sim10$--$20\%$ \citep{Larionov2020}. More turbulent cells would cause greater cancellation of randomly unaligned polarization vectors. Each cell had a diameter of $ \ell=0.04$\,pc and a length of $ \Delta z=0.048$\,pc, the latter value chosen so that the upstream end of each ring of columns was offset by an integer (eight) number of cell lengths (see Fig.\ \ref{fig:TEMZ_sketch}). The magnetic field included a helical and a turbulent component. The latter was assigned via a random number generator and was related to the turbulent field component of nearby cells via the relativistic Kolmogorov spectrum. This relation was accomplished in the manner of \citet{Jones1988}: each cell belonged to four zones: its own, one zone containing 8 ($2\times2\times2$) cells, another zone with 64 ($4\times4\times4$) cells, and one zone with 512 ($8\times8\times8$) cells. 

The code selects the magnitude of the magnetic field, density and the direction of the field, randomly in each zone from a log-normal distribution. The field and density of a cell are then computed as the average of the values in its zones, weighted according to the Kolmogorov spectrum: weight $ w\propto \ell^{2/3}$ for density, magnetic energy density, and square of the turbulent four-velocity. This weighting causes the larger zones to contribute a larger fraction of each quantity. In order to approximate the rotation of turbulent vortices, the code relativistically adds a turbulent circular four-velocity of $0.1c$, rotating about the center of each zone, to the systemic flow velocity \citep[see][]{Calafut2015}.

We assumed that the electrons (including any positrons) at the upstream boundary of the emission region (i.e., just beyond the shock front) had a power-law energy distribution between a minimum energy $\gamma_{\rm min}mc^2$ and maximum energy $\gamma_{\rm max}mc^2$, the latter of which depends on the angle of the magnetic field relative to the shock normal as measured in the plasma frame (see above). This dependence of $\gamma_{\rm max}$ on the magnetic field direction, and the higher radiative energy losses of the higher-energy electrons as the plasma flows downstream of the shock front, means that the electron energies of only a fraction of the cells are high enough to produce synchrotron and Compton radiation at the highest frequencies. These effects cause the slope of the synchrotron and Compton spectra to steepen with frequency, so that, for example, the optical spectrum is steeper than that at infrared frequencies. The combination of the two effects causes the mean polarization and its level of variability to increase with frequency, as has been observed at X-ray and optical frequencies in several blazars \citep[e.g.,][]{Liodakis2022,Marscher2024,2025A&A...695A.217M}.

As the code runs, the plasma advances by one cell length $\Delta z$ per time step. New values of the magnetic field and electron density are introduced at the upstream boundary. These values are derived from the turbulence simulation scheme discussed above, which produces short-term fluctuations, then multiplied by a time-dependent input of total energy density. The latter, which corresponds to variations in the energy injected into the base of the jet, is computed from random sampling of a power spectral density (PSD) in the form of a power law with a slope of $-2$, the value of which is consistent with the optical and X-ray timing analysis of 3C\,279 by \citet{Chatterjee2008}. To apply this, we adapted a subroutine supplied by R.\ Chatterjee, based on the algorithm of \citet{Timmer1995}, which produces a large set of random values $x_i$ between $-1$ and 1 with a power-law PSD. Following \citet{Uttley2005}, the TEMZ code exponentiates each number produced by the subroutine in order to derive a multiplicative factor $\exp(\xi x_i)$ that the code applies to the input value of the energy density, where $\xi$ is a free parameter of the order of unity that determines the ratio of the highest to lowest flux density over time ranges of years. The ratio is observed to be $\sim 100$ in the case of the optical variability of 3C\,279 \citep{Larionov2020}. During the time span of the observations reported here, the range was about one order of magnitude (see Fig.\ref{fig:MWL_LC_2017}).

In the version of the TEMZ code we used for this work, the magnetic field and relativistic electron density of a cell remained constant as the plasma advanced downstream from the shock. However, the energy distribution of the electrons evolved toward lower energies with distance from the shock as they lost energy to synchrotron and Compton radiation (see equations (5) and (6) of \citet{Marscher2014}). The energy-loss rate varied from one cell to the next, commensurate with the differences in the magnetic field and seed photon field among the cells. Adiabatic energy losses were neglected because the jet did not spread appreciably over the length of the computational grid.

The TEMZ code computes the flux density from incoherent synchrotron (including synchrotron self-absorption), synchrotron self-Compton (SSC), and external Compton (EC) radiation. The relevant formulas are given in \citet{Marscher2014}. The main difference of the model for 3C\,279 from the earlier TEMZ versions is that the seed photons for SSC emission produced by a particular cell are from all of the other cells (instead of a Mach disk). The seed photon flux from other cells is suitably time-delayed and Doppler shifted, the latter of which takes into account variations in the turbulent velocities of the different cells. In addition, the seed photons for EC scattering are from a torus containing hot (1200\,K) dust and from broad emission-line clouds. For the latter, we used the prescription of \citet{Tavecchio2008}, approximating the spectrum of the clouds as a blackbody with a temperature of 42,000\,K. 

The peak of the $\gamma$-ray SED of 3C\,279 (Fig. \ref{fig:mwl_sed}) is an order of magnitude higher than that of the IR synchrotron emission, and it reached maximum at an energy of $\sim1$\,GeV. These properties are difficult to reproduce with synchrotron seed photons from the jet \citep{Sikora2009}. 
Compton scattering of photons from the dust torus results in an SED peaking in the 0.01--0.1\,GeV region in a blazar such as 3C\,279 \citep{Sikora2009}, which is also inadequate. The canonical broad emission line region (BLR) is expected to consist of clouds in a shell or torus with an outer boundary $\lesssim0.2$\,pc from the SMBH \citep[see][for a discussion]{Tavecchio2009}.
\citet{Liu2008} showed that $\gamma$ rays observed with energies $\gtrsim30$\,GeV would be heavily depleted by pair production if they were emitted inside such a small dense BLR.

There is, however, another potential source of seed photons for Compton scattering to produce the $\gamma$ rays: polar line-emitting clouds that lie along the outer boundary of the parsec-scale jet. Such clouds have been inferred from emission-line variability in 3C\,279 \citep{Larionov2020} and 4C29.45 \citep{Hallum2022}. A variable component of the Mg~II line is on the red side of the main line, which implies that it is from clouds falling toward the SMBH, possibly as part of a back-flow surrounding the jet \citep{Hallum2022}. The line flux in both objects is proportional to the (beamed) optical flux from the jet, which indicates that the clouds lie along the jet. Since flares are short lived, we adopted a luminosity of the Mg~II line from polar clouds in the low range reported by \citet{Larionov2020} and multiplied this by a factor of $\sim10$ to estimate the luminosity of all lines and continuum \citep[see][]{Peterson1997} from the polar clouds to be $\sim7\times10^{43}$\,erg\,s$^{-1}$. We further adopted a distance of the clouds from the SMBH of $\sim4$\,pc and a distance from the jet axis of $\sim0.3$\,pc, which is somewhat smaller than the values estimated for the more luminous quasar 4C29.45 by \citet{Hallum2022}.

The code calculates the pair-production opacity of VHE $\gamma$ rays as they encounter photons from the line-emitting clouds and the extragalactic background light. These two sources of opacity to $\gamma$ rays dominate in the model. The integrated (over solid angle) intensity of the optical-UV radiation from the clouds is calculated as a function of frequency at the position of every cell. The intensity times the frequency-dependent cross-section \citep{Coppi1990} is integrated over optical-UV frequency to obtain the pair-production optical depth at each VHE energy. This is added to the optical depth from the extragalactic background light as determined by \citep{Dominguez2011}.
A number of input parameters of the code can be set based on observational information. Table \ref{tab:temzparms} lists these parameters and their values.
The mean kinetic power of the simulated jet is $\sim7\times10^{46}$ erg s$^{-1}$ if there are about as many protons as electrons in the jet. If positrons are the main positively-charged component, this is reduced by a factor of $\sim 2$.

\subsection{Model results}

We ran a small number of sample computations of the TEMZ code in order to find values of the free parameters that produced flux, polarization, and SED versus time that roughly resembled the observed behavior of 3C\,279. Table \ref{tab:temzparms} describes these parameters and gives their adopted values.

\begin{table*}[]
    \caption{Parameters of TEMZ Model for 3C~279}
    \label{tab:temzparms}
    \begin{tabular}{lll}
    \multicolumn{3}{c}{\textbf{Parameters based on observations}}\\
         Parameter &Value &  Description \\
         \hline
         $v_{\rm lam}$ & $0.9997c$ & Bulk flow velocity of laminar component of the jet flow (Lorentz factor = 40.8)\\
         $\theta$ & $1.5\degr$ & Angle between jet axis and line of sight (set such that values of $v_{\rm lam}$ and $\theta$ result \\
         && in apparent speed of $37c$ as observed in 2017 in \citet{Larionov2020})\\
         $v_{\rm turb}$ & $0.1c$ & 4-velocity of turbulent rotational component of the jet flow \citep[see][]{Calafut2015}\\
         $s$ & 2.4 & Slope of electron power-law energy distribution immediately past the shock front\\
         $\ell$ & 0.04 pc & Transverse diameter of a cell\\
         $\Delta z$ & 0.048 pc & Length of a cell\\
         $\phi$ & $0.9\degr$  & Intrinsic opening half-angle of jet \citep{2022ApJS..260...12W}\\
         --- & $7\times10^{43}$ erg s$^{-1}$ & Luminosity of thermal emission from polar clouds\\
         --- & 4 pc & Mean distance of polar clouds from SMBH\\
         --- & 0.3 pc & Mean distance of polar clouds from jet axis\\
        \hline
\multicolumn{3}{c}{\textbf{Parameters based on modelling}}\\
         Parameter &Value &  Description \\
         \hline
         $\sigma_B$,$\sigma_n$ & 0.3 & Dispersion-to-mean ratio of turbulent fluctuations in magnetic energy and electron densities\\
         $\xi$ & 1.2 & Parameter in the exponent of the modulation of the input energy\\
         $\langle B \rangle$ & 0.03 G & Mean value of magnetic field immediately upstream of shock\\
         $u_e/u_B$ & 8 & Mean ratio of relativistic electron to magnetic energy densities immediately upstream of shock\\
         $N_{\rm cells}$ & 2880 & Total number of computational cells\\
         $E_{\rm max,min}$ & $2\times10^5 mc^2$ & Highest value of maximum electron energy after plasma crosses shock\\
         $E_{\rm max,max}$ & $1\times10^4 mc^2$ & Lowest value of maximum electron energy after plasma crosses shock\\
         $E_{\rm min}$ & $800 mc^2$ & Minimum electron energy after plasma crosses shock\\
         $\langle B_{\rm ord}/B_{\rm tot}\rangle$ & 0.7 & Mean ratio of ordered helical to total (helical+turbulent) magnetic field\\
         $\psi$ & $70\degr$ & Pitch angle of helical magnetic field\\
        \hline
         \hline
    \end{tabular}
\end{table*}

After setting the above parameters based on the results of the trial runs, we ran the code for 10,000 time steps in order to produce the simulated flux, polarization, and SED at each step to compare them with the data. Because of the randomness inherent in the code (and in some of the actual physical processes), TEMZ cannot reproduce the actual flux and polarization curves. Instead, a successful simulation was one in which the above variations resembled the observed behavior. Since the observations have many temporal gaps, the comparison can only be made rather crudely: the mean and standard deviation of the degree and position angle of optical polarization, range of flux variations during a given time interval at different wavebands, and single-epoch SEDs.

\begin{figure}
\includegraphics[width=0.45\textwidth]{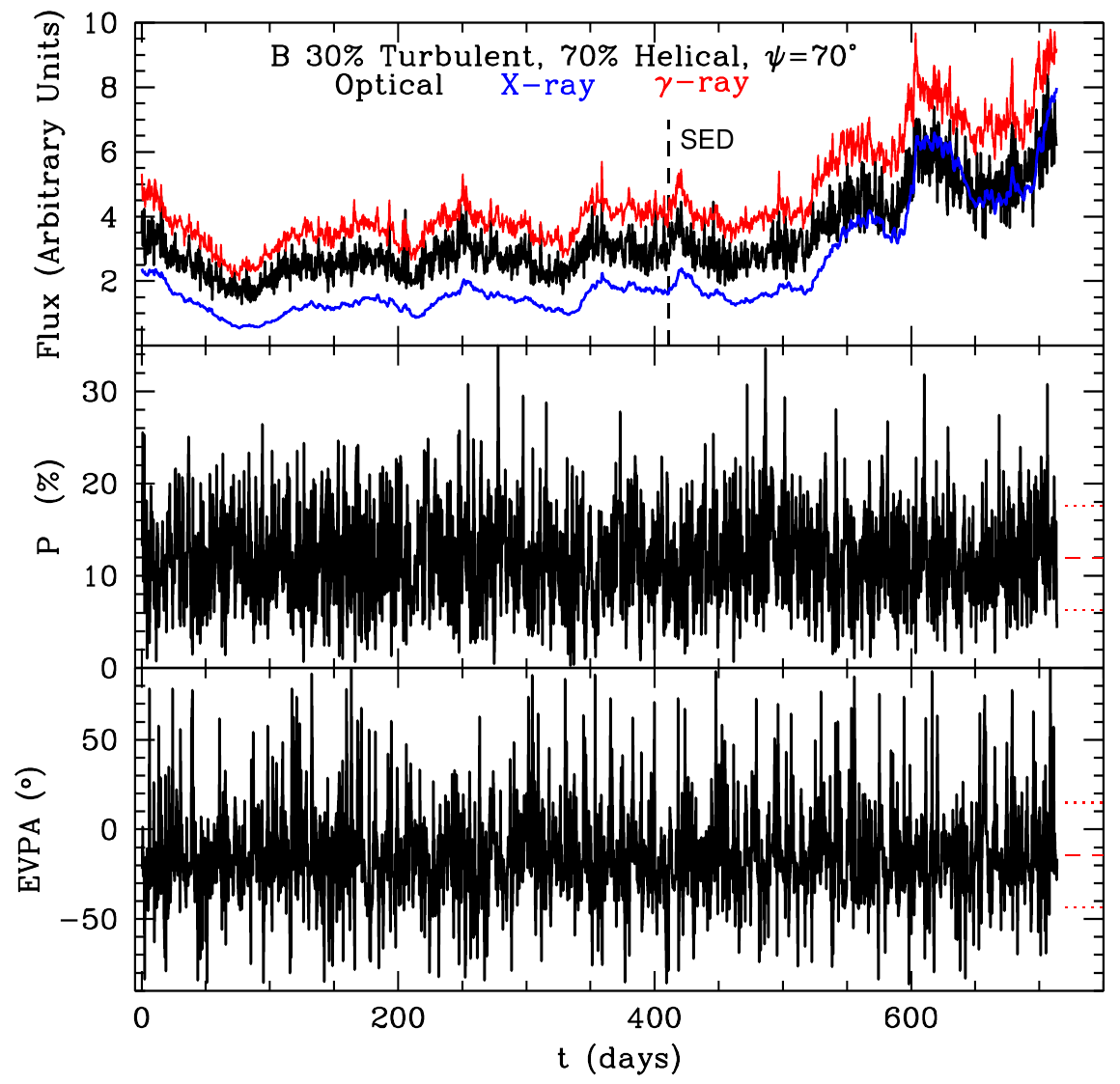}
    \caption{Sample simulated MWL flux and optical polarization degree $P$ and EVPA vs.\ time calculated with the TEMZ model. EVPA = $0\degr$ in the simulation corresponds to the jet direction. The time step corresponding to the SED presented in Fig.\ \ref{fig:sed_3c279_2017TEMZ} is marked in the top panel. The means of $P$ and the EVPA are indicated with dashed lines on the right of the respective panels, while the standard deviations are marked with dotted lines.}
    \label{fig:temz_mwlc}
\end{figure}

\begin{figure}
\includegraphics[width=0.45\textwidth]{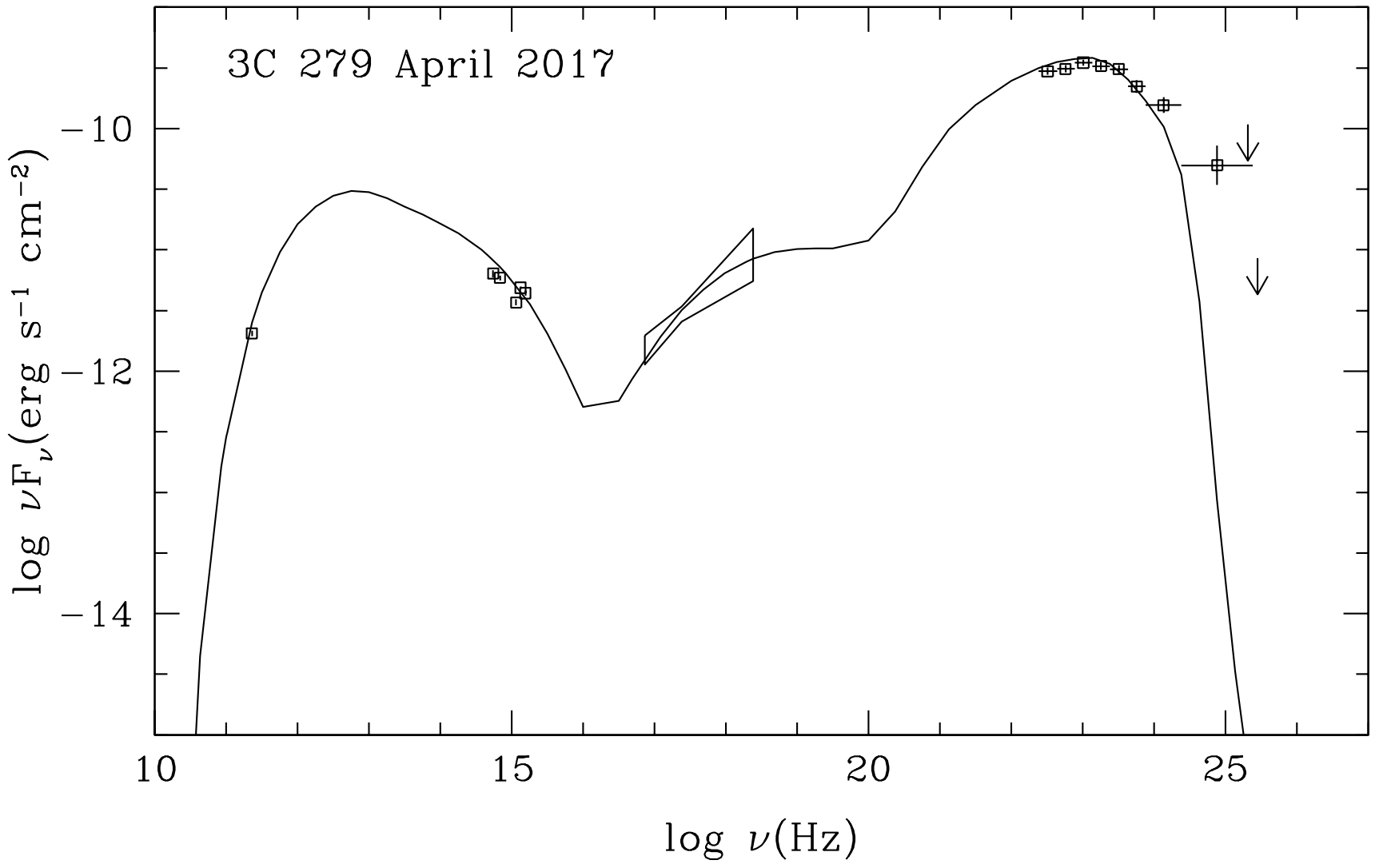}
    \caption{Spectral energy distribution of 3C\,279 in April 2017 (empty squares). Under the assumption that the high-energy emission arises from the core region (C0-0 in the EHT images), only EHT, Swift-UVOT, X-ray, and $\gamma$-ray spectral points are shown. Lower-resolution radio and optical data are excluded as they sample larger emission regions. The solid curve shows a representative SED of component C0-0 from the TEMZ model, selected at the epoch when the simulation reproduced the observed multiwavelength data best (see Fig. \ref{fig:temz_mwlc}).
    \label{fig:sed_3c279_2017TEMZ}}
\end{figure}

Figure~\ref{fig:temz_mwlc} shows the simulated light curves of flux and polarization as a function of time. The simulated fluxes span approximately one order of magnitude, consistent with the variability observed in 3C\,279 (see Fig.~\ref{fig:MWL_LC_2017}). The mean optical EVPA differs only slightly from the jet direction (by $13\degr$) and exhibits a standard deviation of $27\degr$, comparable to the observational data. (The mean EVPA would align exactly with the jet axis if $\psi = 90\degr$.) The mean degree of polarization in the simulation is 12\%, with a standard deviation of 6\%.
These simulated values agree well with the observations, which yield a mean polarization degree of (14 $\pm$ 5)\% and a mean optical EVPA of (46.5 $\pm$ 13.9)\degr, derived from the observational data binned in two-day intervals.

Figure \ref{fig:sed_3c279_2017TEMZ} displays the simulated SED of component C0-0 at a time when the simulation most closely resembled the data. The radio to optical portion is from synchrotron radiation, while SSC emission dominates the X-ray flux. The $\gamma$-ray emission is mainly external Compton of emission-line photons from the polar clouds. The polar clouds contribute to the optical-UV emission lines and to a relatively weak thermal continuum component in the optical–UV range.

\section{Discussion}
\label{sec:discussion}

\subsection{Jet structure}

Microarcsecond-scale angular resolution observations of 3C\,279 with RadioAstron on March 10, 2014, resolved the jet transversely, revealing multiple filaments that were likely produced by plasma instabilities within a kinetically dominated flow \citep{2023NatAs...7.1359F}.
Due to the sparse sampling of the Fourier plane, the 2017 EHT observations \citep{2020A&A...640A..69K} were only able to recover one central filament, shown in Fig.~\ref{fig:composite_radio} as a separated extended jet, which was oriented perpendicularly to the elongated nuclear region. On the other hand, our VLBI observations at longer wavelengths do not resolve the filamentary jet structure at all.
\citet{2023NatAs...7.1359F} suggested that enhanced Doppler boosting might occur when emitting regions traverse such sites where the local jet direction becomes more closely aligned with the line of sight, leading to the observed flux variability. 
\citet{2023NatAs...7.1359F} concluded that resolved bright filaments of the jet of 3C\,279 represent compressed regions with enhanced gas and magnetic pressure and might be associated with the triggering and development of flow instabilities. Current-driven kink or Kelvin–Helmholtz instabilities are the most plausible mechanisms for developing such helical structures.
On the other hand, the elongated complex nuclear structure at 230\,GHz might originate from an extreme bend of the jet.

More recent RadioAstron observations obtained in January 2018 similarly revealed fine-scale structure within the relativistic jet, extending up to a projected distance of approximately 180\,pc from the radio core \citep{2025arXiv250921987T}. Although no filamentary features were detected in these later observations, this might be due to the substantial decline in the total jet intensity and the limited \textit{uv} coverage. The data further suggest a predominantly toroidal magnetic field configuration, consistent with an overall helical magnetic structure.
Independent constraints from multiwavelength variability provide a complementary picture. The behavior of the optical and radio (including 43\,GHz) Stokes parameters \citep{Larionov2020} is also compatible with a predominantly helical magnetic field configuration, or alternatively, with radiating plasma moving on a spiral trajectory. 

Recently, \citet{2025MNRAS.538.2008K} used 15.4\,GHz multi-epoch monitoring data to analyze the dependence of the jet component size, $r$, on the distance from the VLBI core, $d$, for a sample of 447 AGN jets. They concluded that the size of the jet components is a good tracer of jet geometry. Particularly for 3C\,279, \citet{2025MNRAS.538.2008K} obtained the dependence $d\propto r ^ {0.92\pm0.1}$, which implies that the blazar jet follows a conically expanding streamline at scales $\thicksim0.2-2.4$~mas ($\thicksim 7 - 90$~pc deprojected with $\theta=1.5^{\circ}$).
This result is consistent with the estimates of the spatial extension of the intrinsic jet acceleration zone, which are that this zone lies within 100~$\mu$as from the core ($\thicksim25$~pc deprojected) \citep{2020A&A...640A..69K}

Using the 2016--2018 short-segment observations of 3C\,279 at 18\,cm and 6\,cm with RadioAstron (see Sect.~\ref{sec:radioastron}), we fitted the measured visibility amplitudes with three Gaussian components: the largest component, which represented the extended jet; the VLBI core; and a substructure within the core. The latter exhibits a high observed brightness temperature of $T_{\rm b}\thicksim10^{13}$\,K, exceeding that of the core itself. This suggests the presence of more energetic particles in the region.
In the TEMZ scenario of the turbulent plasma crossing a shock, and given the Doppler factor > 30, the rest-frame brightness temperature of a single cell is then $T_{\rm b} \thicksim 3\times 10^{11}$~K. This indicates that any of the compact turbulent cells could have such a high $T_{\rm b}$ at cm wavelengths.

From the 2017 EHT observations, \citet{2020A&A...640A..69K} observed significant day-to-day variations in the closure phases, which appear to be most pronounced in the triangles with the longest baselines. These authors showed that this variation is related to a systematic change in the source structure, which is accompanied by rapid changes in the brightness temperature within the subnuclear structure of the 3C\,279 jet over the time interval of a week. Two inner jet components move non-radially at apparent speeds of $\sim$15\,c and $\sim$20\,c, which more strongly supports the scenario of traveling shocks or instabilities in a bent, possibly rotating jet. 

The components in the jet of 3C\,279 are known to exhibit significantly different speeds and position angles \citep{2004AJ....127.3115J, 2022ApJS..260...12W}, which have been interpreted as evidence of a precessing jet \citep{1998ApJ...496..172A}. Jet-nozzle precession models suggest that the components follow a curved precessing path with a period of $\sim$25\,yr \citep{2011RAA....11...43Q, 2012RAA....12...46Q}.
However, EHT observations \citep{2020A&A...640A..69K} revealed erratic changes in the inner jet position angle in 2011 and 2017, implying a precession period $<$6\,yr if periodic. Shorter-scale quasi-periodic variations of $\sim$3\,yr have also been reported \citep{2021ApJ...923...30L}. Observations at 1.3~mm in 2011 indicated that the inner jet orientation does not fully match precession-model predictions, suggesting an additional erratic process \citep{2013ApJ...772...13L}. A light-curve analysis across radio, optical, and X-ray bands found a period of $P = 130.6 \pm 1.3$ days \citep{2009PASP..121.1172L}.
Long-term PA changes ($>$10\,yr) might still be governed by jet-nozzle precession, consistent with the $\sim$39\degr{} smooth shift observed between extended radio jet regions (Fig.~\ref{fig:composite_radio}). In contrast, shorter-term variations are likely produced by MHD instabilities in disk-jet systems, as suggested by recent 3D MHD simulations \citep{hu2025}.

Similar multi-period jet position angle variations have been observed in M87 \citep{2026ApJ...999..169R, 2023Natur.621..711C}, with an 11\,yr period consistent with Lense–Thirring precession from a tilted accretion disk \citep{2023Natur.621..711C}. Shorter $\sim$1\,yr transverse oscillations \citep{2026ApJ...999..169R} might result from transverse MHD waves (excited by precession, nutation, or magnetic flux eruptions near the central engine) or from jet instabilities such as current-driven or Kelvin–Helmholtz modes. Evidence of the latter has been recently reported within the Bondi radius, revealing helical and elliptical surface modes \citep[][]{2023MNRAS.526.5949N,Kravchenko2026}.
Substantial evidence exists of periodic position-angle variations in other AGN jets \citep[e.g.,][]{2025MNRAS.537..978K, 2025A&A...693A...9L}. Moreover, precessing-jet models successfully reproduce the complex linear polarization profiles observed in AGN jets \citep{2023ARep...67.1275T}, which are inconsistent with simple magnetic field configurations \citep{2023MNRAS.520.6053P}.

\subsection{Origin of the $\gamma$-ray flaring activity}

As discussed in Sect.~\ref{sec:modelling}, the observed spectral and polarization variability can be interpreted within a turbulent plasma jet model. In this framework, turbulent cells crossing a stationary conical shock are compressed, which triggers rapid $\gamma$-ray flares and produces polarization variability, including EVPA rotations and increases in polarized intensity. Our TEMZ simulations reproduced flux variations of roughly one order of magnitude, mean optical polarization degrees of 12\% ($\sigma \approx 6\%$), and EVPA deviations of 13\degr{} from the jet axis with $\sim$27\degr{}, consistent with the observed values of $14\pm5\%$ and $46.5 \pm13.9$\degr{}, see Fig. \ref{fig:MWL_LC_2017}.

Magnetic reconnection near the collimation zone provides an additional dissipation mechanism \citep{2020NatCo..11.4176S}, where kink or shear-flow instabilities create filamentary structures that efficiently convert magnetic energy. Relativistic particles accelerated in these sites can produce $\gamma$-rays via inverse Compton scattering. For typical fields $B \sim 0.1$–1~G and bulk Lorentz factors $\Gamma \sim 20$, this process reproduces the observed $\gamma$-ray fluxes and one- to two-day timescales, while higher magnetic fields and Lorentz factors can produce even shorter flares \citep[e.g.,][]{giannios2013,petropoulou2016}.

Further constraints on the high-energy emission mechanism come from more recent IXPE observations of 3C\,279 \citep{2024ApJ...972...74M}. The X-ray spectrum was fitted with a power law ($\Gamma = 1.79 \pm 0.04$) and Galactic absorption of $N_{\rm H} = 2.25 \times 10^{20}\,{\rm cm^{-2}}$. The X-ray polarization degree was constrained to be $<12.7\%$, compared to $\sim$4\% in the millimeter and $\sim$12\% in the optical and IR bands. While the current IXPE constraint does not by itself rule out a hadronic origin of the X-rays, it is compatible with leptonic models in which the X-ray and $\gamma$-ray emission is dominated by inverse-Compton scattering of synchrotron or external photons by relativistic electrons in the jet.
Similarly, during a TeV flare of the blazar Mrk\,421 observed jointly with IXPE, strong and erratic X-ray polarization variability was detected, while the optical polarization remained nearly constant \citep{2025A&A...695A.217M}. This behavior closely resembles the $\gamma$-ray flaring state of 3C\,279 in April 2017, albeit at higher energies given the high-synchrotron-peak nature of Mrk\,421 compared to the low-synchrotron-peak classification of 3C\,279. In contrast, the flare in 3C\,279 is more pronounced in the MeV--GeV band, and the polarization variability is observed in the optical.

The kinematic analysis by \citet{2022ApJS..260...12W} indicated that a superluminal knot (C37) was ejected at $T_0 = 2017.253 \pm 0.035$ (April 2 $\pm$ 13 days) with an apparent velocity of $(25 \pm 2)c$, coinciding with $\gamma$-ray flaring activity in late March –- early April (Fig.~\ref{fig:MWL_LC_2017}). The optical peak at the end of March, along with enhanced $\gamma$-ray flux, likely reflects an increase in the magnetic field strength and radiating electron density \citep{Larionov2020}. A concurrent rise in the radio-core polarization was observed in April–May 2017 (Fig.~\ref{fig:vlba_bu}).
These observations are consistent with a shock-in-jet scenario, where a disturbance near the jet base propagates downstream, becoming a superluminal radio knot. The high apparent speed of C37 suggests either an increase in the bulk Lorentz factor $\Gamma$ or a change in jet orientation \citep{Larionov2020}. Doppler factor variations, coupled with changes in the emitting electron population, are consistent with emission modeling of previous flares \citep{2015ApJ...807...79H,Larionov2020,2025ApJ...989..125M}.

Finally, the unprecedented optical flare (a synchrotron flux increase of $\sim20 \times$ above the low state), together with contemporaneous UV and X-ray enhancements observed by Swift during the onset of the $\gamma$-ray flare (see Fig. \ref{fig:MWL_LC_2017}), supports a cospatial origin of the emission across these bands. Flux variability is likely driven by knot evolution and structural changes in the jet. This interpretation is consistent with the simultaneous $\gamma$-ray flare and optical polarization event reported by \citet{2010Natur.463..919A} and with zero-lag optical-UV–$\gamma$-ray correlations observed during flares \citep{2017MNRAS.464..418R,2020ApJ...890..164P}.

\section{Summary}
\label{sec:summary}
We presented the results of the first MWL observational campaign on 3C\,279 together with the highest ever resolution millimeter-VLBI images using the EHT from its April 2017 campaign. 
This includes one of the most extensive quasi-simultaneous broadband spectra, along with details of the individual observations and light curves aimed at investigating its emission variability and constraining the origin of the $\gamma$-ray emission in this source.
The primary outcome of this campaign is represented by the rich legacy dataset, for which all data and some analysis scripts are available to the community via a \textsc{Cyverse} repository (see Appendix \ref{appendix:mwl_sed}). 
While we defer a detailed modeling to future works by the broader community, we draw preliminary conclusions using a heuristic approach to the SED modeling.

The main results of the 2017 EHT–MWL observational campaign on 3C\,279 can be summarised as follows:
\begin{itemize}
    \item EHT observations of the innermost core region (component C0–0; \citealt{2020A&A...640A..69K}) reveal a clear flux increase from $\sim$0.5\,Jy to $\sim$0.95\,Jy between April 5–6 and 10–11.
    \item VLBA observations reveal a significant enhancement of the core flux density and polarized emission during April–May. \citet{2022ApJS..260...12W} reported the contemporaneous ejection of a superluminal knot at $T_0 = 2017.253 \pm 0.035$ (April 2 $\pm$ 13 days), presenting a high apparent speed of (25 $\pm$ 2)\,c.
    \item An unprecedented UV–optical flare, with synchrotron flux increases by a factor of $\sim20$ relative to the low state, was detected around March 21–22 , followed by a low-emission state in April. Strong optical polarization variability (degree 5–25 \%, angle 20°–80°) closely followed the optical flux evolution.
    \item In the X-ray band, \textit{Swift}-XRT observations indicate a low X-ray emission state, with a flux below the long-term average.
    \item High-energy $\gamma$-ray flaring activity was observed between late March and early April. It ceased before the end of the EHT observing window, while no significant VHE emission was detected.
    \item The broadband variability in the SED and polarization can be reproduced with a jet model involving turbulent plasma, as simulated using the TEMZ code (see Sect.~\ref{sec:modelling}). In this scenario, turbulent cells crossing a stationary conical shock undergo compression capable of triggering the observed $\gamma$-ray flares.
    \item The high observed brightness temperature of $T_{\rm b}\thicksim10^{13}$\,K obtained from the RadioAstron observations at 18\,cm and 6\,cm  supports the TEMZ scenario and existence of the compact components in the 3C\,279 jet.
    \item An alternative scenario suggested by the observations is that the $\gamma$-ray flare might be associated with a moving shock-in-jet, possibly linked to the superluminal component ejected during the same period.
\end{itemize}

The results of this broadband MWL campaign on 3C\,279 provide a valuable opportunity to probe the origin of the $\gamma$-ray emission during a $\gamma$-ray flaring episode. They underscore the crucial role of coordinated densely sampled MWL observations in capturing and characterizing the spectral variability of 3C\,279, which likely spans  multiple temporal scales. Furthermore, the publicly released datasets from this campaign constitute a rich resource for future studies.

The EHT–MWL observational campaigns conducted with increasingly sensitive arrays in 2018, 2021, 2022, 2024, 2025, and those planned for the coming years, will provide crucial new insights into the physical processes at play in the archetypal blazar 3C\,279. They include improved constraints on the jet structure and the origin and mechanisms that produce $\gamma$-ray emission.
The synergy between high-resolution EHT mm-VLBI imaging and broadband SED information will enable a deeper understanding of the emission mechanisms that power AGNs.

\section*{Data availability}
\label{sec:appendix-supplementary}

Data products presented in this paper are available for download through the EHT Collaboration Data Webpage (\url{https://eventhorizontelescope.org/for-astronomers/data}), or \textit{Fermi}-LAT Collaboration Publication Webpage (\url{https://www-glast.stanford.edu/cgi-bin/pubpub}). 
The repository contains the following data products:
\begin{itemize}
    \item Broadband spectrum table (see Table \ref{tab:sed2017}) with frequency, flux density, its uncertainty, and instrument index (format: DAT)
    \item Multi-wavelength light curve (see Fig. \ref{fig:MWL_LC_2017}) table with frequency, flux density, its uncertainty, and instrument index (format: DAT)
    \item Sampled posterior distributions of the SED broadband spectral model described in Sect. \ref{sec:modelling} (format:DAT).
\end{itemize}

\begin{acknowledgements}
The full acknowledgements are available in Appendix \ref{appendix:ackn}.
\end{acknowledgements}

%

\bibliography{main_bib}

\begin{appendix} 
\nolinenumbers
\section{Acknowledgements}
\label{appendix:ackn}

\begin{acknowledgements}

G.P acknowledges the CINECA award under the ISCRA initiative, for the availability of high performance computing resources and support. G.P. acknowledge also partial support by ICSC – Centro Nazionale di Ricerca in High Performance Computing, Big Data and Quantum Computing, funded by European Union – NextGenerationEU.

\\
The Event Horizon Telescope Collaboration thanks the following organizations and programs: the Academia Sinica; the Research Council of Finland (project 362572); the Agencia Nacional de Investigaci\'{o}n y Desarrollo (ANID), Chile via NCN$19\_058$ (TITANs), Fondecyt 1221421 and BASAL FB210003; the Alexander von Humboldt Stiftung (including the Feodor Lynen Fellowship); an Alfred P. Sloan Research Fellowship; Allegro, the European ALMA Regional Centre node in the Netherlands, the NL astronomy research network NOVA and the astronomy institutes of the University of Amsterdam, Leiden University, and Radboud University; the ALMA North America Development Fund; the Astrophysics and High Energy Physics programme by MCIN (with funding from European Union NextGenerationEU, PRTR-C17I1); the Black Hole Initiative, which is funded by grants from the John Templeton Foundation (60477, 61497, 62286) and the Gordon and Betty Moore Foundation (Grants GBMF-8273, GBMF12987) -- although the opinions expressed in this work are those of the authors and do not necessarily reflect the views of these Foundations; the Black Hole Initiative, which is funded by grants from the John Templeton Foundation and the Gordon and Betty Moore Foundation (although the opinions expressed in this work are those of the author(s) and do not necessarily reflect the views of these Foundations); the Brinson Foundation; the Canada Research Chairs (CRC) program; Chandra DD7-18089X and TM6-17006X; the China Scholarship
Council; the China Postdoctoral Science Foundation fellowships (2020M671266, 2022M712084); ANID through Fondecyt Postdoctorado (project 3250762); Conicyt through Fondecyt Postdoctorado (project 3220195); Consejo Nacional de Humanidades, Ciencia y Tecnología (CONAHCYT, Mexico, projects U0004-246083, U0004-259839, F0003-272050, M0037-279006, F0003-281692, 104497, 275201, 263356, CBF2023-2024-1102, 257435); the Colfuturo Scholarship; the Consejo Superior de Investigaciones  ient\'{i}ficas (grant 2019AEP112); the Delaney Family via the Delaney Family John A. Wheeler Chair at Perimeter Institute; Dirección General de Asuntos del Personal Académico-Universidad Nacional Autónoma de México (DGAPA-UNAM, projects IN112820 and IN108324); the Dutch Research Council (NWO) for the VICI award (grant 639.043.513), the grant OCENW.KLEIN.113, and the Dutch Black Hole Consortium (with project No. NWA 1292.19.202) of the research programme the National Science Agenda; the Dutch National Supercomputers, Cartesius and Snellius  (NWO grant 2021.013); the EACOA Fellowship awarded by the East Asia Core Observatories Association, which consists of the Academia Sinica Institute of Astronomy and Astrophysics, the National Astronomical Observatory of Japan, Center for Astronomical Mega-Science, Chinese Academy of Sciences, and the Korea Astronomy and Space Science Institute; the European Research Council (ERC) Synergy Grant ``BlackHoleCam: Imaging the Event Horizon of Black Holes'' (grant 610058) and Synergy Grant ``BlackHolistic:  Colour Movies of Black Holes: Understanding Black Hole Astrophysics from the Event Horizon to Galactic Scales'' (grant 10107164); the European Union Horizon 2020 research and innovation programme under grant agreements BlackHolistic (No. 101071643), RadioNet (No. 730562), M2FINDERS (No. 101018682); the European Research Council for advanced grant ``JETSET: Launching, propagation and emission of relativistic jets from binary mergers and across mass scales'' (grant No. 884631); the European Horizon Europe staff exchange (SE) programme HORIZON-MSCA-2021-SE-01 grant NewFunFiCO (No. 10108625); the Horizon ERC Grants 2021 programme under grant agreement No. 101040021; the FAPESP (Funda\c{c}\~ao de Amparo \'a Pesquisa do Estado de S\~ao Paulo) under grant 2021/01183-8; the Fondes de Recherche Nature et Technologies (FRQNT); the Fondo CAS-ANID folio CAS220010; the Generalitat Valenciana (grants APOSTD/2018/177 and  ASFAE/2022/018) and GenT Program (project CIDEGENT/2018/021); the Gordon and Betty Moore Foundation (GBMF-3561, GBMF-5278, GBMF-10423); the Institute for Advanced Study; the ICSC – Centro Nazionale di Ricerca in High Performance Computing, Big Data and Quantum Computing, funded by European Union – NextGenerationEU; the Istituto Nazionale di Fisica Nucleare (INFN) sezione di Napoli, iniziative specifiche TEONGRAV; the International Max Planck Research School for Astronomy and Astrophysics at the Universities of Bonn and Cologne; the Italian Ministry of University and Research (MUR)– Project CUP F53D23001260001, funded by the European Union – NextGenerationEU; Deutsche Forschungsgemeinschaft (DFG) research grant ``Jet physics on horizon scales and beyond'' (grant No. 443220636) and DFG research grant 443220636; Joint Columbia/Flatiron Postdoctoral Fellowship (research at the Flatiron Institute is supported by the Simons Foundation); the Japan Ministry of Education, Culture, Sports, Science and Technology (MEXT; grant JPMXP1020200109); the Japan Society for the Promotion of Science (JSPS) Grant-in-Aid for JSPS Research Fellowship (JP17J08829); the Joint Institute for Computational Fundamental Science, Japan; the Key Research Program of Frontier Sciences, Chinese Academy of Sciences (CAS, grants QYZDJ-SSW-SLH057, QYZDJSSW-SYS008, ZDBS-LY-SLH011);  the Leverhulme Trust Early Career Research Fellowship; the Max-Planck-Gesellschaft (MPG); the Max Planck Partner Group of the MPG and the CAS; the MEXT/JSPS KAKENHI (grants 18KK0090, JP21H01137, JP18H03721, JP18K13594, 18K03709, JP19K14761, 18H01245, 25120007, 19H01943, 21H01137, 21H04488, 22H00157, 23K03453); the MICINN Research Projects PID2019-108995GB-C22, PID2022-140888NB-C22; the MIT International Science and Technology Initiatives (MISTI) Funds;  the Ministry of Science and Technology (MOST) of Taiwan (103-2119-M-001-010-MY2, 105-2112-M-001-025-MY3, 105-2119-M-001-042, 106-2112-M-001-011, 106-2119-M-001-013, 106-2119-M-001-027, 106-2923-M-001-005, 107-2119-M-001-017, 107-2119-M-001-020, 107-2119-M-001-041, 107-2119-M-110-005, 107-2923-M-001-009, 108-2112-M-001-048, 108-2112-M-001-051, 108-2923-M-001-002, 109-2112-M-001-025, 109-2124-M-001-005, 109-2923-M-001-001, 110-2112-M-001-033, 110-2124-M-001-007 and 110-2923-M-001-001); the National Science and Technology Council (NSTC) of Taiwan (111-2124-M-001-005, 112-2124-M-001-014,  112-2112-M-003-010-MY3, and 113-2124-M-001-008); the Ministry of Education (MoE) of Taiwan Yushan Young Scholar Program; the Physics Division, National Center for Theoretical Sciences of Taiwan; the National Aeronautics and Space Administration (NASA, Fermi Guest Investigator grant 80NSSC23K1508, NASA Astrophysics Theory Program grant 80NSSC20K0527, NASA NuSTAR award  80NSSC20K0645); NASA Hubble Fellowship Program Einstein Fellowship; NASA Hubble Fellowship  grants HST-HF2-51431.001-A, HST-HF2-51482.001-A, HST-HF2-51539.001-A, HST-HF2-51552.001A awarded  by the Space Telescope Science Institute, which is operated by the Association of Universities for Research in Astronomy, Inc., for NASA, under contract NAS5-26555; the National Institute of Natural Sciences (NINS) of Japan; the National Key Research and Development Program of China (grant 2016YFA0400704, 2017YFA0402703, 2016YFA0400702); the National Science and Technology Council (NSTC, grants NSTC 111-2112-M-001 -041, NSTC 111-2124-M-001-005, NSTC 112-2124-M-001-014); the US National Science Foundation (NSF, grants AST-0096454, AST-0352953, AST-0521233, AST-0705062, AST-0905844, AST-0922984, AST-1126433, OIA-1126433, AST-1140030, DGE-1144085, AST-1207704, AST-1207730, AST-1207752, MRI-1228509, OPP-1248097, AST-1310896, AST-1440254, AST-1555365, AST-1614868, AST-1615796, AST-1715061, AST-1716327,  AST-1726637, OISE-1743747, AST-1743747, AST-1816420, AST-1935980, AST-1952099, AST-2034306,  AST-2205908, AST-2307887, AST-2407810); NSF Astronomy and Astrophysics Postdoctoral Fellowship (AST-1903847); the Natural Science Foundation of China (grants 11650110427, 10625314, 11721303, 11725312, 11873028, 11933007, 11991052, 11991053, 12192220, 12192223, 12273022, 12325302, 12303021);  the Natural Sciences and Engineering Research Council of Canada (NSERC);  the National Research Foundation of Korea (the Global PhD Fellowship Grant: grants NRF-2015H1A2A1033752; the Korea Research Fellowship Program: NRF-2015H1D3A1066561; Brain Pool Program: RS-2024-00407499;  Basic Research Support Grant 2019R1F1A1059721, 2021R1A6A3A01086420, 2022R1C1C1005255, RS-2022-NR071771, RS-2025-16067786); the POSCO Science Fellowship of the POSCO TJ Park Foundation; NOIRLab, which is managed by the Association of Universities for Research in Astronomy (AURA) under a cooperative agreement with the National Science Foundation;  the A.G. Leventis Foundation;  Onsala Space Observatory (OSO) national infrastructure, for the provisioning of its facilities/observational support (OSO receives funding through the Swedish Research Council under grant 2017-00648);  the Perimeter Institute for Theoretical Physics (research at Perimeter Institute is supported by the Government of Canada through the Department of Innovation, Science and Economic Development and by the Province of Ontario through the Ministry of Research, Innovation and Science); the Portuguese Foundation for Science and Technology (FCT) grants (Individual CEEC program – 5th edition, CIDMA through the FCT Multi-Annual Financing Program for R\&D Units UID/04106, CERN/FIS-PAR/0024/2021, 2022.04560.PTDC); the Princeton Gravity Initiative; the Spanish Ministerio de Ciencia, Innovaci\'{o}n  y Universidades (grants PID2022-140888NB-C21, PID2022-140888NB-C22, PID2023-147883NB-C21, RYC2023-042988-I); the Severo Ochoa grant CEX2021-001131-S funded by MICIU/AEI/10.13039/501100011033; The European Union’s Horizon Europe research and innovation program under grant agreement No. 101093934 (RADIOBLOCKS); The European Union “NextGenerationEU”, the Recovery, Transformation and Resilience Plan, the CUII of the Andalusian Regional Government and the Spanish CSIC through grant AST22\_00001\_Subproject\_10; ``la Caixa'' Foundation (ID 100010434) through fellowship codes LCF/BQ/DI22/11940027 and LCF/BQ/DI22/11940030; the University of Pretoria for financial aid in the provision of the new Cluster Server nodes and SuperMicro (USA) for a SEEDING GRANT approved toward these nodes in 2020; the Shanghai Municipality orientation program of basic research for international scientists (grant no. 22JC1410600); the Shanghai Pilot Program for Basic Research, Chinese Academy of Science, Shanghai Branch (JCYJ-SHFY-2021-013); the Simons Foundation (grant 00001470); the Spanish Ministry for Science and Innovation grant CEX2021-001131-S funded by MCIN/AEI/10.13039/501100011033; the Spinoza Prize SPI 78-409; the South African Research Chairs Initiative, through the South African Radio Astronomy Observatory (SARAO, grant ID 77948),  which is a facility of the National Research Foundation (NRF), an agency of the Department of Science and Innovation (DSI) of South Africa; the Swedish Research Council (VR); the Taplin Fellowship; the Toray Science Foundation; the UK Science and Technology Facilities Council (grant no. ST/X508329/1); the US Department of Energy (USDOE) through the Los Alamos National Laboratory (operated by Triad National Security, LLC, for the National Nuclear Security Administration of the USDOE, contract 89233218CNA000001); and the YCAA Prize Postdoctoral Fellowship. This work was also supported by the National Research Foundation of Korea (NRF) grant funded by the Korea government(MSIT) (RS-2024-00449206). We acknowledge support from the Coordenação de Aperfeiçoamento de Pessoal de Nível Superior (CAPES) of Brazil through PROEX grant number 88887.845378/2023-00. We acknowledge financial support from Millenium Nucleus NCN23\_002 (TITANs) and Comité Mixto ESO-Chile.
We thank the staff at the participating observatories, correlation centers, and institutions for their enthusiastic support. This paper makes use of the following ALMA data: ADS/JAO.ALMA\#2017.1.00841.V and ADS/JAO.ALMA\#2019.1.01797.V. ALMA is a partnership of the European Southern Observatory (ESO; Europe, representing its member states), NSF, andNational Institutes of Natural Sciences of Japan, together with National Research Council (Canada), Ministry of Science and Technology (MOST; Taiwan), Academia Sinica Institute of Astronomy and Astrophysics (ASIAA; Taiwan), and Korea Astronomy and Space Science Institute (KASI; Republic of Korea), in cooperation with the Republic of Chile. The Joint ALMA Observatory is operated by ESO, Associated Universities, Inc. (AUI)/NRAO, and the National Astronomical Observatory of Japan (NAOJ). The NRAO is a facility of the NSF operated under cooperative agreement by AUI. This research used resources of the Oak Ridge Leadership Computing Facility at the Oak Ridge National Laboratory, which is supported by the Office of Science of the U.S. Department of Energy under contract No. DE-AC05-00OR22725; the ASTROVIVES FEDER infrastructure, with project code IDIFEDER-2021-086; the computing cluster of Shanghai VLBI correlator supported by the Special Fund for Astronomy from the Ministry of Finance in China. We also thank the Center for Computational Astrophysics, National Astronomical Observatory of Japan. This work was supported by FAPESP (Fundacao de Amparo a Pesquisa do Estado de Sao Paulo) under grant 2021/01183-8.
APEX is a collaboration between the Max-Planck-Institut f{\"u}r Radioastronomie (Germany), ESO, and the Onsala Space Observatory (Sweden). The SMA is a joint project between the SAO and ASIAA and is funded by the Smithsonian Institution and the Academia Sinica. The JCMT is operated by the East Asian Observatory on behalf of the NAOJ, ASIAA, and KASI, as well as the Ministry of Finance of China, Chinese Academy of Sciences, and the National Key Research and Development Program (No. 2017YFA0402700) of China and Natural Science Foundation of China grant 11873028. Additional funding support for the JCMT is provided by the Science and Technologies Facility Council (UK) and participating universities in the UK and Canada. The LMT is a project operated by the Instituto Nacional de Astr\'{o}fisica, \'{O}ptica, y Electr\'{o}nica (Mexico) and the University of Massachusetts at Amherst (USA). The IRAM 30 m telescope on Pico Veleta, Spain and the NOEMA interferometer on Plateau de Bure, France are operated by IRAM and supported by CNRS (Centre National de la Recherche Scientifique, France), MPG (Max-Planck-Gesellschaft, Germany), and IGN (Instituto Geográfico Nacional, Spain). The SMT is operated by the Arizona Radio Observatory, a part of the Steward Observatory of the University of Arizona, with financial support of operations from the State of Arizona and financial support for instrumentation development from the NSF. Support for SPT participation in the EHT is provided by the National Science Foundation through award OPP-1852617 to the University of Chicago. Partial support is also provided by the Kavli Institute of Cosmological Physics at the University of Chicago. The SPT hydrogen maser was provided on loan from the GLT, courtesy of ASIAA.
This work used the Extreme Science and Engineering Discovery Environment (XSEDE), supported by NSF grant ACI-1548562, and CyVerse, supported by NSF grants DBI-0735191, DBI-1265383, and DBI-1743442. XSEDE Stampede2 resource at TACC was allocated through TG-AST170024 and TG-AST080026N. XSEDE JetStream resource at PTI and TACC was allocated through AST170028. This research is part of the Frontera computing project at the Texas Advanced  Computing Center through the Frontera Large-Scale Community Partnerships allocation AST20023. Frontera is made possible by National Science Foundation award OAC-1818253. This research was done using services provided by the OSG Consortium~\citep{osg07,osg09}, which is supported by the National Science Foundation award Nos. 2030508 and 1836650. Additional work used ABACUS2.0, which is part of the eScience center at Southern Denmark University, and the Kultrun Astronomy Hybrid Cluster (projects Conicyt Programa de Astronomia Fondo Quimal QUIMAL170001, Conicyt PIA ACT172033, Fondecyt Iniciacion 11170268, Quimal 220002). Simulations were also performed on the SuperMUC cluster at the LRZ in Garching, on the LOEWE cluster in CSC in Frankfurt, on the HazelHen cluster at the HLRS in Stuttgart, and on the Pi2.0 and Siyuan Mark-I at Shanghai Jiao Tong University. The computer resources of the Finnish IT Center for Science (CSC) and the Finnish Computing  Competence Infrastructure (FCCI) project are acknowledged. This research was enabled in part by support provided by Compute Ontario (http://computeontario.ca), Calcul Quebec (http://www.calculquebec.ca), and the Digital Research Alliance of Canada (https://alliancecan.ca/en). 
The EHTC has received generous donations of FPGA chips from Xilinx Inc., under the Xilinx University Program. The EHTC has benefited from technology shared under open-source license by the Collaboration for Astronomy Signal Processing and Electronics Research (CASPER). The EHT project is grateful to T4Science and Microsemi for their assistance with hydrogen masers. This research has made use of NASA's Astrophysics Data System. We gratefully acknowledge the support provided by the extended staff of the ALMA, from the inception of the ALMA Phasing Project through the observational campaigns of 2017 and 2018. We would like to thank A. Deller and W. Brisken for EHT-specific support with the use of DiFX. We thank Martin Shepherd for the addition of extra features in the Difmap software  that were used for the CLEAN imaging results presented in this paper. We acknowledge the significance that Maunakea, where the SMA and JCMT EHT stations are located, has for the indigenous Hawaiian people.

The \textit{Fermi} LAT Collaboration acknowledges generous ongoing support from a number of agencies and institutes that have supported the development and operation of the LAT as well as scientific data analysis. These include the National Aeronautics and Space Administration and the Department of Energy in the United States, the Commissariat à l’Energie Atomique and the Centre National de la Recherche Scientifique / Institut National de Physique Nucléaire et de Physique des Particules in France, the Agenzia Spaziale Italiana and the Istituto Nazionale di Fisica Nucleare in Italy, the Ministry of Education, Culture, Sports, Science and Technology (MEXT), High Energy Accelerator Research Organization (KEK) and Japan Aerospace Exploration Agency (JAXA) in Japan, and the K. A.Wallenberg Foundation, the Swedish Research Council and the Swedish National Space Board in Sweden. Additional support for science analysis during the operations phase is gratefully acknowledged from the Istituto Nazionale di Astrofisica in Italy and the Centre National d’Etudes Spatiales in France. This work is performed in part under DOE Contract DE-AC02-76SF00515.
\\
This research has made use of data obtained with the Global Millimeter VLBI Array (GMVA), which consists of telescopes operated by the MPIfR, IRAM, Onsala, Metsahovi, Yebes, the Korean VLBI Network, the Greenland Telescope, the Green Bank Observatory and the Very Long Baseline Array (VLBA). The VLBA and the GBT are facilities of the National Science Foundation operated under cooperative agreement by Associated Universities, Inc. The data were correlated at the correlator of the MPIfR in Bonn, Germany.
\\
We would like to thank the Instituto de Astrof\'{\i}sica de Canarias for the excellent working conditions at the Observatorio del Roque de los Muchachos in La Palma. The financial support of the German BMFTR, MPG and HGF; the Italian INFN and INAF; the Swiss National Fund SNF; the grants PID2022-136828NB-C41, PID2022-137810NB-C22, PID2022-138172NB-C41, PID2022-138172NB-C42, PID2022-138172NB-C43, PID2022-139117NB-C41, PID2022-139117NB-C42, PID2022-139117NB-C43, PID2022-139117NB-C44, CNS2023-144504 funded by the Spanish MCIN/AEI/ 10.13039/501100011033 and "ERDF A way of making Europe"; the Indian Department of Atomic Energy; the Japanese ICRR, the University of Tokyo, JSPS, and MEXT; the Bulgarian Ministry of Education and Science, National RI Roadmap Project DO1-400/18.12.2020 and the Academy of Finland grant nr. 320045 is gratefully acknowledged. This work has also been supported by Centros de Excelencia ``Severo Ochoa'' y Unidades ``Mar\'{\i}a de Maeztu'' program of the Spanish MCIN/AEI/ 10.13039/501100011033 (CEX2019-000918-M, CEX2021-001131-S, CEX2024001442-S), by AST22\_00001\_9 with funding from NextGenerationEU funds and by the CERCA institution and grants 2021SGR00426, 2021SGR00607 and 2021SGR00773 of the Generalitat de Catalunya; by the Croatian Science Foundation (HrZZ) Project IP-2022-10-4595 and the University of Rijeka Project uniri-prirod-18-48; by the Deutsche Forschungsgemeinschaft (SFB1491) and by the Lamarr-Institute for Machine Learning and Artificial Intelligence; by the Polish Ministry of Science and Higher Education grant No. 2025/WK/04; by the European Union (ERC, MicroStars, 101076533); and by the Brazilian MCTIC, the CNPq Productivity Grant 309053/2022-6 and FAPERJ Grants E-26/200.532/2023 and E-26/211.342/2021.
\\
We thank the \HESS Collaboration for providing the \HESS data that have been analysed for this publication.
\\
\end{acknowledgements}

\section{Observations description}
\label{appendix:observations}
This section provides a detailed description of the MWL observations carried out during the EHT campaigns on 3C\,279 in April 2017. The measurements contributing to the quasi-simultaneous spectrum of 3C\,279 in 2017 (see Fig.~\ref{fig:mwl_sed}) are summarized in Table~\ref{tab:sed2017}.

\subsection{Radio observations}
\label{sec:radio_observations}

\subsubsection{RadioAstron - 18 and 6 cm space VLBI}
\label{sec:radioastron}
RadioAstron was a mission comprising a parabolic 10-m Space Radio Telescope on a highly elliptical orbit around Earth, observing together with ground-based radio telescopes. 3C\,279 was monitored during the RadioAstron mission \citep{2017SoSyR..51..535K} at wavelengths $\lambda=18$\,cm (1.7\,GHz; 58 fringe detections on space-ground baselines longer than 1~Earth diameter), $\lambda=6$\,cm (4.8\,GHz; 226 detections), and $\lambda=1.3$\,cm (22\,GHz; 9 detections). In the period 2016--2018, there were 7 and 12 non-imaging experiments conducted at 18\,cm and 6\,cm delivering, in total, 13 and 17 space-ground fringe detections, respectively. The longest baseline with fringe detection was 25.3~Earth diameters (1.8~G$\lambda$, 322599\,km) at 18\,cm and 23.2~Earth diameters (4.8~G$\lambda$, 295390\,km) at 6\,cm.

At 18\,cm, the ground network of participating telescopes consisted of Effelsberg (100\,m), Green Bank (110\,m), Hartbeesthoek (26\,m), Mopra (22\,m), Medicina (32\,m), Torun (32\,m), and Irbene (16\,m). The first four provided fringe detections on the baselines with the space antenna. 
At 6\,cm, participating antennas included Effelsberg, Hartbeesthoek, Torun, Mopra, Badary (32\,m), Warkworth (30\,m), Noto (30\,m), Westerbork (single antenna, 25\,m), and Irbene. The first five provided fringe detections with the space antenna. 

The amplitude of the signal was calibrated using system temperatures measured at individual telescopes. All subsequent analysis of RadioAstron data was performed with amplitude data only. At both frequencies, we fitted measured amplitudes using three Gaussian components. Due to a scarcity of ground-space fringe detections, we combined all the data taken between 2016 and 2018. The robustness of fits and uncertainty estimates were performed using bootstrapping. In each of 1000 realizations, a random 10\% of the data was removed and the fitting procedure was performed again.

Parameters of the Gaussian model components are provided in Table~\ref{tab:ra_modelfit}. Since we used only amplitudes, we can robustly constrain only size and flux of the components, but not their relative positions. However, at both wavelengths, it is clear that the largest components (c1 and l1) describe extended jet, the next ones (c2 and l2) describe the overall VLBI core, while the smallest and the dimmest ones (c3 and l3) describe some fine structure within the cores. These components show the highest observed brightness temperature, surpassing that of the core. This can indicate the presence of more energetic particles emitting in these regions, which might be associated with either a fast jet spine or regions of particle re-acceleration. It should be noted that even the core is located well downstream of the central black hole, at distances as great as tens of parsecs.

\begin{table}[]
    \centering
    \caption{Gaussian components fitted to 2016-2018 ground-space VLBI imaging data including RadioAstron at 6\,cm and 18\,cm.}
    \label{tab:ra_modelfit}
    \begin{tabular}{cccc}
        \hline
        \hline
         Comp & Flux & Size FWHM & $T_\mathrm{b}$ \\
         & [Jy] & [mas] & [K] \\
         \hline
         \multicolumn{4}{c}{6 cm} \\
         \hline
         c1 & $5.19\pm0.49$ & $3.3\pm0.8$   & $(2.3\pm1.2)\times10^{10} $\\
         c2 & $3.86\pm0.25$ & $0.21\pm0.01$ & $(3.4\pm0.5)\times10^{12}$ \\
         c3 & $0.46\pm0.03$ & $0.11\pm0.03$ & $(8.0\pm0.9)\times10^{12}$\\
         \hline
        \multicolumn{4}{c}{18 cm} \\
        \hline
         l1 & $10.53\pm0.16$ & $22.1\pm0.4$ & $(9.5\pm0.4)\times10^{9}$ \\
         l2 & $3.37\pm0.08$& $0.5\pm0.0$& $(5.1\pm0.4)\times10^{12}$\\
         l3 & $0.32\pm0.06$ & $0.1\pm0.0$& $(6.5\pm1.9)\times10^{12}$\\
         \hline
    \end{tabular}
\end{table}

\subsubsection{VLBA}
\label{sec:radio-vlba}

\begin{figure}[t]
    \centering
    \includegraphics[width=0.45\textwidth]{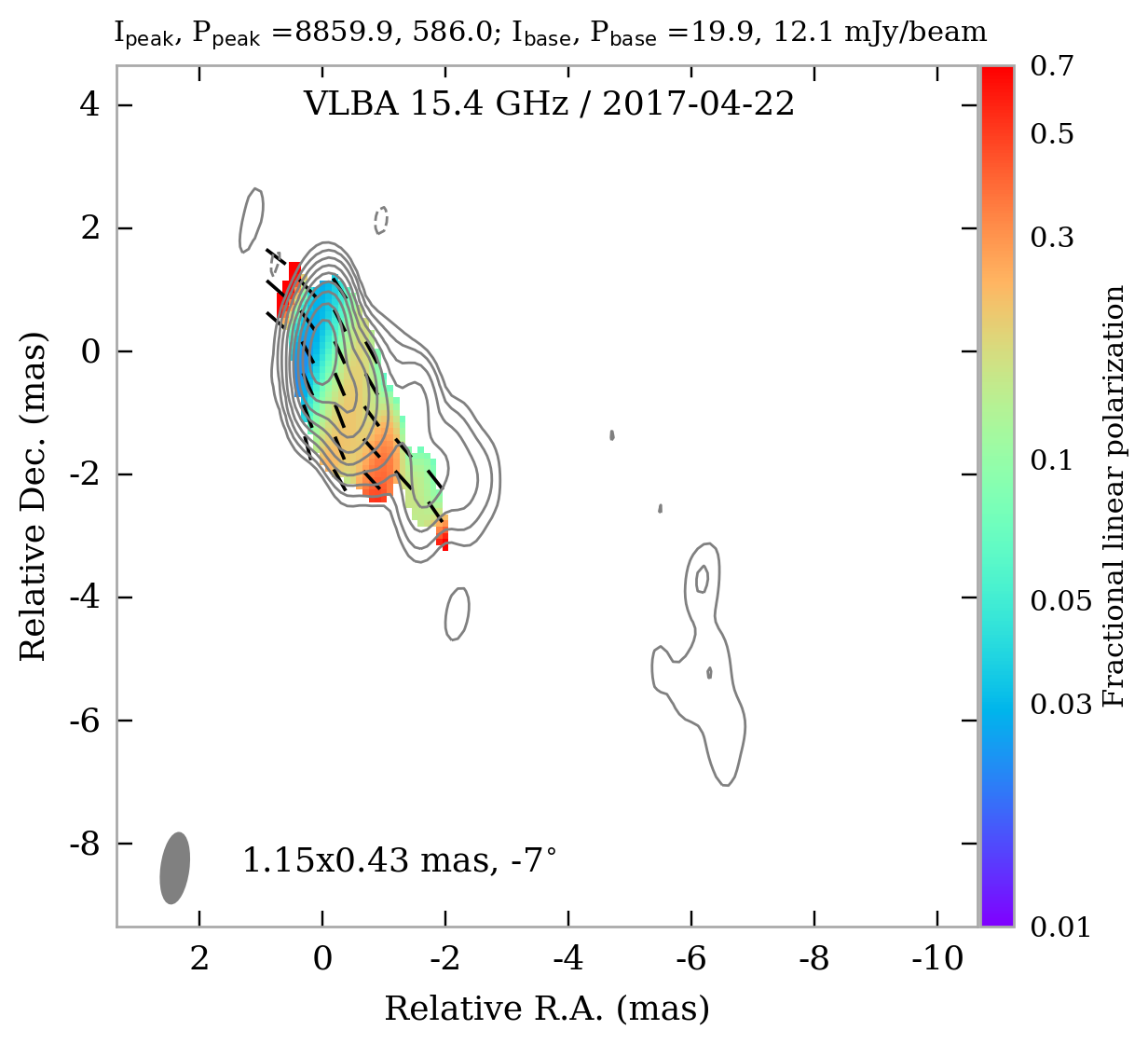}\\
    \caption{VLBA MOJAVE linear (contours) and polarized (color scale) intensity image at 15\,GHz  of the 3C\,279 jet on April 22, 2017. Black line segments indicate orientation of polarization vectors (not corrected for Faraday rotation).}
    \label{fig:mojave_image}
\end{figure}

\paragraph{VLBA 2\,cm-VLBI}
Very Long Baseline Array (VLBA) total and polarized intensity images at a wavelength of 2\,cm were obtained from the MOJAVE program\footnote{\url{https://www.cv.nrao.edu/MOJAVE/}} \citep{2018ApJS..234...12L}.
A VLBA 15\,GHz image from 2017 Apr 22, quasi-simultaneous with our campaign, is shown in Fig.~\ref{fig:mojave_image}.

\begin{figure}[t]
    \centering    
    \includegraphics[width=0.45\textwidth]{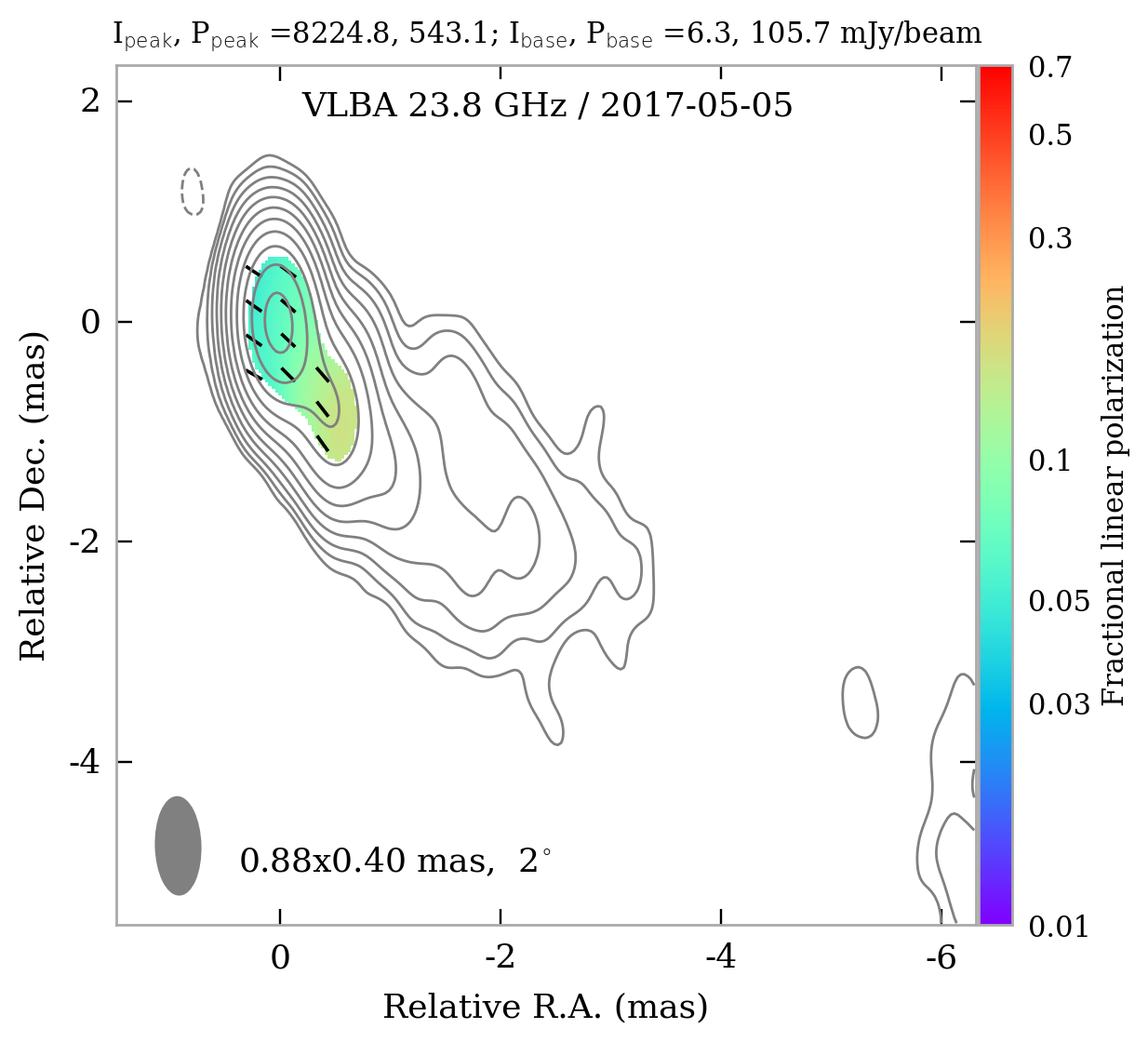}
    \caption{VLBA naturally weighted image at  1.2\,cm (24\,GHz) of the 3C\,279 jet from May 5, 2017. Black line segments indicate orientation of polarization vectors (not corrected for Faraday rotation).}
    \label{fig:img_24ghz}
\end{figure}

\paragraph{VLBA - 1.2\,cm VLBI}
3C\,279 was used as a calibrator for a VLBA observation (program BG251A) at 1.2\,cm (24\,GHz) on May 5, 2017.The detailed description of the observational setup and calibration procedures is given in \citet{2020A&A...637L...6K}. The total and linearly polarized intensity image is displayed in Fig.~\ref{fig:img_24ghz}.

\paragraph{VLBA - 7\,mm VLBI}

VLBA observations at 7\,mm (43\,GHz) included in the present study were obtained within the VLBA-BU-BLAZAR program\footnote{www.bu.edu/blazars/BEAM-ME.html} \citep{2017ApJ...846...98J} with a roughly monthly cadence from February 4 to June 8, 2017.
The parameters of the maps shown in Fig.~\ref{fig:vlba_bu} are given in Table~\ref{tab:MapParm_7mm} as follows: 1 - epoch of observation; 2 - MJD; 3 - total intensity peak; 4 - beam size; 5 - beam position angle; 6 - total intensity noise level.

\begin{table}
\scriptsize
\setlength{\tabcolsep}{3pt}
\caption{Parameters of the 43\,GHz images presented in Figure~\ref{fig:vlba_bu}.}
\label{tab:MapParm_7mm}
\small
\centering
\begin{tabular}{rlllrr}
\hline\hline
\small
Date& MJD & I-peak & Beam & Beam PA& RMS  \\
    &     & [Jy/beam] & [mas$^2$] & [deg] & [mJy/beam] \\
(1) & (2) &  (3) & (4)& (5)& (6) \\   
\hline
Feb. 4&57788&6.56&0.50$\times$0.14&$-$11.5&6 \\
Mar. 19&57831&5.46&0.39$\times$0.16&$-$4.5&6 \\
Apr. 16&57859&5.79&0.42$\times$0.16&$-$10.1&4 \\
May 13&57886&9.36&0.37$\times$0.15&$-$4.6&16 \\
Jun. 8&57912&10.5&0.39$\times$0.15&$-$4.9&9 \\
\hline
\end{tabular}
\end{table}

\subsubsection{KVN - 13.6, 7, 3.5 and 2.3\,mm VLBI}
\label{sec:radio-kvn}
3C\,279 was monitored with the KVN as part of the Interferometric Monitoring of GAmma-ray Bright Active galactic nuclei \citep[iMOGABA][]{2016ApJS..227....8L} program. This consisted of monthly, simultaneous 22, 43, 86, and 129\,GHz, single polarization (LCP) observations. The data were calibrated using a modified version of the KVN pipeline \citep{2016JKAS...49..137H, 2024A&A...692A.140A}. We coherently averaged the 22 and 43\,GHz data data over 30~seconds, whereas we averaged the 86 and 129\,GHz data over 10~seconds prior to imaging. Imaging and phase-only self calibration were conducted in \texttt{Difmap} \citep{1997ASPC..125...77S}. Afterwards, the CLEAN components within a 2\,mas field of view of the core were summed to obtain the compact flux density at 22 and 43\,GHz. At 86 and 129\,GHz, the CLEAN components within a 1\,mas field of view were used. Uncertainties of 10, 20, and 30\% were adopted for the flux density measurements at 43, 86, and 129\,GHz, respectively. Uncertainties of 10\% to 15\% were adopted for measurements at 22\,GHz, depending on whether the specific observation was affected by additional flux loss \citep[][]{2024A&A...692A.140A}.

\begin{figure}
    \centering
\includegraphics[width=\columnwidth]{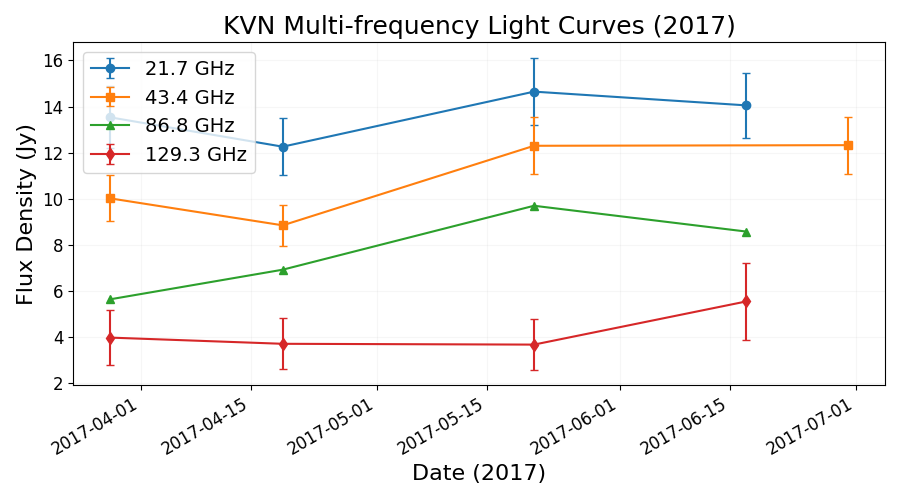}
    \caption{KVN multi-frequency light curves showing flux density variations observed between March and June 2017.}
    \label{fig:kvn_lc}
\end{figure}

KVN multi-frequency (22, 43, 86, and 129\,GHz) observations of the milliarcsecond-scale core reveal flux density variability, with the emission increasing during April–May (see Fig.~\ref{fig:kvn_lc}).

\subsubsection{GMVA - 3\,mm VLBI}
\label{sec:radio-gmva}
Quasi-simultaneous observations with the Global Millimeter VLBI Array (GMVA) were conducted on April 1 2017, as part of the GMVA blazar monitoring program\footnote{\url{https://www.bu.edu/blazars/vlbi3mm/index.html}}.The resulting image of the source is presented in the middle panel of Fig.~\ref{fig:composite_radio}.

\subsubsection{ALMA - 3.3, 2.9, 1.3 and 0.8\,mm}
\label{sec:radio-alma}
Archival observations of 3C\,279 with the Atacama Large Millimeter/submillimeter Array (ALMA) at 91.5, 103.5, 233, and 343\,GHz were obtained from \citet{2021ApJ...910L..14G} and from the ALMA Calibrator Source Catalogue\footnote{\url{https://almascience.eso.org/alma-data/calibrator-catalogue}}. 

\subsubsection{SMA - 1.3 and 0.8\,mm}
\label{sec:radio-sma}
Light curves of 3C\,279 at 1.3\,mm and 0.8\,mm (230 and 345\,GHz) with the Submillimeter Aray (SMA) were obtained from the SMA calibrator list\footnote{\url{http://sma1.sma.hawaii.edu/callist/callist.html?data=1256-057}}.

\subsubsection{EHT - 1.3\,mm VLBI}
\label{sec:radio-eht}
The source 3C\,279 was observed in 2017 by the EHT array on April 5, 6, 10, and 11. Details of the observations, calibration, and imaging can be found from \citep{PaperII,PaperIII}, and \citep{PaperIV}.
The source image on April 11 is shown in the rightmost panel of Fig. \ref{fig:composite_radio}.

\subsection{Optical-UV data}
\label{sec:optical}

\subsubsection{Kanata/HONIR}
\label{sec:optical-kanata}
The HONIR instrument \citep{2014SPIE.9147E..4OA}, mounted on the Kanata telescope at the Higashi-Hiroshima Observatory, was used for the observations. Standard data reduction procedures were applied, including bias or dark subtraction, flat-field correction, removal of cosmic-ray contamination, sky-background subtraction, and stacking of the images to improve the signal-to-noise ratio.
Photometric magnitudes of the target were measured through an aperture photometry using \texttt{SExtractor} \citep{1996A&AS..117..393B}. 
The typical image size of a point source, defined by the full width at half maximum (FWHM) of its point spread function, ranged from 1–3\arcsec\ depending on observational conditions. The aperture diameter was set to 2.8 times the FWHM. Object fluxes were converted to magnitudes by comparison with nearby stars, using the Pan-STARRS1 catalog \citep{2020ApJS..251....7F} for optical bands (after filter conversion; \citealt{2012ApJ...750...99T}) and the 2MASS catalog \citep{2006AJ....131.1163S} for NIR bands, with corrections for Galactic extinction \citep{2011ApJ...737..103S}.
HONIR provides independent linear polarization components, split by a Wollaston prism, with a half-wave plate (HWP) in front to rotate the plane of polarization. Four pairs of polarized images in $R_{\rm C}$ and \emph{J} bands were obtained at HWP angles of 0$^\circ$, 45$^\circ$, 22.5$^\circ$, and 67.5$^\circ$. Integrated target fluxes were measured via aperture photometry, and normalized Stokes parameters were calculated following \citet{1999PASP..111..898K}.

\subsubsection{Optical Photometric and Polarimetric Data}
\label{sec:optical-perkins}
Optical polarization observations of 3C\,279 in \emph{R} band, along with photometric estimates, were carried out during the EHT 2017 campaign at four telescopes: the 1.8~m Perkins telescope (PTO, Flagstaff, AZ, USA), the 40~cm LX-200 telescope (St.Petersburg, Russia), the 70\,cm AZT-8 telescope (Crimea Astrophysical Obs.), and  the 1.54\,m Kuiper telescope (Steward Obs., Tuscon, AZ, USA). The Perkins telescope is equipped with the PRISM camera, which includes a polarimeter with a rotating half-wave plate. For details of observations and data reduction see \citet{Jorstad2010}. Polarization observations at the LX-200 and AZT-8 telescopes were performed in the same manner, each using an identical photometer-polarimeter, with two Savart plates rotated by $45^\circ$ relative to each other. Swapping the plates allows one to obtain a normalized Stokes parameter, either \emph{q} or \emph{u} (for more detail see \citealt{Larionov2008}). Observations at the Kuiper telescope were performed using the CCD Imaging and Spectro-polarimeter \citep{Schmidt1992}, yielding spectra that span the range of 4000–7550~\AA\, with a dispersion of 4~\AA\ per pixel. In polarization mode, full-resolution Stokes spectra were obtained to calculate the linear polarization parameters within 5000–7000~\AA, close to \emph{R} band. Details of the spectropolarimetric data reduction can be found in \citet{Smith2016}.

Photometric measurements of 3C\,279 were also obtained between 2015 and 2024 using optical and near-infrared band-pass filters: \emph{V} and $R_{\rm C}$ bands from the Johnson-Cousins photometric system, and \emph{J} and $K_{\rm S}$ bands from the Mauna Kea Observatories Near-Infrared photometric system \citep{2002PASP..114..180T}. 

\subsubsection{Swift-UVOT}
\label{sec:swift_uvot}
The UV--Optical Telescope \citep[UVOT,][]{Roming2005} of the Neil Gehrels \emph{Swift}\ Observatory provides UV and optical data in 6 bands - uvw1, uvm2, uvw2, u, b, and v. We retrieved the data from the \emph{Swift}\ Archive and reduced them with v6.31.1 of the \texttt{HEAsoft} software and CALDB v20221216. For photometric analysis we used a 5\arcsec\ circular region centered on the source and a 20\arcsec\ circular aperture on a source-free region of the image to represent the background and processed the data with the \texttt{uvotsource} task, setting the detection significance level to $5\sigma$. None of the observations suffered a high coincidence loss. We used the count-rate-to-flux conversion factors reported by \citet{Breeveld2011} for $\gamma$-ray burst models, which correspond to continuum spectra similar to those of blazars. For the UV data, we de-reddened the magnitudes using the \citet{Fitzpatrick1999} interstellar extinction curve with an R$_{\texttt{v}}$=3.1 and $A_\lambda$ values from \citet{Schlafly2011}. Optical magnitudes were corrected for Galactic extinction using the \citet{Schlafly2011} values. The derived UVOT flux densities obtained during the EHT 2017 campaign are plotted in Fig. \ref{fig:MWL_LC_2017}. 

\subsection{X-ray observations: Swift-XRT}
\label{sec:xray-swiftxrt}

During the 2017 EHT campaign we obtained measurements with the Neil Gehrels \emph{Swift}\ Observatory X-ray Telescope \citep[XRT,][]{Burrows2005} over the 0.3-10\,keV photon energy range. The observations during the 2017 campaign were mostly quasi-simultaneous, on March 26 and 27 and on April 8, 12, and 13. All observations were made in photon counting mode. We used v6.31.1 of the \texttt{HEAsoft} package and CALDB v20221216 to process the data. A circular source region with a 50\arcsec\ radius and an annular background region with inner radius 75\arcsec\ and outer radius 100\arcsec, both centered on 3C\,279, were employed for data processing. The latter includes: cleaning the data,  creating an exposure map with the \texttt{xrtpipeline} tool, extracting the image and spectra using \texttt{XSELECT}, and generating the ancillary response file with \texttt{xrtmkarf}. We used Cash statistics \citep{Cash1979, Humphrey2009} in the form of the modified C-statistic \texttt{cstat} to fit the data in \texttt{XSPEC}, and grouped the data by single photons in \texttt{grppha}. The spectrum was then fit in \texttt{XSPEC} using an absorbed simple power law with Galactic column density $N_{\text{H}} = 2.2\times10^{20}$ cm$^{-2}$, as provided by the NASA HEASARC TOOL page. 

Figure \ref{fig:MWL_LC_2017} presents derived fluxes of 3C\,279 at 0.3-10\,keV. The spectral fit yields a photon index of $\Gamma=1.5\pm0.1$ during the 2017 EHT campaign. 
In particular, for the observations performed on April 8 (MJD 57851.7), which were used for the spectral modeling, the source presents a photon flux of $\Phi=(5.01\pm0.57)\times10^{-3}\,\mathrm{ph}\,\mathrm{cm}^2\,\mathrm{s}^{-1}$ and a photon index of $\Gamma=1.46\pm0.19$.

\subsection{Gamma-ray data}

\subsubsection{\textit{Fermi}-LAT observations}
\label{sec:gamma-fermi}

The \textit{Fermi}-LAT analysis of $\gamma$-ray emission of the blazar 3C\,279, presented in this work, was perfomed using 16 years of observations (2008 August 4–2024 August 4) and, separately with higher time resolution, a 1,5-month interval around the 2017 EHT campaign (March 15 – April 30, 2017). We selected P8R3 source class events \citep{2018arXiv181011394B} within a 15$^{\circ}$ ROI centered on 3C\,279, restricting the energy range to 100\,MeV–1\,TeV for optimal data quality \citep[although the source is detected also below 100\,MeV;][]{2018A&A...618A..22P}.
The analysis, including model optimization, search for new sources, source localization, spectral, and variability studies, was carried out with \texttt{Fermipy}\footnote{http://fermipy.readthedocs.io/en/latest/} \citep{2017arXiv170709551W} (v1.3.1) and the P8R3\_Source\_V3 IRFs. Maps were produced with 0.1$^{\circ}$ pixels. To minimize contamination from Earth-limb emission \citep{2009PhRvD..80l2004A}, below 300 MeV we excluded events with zenith angles $z>85^{\circ}$ and photons from PSF0 and PSF1 event types. Between 300 MeV and 1 GeV we excluded events with $z>85^{\circ}$ and photons from the PSF0 event type. While above 1 GeV we used all events with $z<105^{\circ}$. The model included all 4FGL-DR4 sources \citep{2020ApJS..247...33A,2022ApJS..260...53A} within 20$^{\circ}$ of 3C\,279, along with Galactic diffuse and isotropic emission templates\footnote{https://fermi.gsfc.nasa.gov/ssc/data/access/lat/BackgroundModels.html}. The spectral normalization parameter was left free for sources within 4$^{\circ}$, while it was kept fixed otherwise. The spectral index was freed for sources within 2$^{\circ}$ and fixed otherwise.
For variability studies during March-April 2017, the data were binned into 3-hour intervals. Light curves were derived by fixing the photon index to the 16-year best-fit value and allowing only the normalization to vary; 95\% upper limits were computed for bins with $<3 \sigma$ significance.
The source localization analysis yielded a highly significant ($>500\sigma$) detection at (R.A., dec.) = ($194.045^{\circ}\pm0.001^{\circ}, -5.788^{\circ}\pm0.001^{\circ}$), with $R_{95} = 0.003^{\circ}$, consistent with 4FGL\,J1256.1$-$0547 (4FGL-DR4), identified as 3C\,279. 
The $\gamma$-ray spectrum of 3C\,279 was modeled using a log-parabolic function, as adopted in the 4FGL catalog for the entire 16-year dataset and for the periods of April 5–11, 2017, and April 5–6, 2017.

\begin{figure}[h]
\centering
\includegraphics[width=\columnwidth]{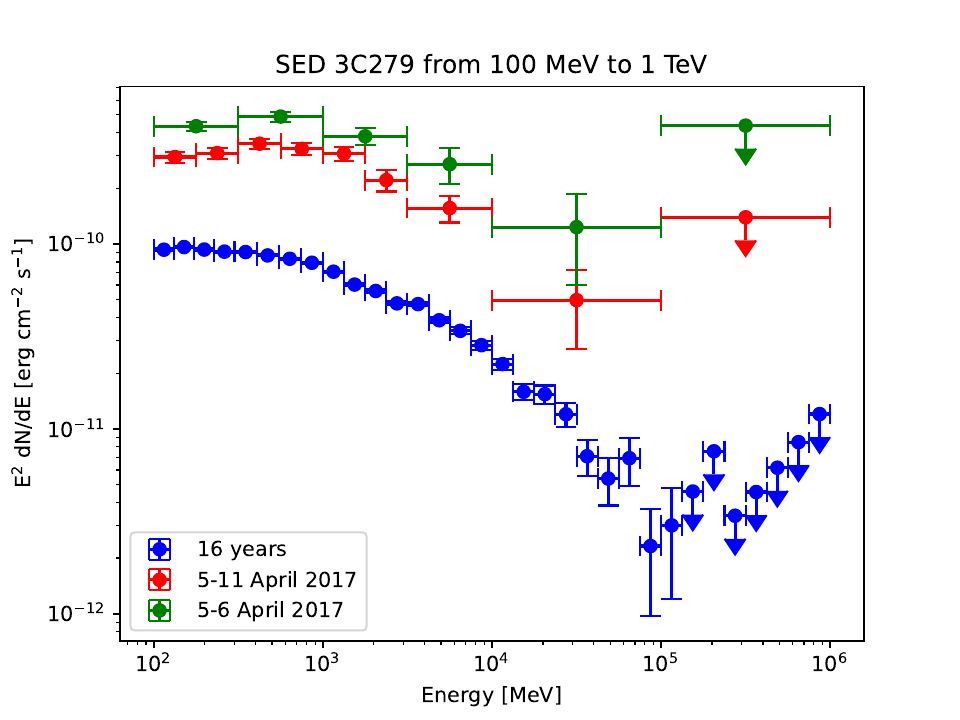}
\caption{\small \label{sed_fermi} 3C\,279 SED from \textit{Fermi}-LAT obtained using 16 years (2008 August 4–2024 August 4, blue points), the data of April 5-11, 2017 (red points), and April 5-6, 2017 (green points). A 1-$\sigma$ upper limit is reported for bins with $<3 \sigma$ significance.}
\end{figure}

\begin{figure}[h]
\centering
\includegraphics[width=\columnwidth]{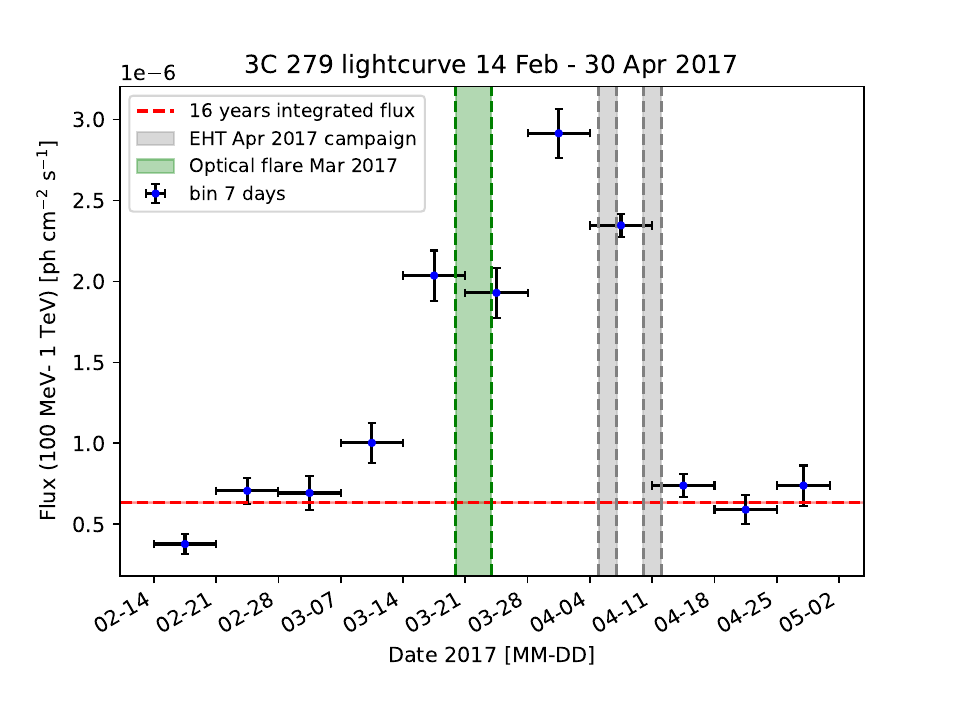}
\caption{\small \label{fig:fermi_lc} \textit{Fermi}-LAT 7-day binned light curve of 3C\,279 between February 14 and April 30, 2017. The dashed red line represents the 16-year-averaged flux $\phi_{E>100 \, \mathrm{MeV}} = (6.32 \pm 0.03) \times 10^{-7}\ \mathrm{ph\ cm^{-2}\ s^{-1}}$. The vertical green band indicates the duration of the record optical flare, and the vertical gray bands mark the EHT observing dates.}
\end{figure}

Figure~\ref{sed_fermi} presents the SED derived from 16 years of \textit{Fermi}-LAT observations, along with the SED for April 5-11, 2017, contemporaneous with the EHT campaign.
The $\gamma$-ray emission variability of 3C\,279 (see the two light curves reported in Fig.~\ref{fig:MWL_LC_2017} and Fig.~\ref{fig:fermi_lc}) indicates that the source was in a high-activity state during the EHT observing campaign. Specifically, contemporaneous with the first two EHT observations, 3C\,279 exhibited enhanced emission between April 5 and 6, reaching a flux of $\phi_{\mathrm{April\,5-6}, E>100 \, \mathrm{MeV}} = (2.90 \pm 0.12) \times 10^{-6}\ \mathrm{ph\ cm^{-2}\ s^{-1}}$. Following this peak, the flux declined to a level consistent with the 16-year \textit{Fermi}-LAT average of $\phi_{\mathrm{2008-2024},E>100 \, \mathrm{MeV}} = (6.32 \pm 0.03) \times 10^{-7}\ \mathrm{ph\ cm^{-2}\ s^{-1}}$. Moreover, during the raising phase of the flaring episode a record optical flare occurred (see Sect.~\ref{sec:mwl_lc}).

\subsubsection{H.E.S.S. observations}
\label{sec:gamma-hess}
The High Energy Stereoscopic System (\HESS) is a hybrid array of Imaging Atmospheric Cherenkov Telescopes (IACTs) located in the Khomas Highlands in Namibia, at an altitude of about 1800\,m above sea level. It is composed of four (CT1-4) Cherenkov telescopes of 12\,m diameter that have been observing the VHE $\gamma$-ray sky for more than 20 years, complemented by a fifth 28\,m telescope (CT5) since 2012. The \HESS array has a sensitivity of $\sim$1\% of Crab Nebula's flux above 100\,GeV, over 25\,h of observations under low zenith angles ($\sim$20$^\circ$) \citep{HESS_Collaboration_2006}.

\HESS observed 3C\,279 during the night of March 24 to 25, 2017, which constitutes the closest \HESS observation epoch to the 2017 coordinated EHT-MWL campaign.
Only data from the CT1-4 telescopes could be analyzed, since CT5 suffered from hardware problems throughout the entire observation night. The analysis was performed using two independent techniques: \textit{Model analysis} \citep{de_Naurois_Rolland_2009} and Image Pixel-wise fit for Atmospheric Cherenkov Telescopes (\textit{ImPACT}) \citep{Parsons_Hinton_2014}) chains, which agreed with each other.
High-level data products containing already reconstructed events and Instrument Response Functions were processed using \textit{Gammapy} \citep{Donath_2023}(version 1.2)\footnote{\textit{Gammapy} is an open-source Python package for $\gamma$-ray data analysis (v. 1.2, \url{https://doi.org/10.5281/zenodo.10726484})}. Moreover, contrary to the usual "wobble" observing mode, which permits simultaneous background measurements \citep{Berge_2007}, the late March 2017 observations were taken directly on the source position, requiring 3D background subtraction, for which a template background model was generated using the BAccMod module\footnote{\url{https://github.com/mdebony/BAccMod}}.

3C\,279 was not detected in a total of $\sim$4.2 hours of livetime. Nightly-averaged flux upper limits were derived above 100\,GeV (mean energy threshold of $\sim$105\,GeV), at 95\% confidence level, using a typical photon index of 4.0 \citep{2019A&A...627A.159H} and the likelihood profile method implemented in \textit{Gammapy}. They are included in Fig. \ref{fig:MWL_LC_2017}, while differential upper limits are presented in Fig. \ref{fig:mwl_sed}.

\subsubsection{MAGIC observations}
\label{sec:gamma-magic}
The Major Atmospheric Gamma Imaging Cherenkov (MAGIC) stereoscopic telescope system consists of two 17\,m diameter IACTs located at the Roque de los Muchachos Observatory on the Canary Island of La Palma ($28.7^{\circ}$\,N, $17.9^{\circ}$\,W), at an altitude of 2200\,m above sea level. Designed to detect air Cherenkov showers initiated by $\gamma$ rays, MAGIC is sensitive to energies ranging from approximately 50\,GeV to 50\,TeV. For low-zenith-angle observations ($15^\circ < zd < 30^\circ$), its sensitivity above 104\,GeV is $(1.445 \pm 0.015)\,\%$ of the Crab Nebula's flux over 50\,h of observation for a point-like source~\citep{2016APh....72...76A}.

The observations were performed in wobble mode, using 4 symmetrical positions w.r.t. the pointing position located $0.4^\circ$ away from the source, allowing us to collect simultaneously the signal and background~\citep{1994APh.....2..137F}. The observations presented in this work were taken during the 2017 campaign as reported in Table~\ref{tab:sed2017}. A total of 8.51\,h were collected during 2017 after quality cuts. The observations were performed within a zenith angle range from $30^{\circ}$ to $52^{\circ}$. The energy threshold of the analysis, calculated as the peak of the number of events for these observation conditions and a photon index of $4.0$, is located at approximately 100\,GeV. The analysis was performed using the standard MAGIC analysis framework MARS. A full description of the MAGIC analysis and systematic uncertainties estimation can be found in~\citet{2016APh....72...76A} and references therein.

3C\,279 was not detected above the standard $5\,\sigma$ significance level in the VHE band by the MAGIC telescopes during the 2017 campaign. The nightly LC was derived, yielding a single flux data point with a significance of $2.5\,\sigma$. For the remaining nights, the measured significances were below $2\,\sigma$, and flux upper limits were therefore computed. The 95\% confidence level U.L.s were calculated above an energy threshold of $100\,\mathrm{GeV}$, defined as the peak of the event distribution, assuming a power-law spectral model with a photon index of $4.0$, consistent with previous MAGIC detections of the source~\citep{2008Sci...320.1752M}. The resulting light curve is shown in Fig.~\ref{fig:MWL_LC_2017}.
A spectral analysis was also performed for the MWL SED. Since no significant VHE detection was achieved, only 95\% confidence level upper limits are reported for energies above $74\,\mathrm{GeV}$.

\section{Quasi-simultaneous multi-wavelength spectrum}
\label{appendix:mwl_sed}

This section contains the MWL SED for the 3C\,279 inner jet around the EHT observing campaign in 2017 (see Table \ref{tab:sed2017}). To select the closest-in-time MWL data, the EHT observation on April 10 was adopted as a reference, since the only X-ray observation within the EHT observing window was on April 8. For the radio data included in the MWL SED, we considered the innermost region accessible at each observing frequency, corresponding either to the compact core alone (C0–0 component for the EHT; see \citealt{2020A&A...640A..69K}) or, for non-EHT observations, to the combined emission from the core and the inner jet.

\clearpage 
\onecolumn 
\scriptsize
\begin{longtable}{cccccccc}
\caption{Spectral Energy Distribution for 3C\,279 during the 2017 EHT-MWL campaign.} 
\label{tab:sed2017}\\
\hline\hline
\small
Observatory & Band & $\nu$ & Angular Scale\footnotemark[1] & Obs. date &  Flux  & $\nu\,F_{\nu}$ & Sect.\\
 &  & [Hz] & [$\arcsec$] & [MJD] & [Jy] & [$\times10^{-12}$ erg cm$^{-2}$ s$^{-1}$] & \\
\hline
\endfirsthead
\endhead
\endfoot
\hline
\endlastfoot
RadioAstron & 1.67\,GHz & $1.67\times10^{9}$ & $5.0\times10^{-4}$ & 2016—2018 & $3.37\pm 0.08$  & $ (5.63\pm 0.14)\times10^{-2} $  & \ref{sec:radioastron}. \\
RadioAstron & 5\,GHz & $5\times10^{9}$ & $2.1\times10^{-4}$ & 2016—2018 & $3.86\pm 0.25$  & $(1.93\pm 0.12)\times10^{-1}$  & \ref{sec:radioastron}. \\
VLBA & 15\,GHz & $1.5\times10^{10}$ & $1.0\times10^{-3}$ & 22-04-2017 / 57865 & $14.52\pm1.25$  & $ 2.18\pm 0.11 $  & \ref{sec:radio-vlba}. \\
VLBA & 24\,GHz & $2.4\times10^{10}$ & $5.9\times10^{-4}$ & 05-05-2017 / 57878 & $7.15\pm0.52$  & $ 1.72\pm 0.13 $  & \ref{sec:radio-vlba}. \\
VLBA & 43\,GHz & $4.3\times10^{10}$ & $1.3\times10^{-4}$ & 16-04-2017 / 57859 & $6.16\pm0.49$  & $ 2.65\pm 0.21 $ & \ref{sec:radio-vlba}. \\
KVN K & 22\,GHz & $2.2\times10^{10}$ & $4.3\times10^{-3}$ & 18-04-2017 / 57861 & $12.25\pm1.22$  & $ 2.70\pm 0.27 $  & \ref{sec:radio-kvn}. \\
KVN Q & 43\,GHz & $4.3\times10^{10}$ & $2.2\times10^{-3}$ & 18-04-2017 / 57861 & $8.84\pm0.88$  & $ 3.80\pm 0.38 $ & \ref{sec:radio-kvn}. \\
KVN W & 86\,GHz & $8.6\times10^{10}$ & $1.1\times10^{-3}$ & 18-04-2017 / 57861 & $6.91\pm0.69$  & $ 5.94\pm 0.59 $ & \ref{sec:radio-kvn}. \\
KVN D & 129\,GHz & $1.29\times10^{11}$ & $0.8\times10^{-3}$ & 18-04-2017 / 57861 & $3.69\pm1.11$  & $ 4.75\pm 1.43 $ & \ref{sec:radio-kvn}. \\
GMVA & 86\,GHz & $8.6\times10^{11}$ & $2.0\times10^{-4}$ & 01-04-2017 / 57844 & $5.12\pm0.77$  & $ 4.42\pm 0.66 $ & \ref{sec:radio-gmva}. \\
ALMA & 91.5\,GHz & $9.3\times10^{10}$ & $\simeq4$ & 04-04-2017 / 57847 & $12.93 \pm 0.65 $  & $ 11.83\pm 0.60 $ & \ref{sec:radio-alma}. \\
ALMA & 103.5\,GHz & $1.03\times10^{11}$ & $\simeq3$ & 13-04-2017 / 57856 & $12.17 \pm 0.7 $  & $ 12.59\pm 0.72 $ & \ref{sec:radio-alma}. \\
ALMA & 221\,GHz & $2.21\times10^{11}$ & $\simeq1$ & 06-04-2017 / 57849 & $9.36\pm0.94$  & $ 20.70\pm 2.10 $ & \ref{sec:radio-alma}. \\
ALMA & 343\,GHz & $3.43\times10^{11}$ & $\simeq0.4$ & 13--04-2017 / 57856 & $6.30\pm0.07$  & $ 21.61\pm 0.24 $ & \ref{sec:radio-alma}. \\
SMA & 235\,GHz & $2.35\times10^{11}$ & $\sim3$ & 03-04-2017 / 57846 & $8.55\pm0.45$  & $ 20.1\pm 1.1 $ & \ref{sec:radio-sma}. \\
SMA & 261\,GHz & $2.61\times10^{11}$ & $\sim3$ & 03-04-2017 / 57846 & $7.79\pm0.45$  & $ 20.3\pm 1.2 $ & \ref{sec:radio-sma}. \\
EHT & 230\,GHz & $2.3\times10^{11}$ & $2.4\times10^{-5}$ & 10-04-2017 / 57853 & $ 0.89\pm 0.05 $ & $ 2.05\pm 0.12 $ & \ref{sec:radio-eht}. \\
Kanata & 1220\,nm & $2.4\times10^{14}$ & $\simeq$6 & 22-03-2017 / 57834 & $<40.8 \times 10^{-3}$  & $<97.9$ & \ref{sec:optical-kanata}. \\
Kanata & 658\,nm & $4.5\times10^{14}$ & $\simeq$6 & 22-03-2017 / 57834 & $<21.6 \times 10^{-3}$  & $<97.2$ & \ref{sec:optical-kanata}. \\
Kanata & 551\,nm & $5.4\times10^{14}$ & $\simeq$6 & 18-03-2017 / 57830 & $<9.70 \times 10^{-3}$  & $<52.4$ & \ref{sec:optical-kanata}. \\
Perkins I & 806\,nm & $3.7\times10^{14}$ & $\simeq$3 & 31-03-2017 / 57843 & $(8.51\pm0.16)\times 10^{-3}$  & $31.7\pm0.6$ & \ref{sec:optical-perkins}. \\
Perkins R & 658\,nm & $4.5\times10^{14}$ & $\simeq$3 & 31-03-2017 / 57843 & $(6.15\pm0.07)\times 10^{-3}$  & $28.0\pm0.3$ & \ref{sec:optical-perkins}. \\
Perkins V & 551\,nm & $5.4\times10^{14}$ & $\simeq$3 & 31-03-2017 / 57843 & $(4.39\pm0.09)\times 10^{-3}$  & $23.9\pm0.5$ & \ref{sec:optical-perkins}. \\
Perkins B & 445\,nm & $6.7\times10^{14}$ & $\simeq$3 & 31-03-2017 / 57843 & $(2.98\pm0.10)\times 10^{-3}$  & $20.3\pm0.7$ & \ref{sec:optical-perkins}. \\
\emph{Swift}-UVOT& 546.8\,nm  & $5.48\times10^{14}$ & $\simeq$3 & 14-04-2017 / 57857 & $(1.16\pm0.09)\times10^{-3}$ & $6.36\pm0.49$ & \ref{sec:swift_uvot}. \\
\emph{Swift}-UVOT& 439.2\,nm  & $6.83\times10^{14}$ & $\simeq$3 & 14-04-2017 / 57857 & $(8.56\pm0.41)\times10^{-4}$ & $5.91\pm0.28$ & \ref{sec:swift_uvot}. \\
\emph{Swift}-UVOT& 346.5\,nm  & $8.65\times10^{14}$ & $\simeq$3 & 27-04-2017 / 57839 & $(1.00\pm0.04)\times10^{-3}$ & $8.65\pm0.35$ & \ref{sec:swift_uvot}. \\
\emph{Swift}-UVOT& 260.0\,nm  & $1.15\times10^{15}$ & $\simeq$3 & 14-04-2017 / 57857 & $(4.32\pm0.22)\times10^{-4}$ & $4.97\pm0.25$ & \ref{sec:swift_uvot}. \\
\emph{Swift}-UVOT& 224.6\,nm  & $1.33\times10^{15}$ & $\simeq$3 & 14-04-2017 / 57857 & $(3.22\pm0.19)\times10^{-4}$ & $4.28\pm0.25$ & \ref{sec:swift_uvot}. \\
\emph{Swift}-UVOT& 192.8\,nm  & $1.55\times10^{15}$ & $\simeq$3 & 14-04-2017 / 57857 & $(2.84\pm0.12)\times10^{-4}$ & $4.40\pm0.19$ & \ref{sec:swift_uvot}. \\
\emph{Swift}-XRT & 0.3-10\,keV  & $4.19\times10^{17}$ & $\simeq$75 & 08-04-2017 / 57851 &  $(6.7\pm 1.3)\times10^{-7}$\footnotemark[2] & $15.8\pm3.1$\footnotemark[2] & \ref{sec:xray-swiftxrt}. \\
\emph{Fermi}-LAT & 0.1\, - \,0.18 GeV & $3.22\times10^{22}$ & $\simeq$7000 & 05--11-04-2017 / 57848–57854 & $(9.16\pm 0.65)\times  10^{-10} $  & $295.5\pm20.9$ & \ref{sec:gamma-fermi}. \\
\emph{Fermi}-LAT & 0.18\, - \,0.32 GeV & $5.73\times10^{22}$ & $\simeq$7000 & 05--11-04-2017 / 57848–57854 & $(5.41\pm 0.38)\times  10^{-10} $  & $310.2\pm21.9$ & \ref{sec:gamma-fermi}. \\
\emph{Fermi}-LAT & 0.32\, - \,0.56 GeV & $1.02\times10^{23}$  & $\simeq$4500 & 05--11-04-2017 / 57848–57854 & $ (3.42\pm 0.22)\times 10^{-10}$  & $349.0\pm22.4$ & \ref{sec:gamma-fermi}. \\
\emph{Fermi}-LAT & 0.56\, - \,1.00 GeV & $1.81\times10^{23}$  & $\simeq$2800 & 05--11-04-2017 /57848–57854 & $(1.80\pm0.14)\times 10^{-10}$  & $327.1\pm25.3$ & \ref{sec:gamma-fermi}. \\
\emph{Fermi}-LAT & 1.00\, - \,1.78 GeV & $3.22\times10^{23}$  & $\simeq$2000 & 05--11-04-2017 / 57848–57854 & $(9.56\pm0.83)\times 10^{-11}$  & $308.4\pm26.8$ & \ref{sec:gamma-fermi}. \\
\emph{Fermi}-LAT & 1.78\, - \,3.16 GeV & $5.73\times10^{23}$  & $\simeq$1100 & 05--11-04-2017 / 57848–57854 & $(3.87\pm0.51)\times 10^{-11}$  & $221.7\pm29.3$ & \ref{sec:gamma-fermi}. \\
\emph{Fermi}-LAT & 3.16\, - \,10 GeV & $1.36\times10^{24}$  & $\simeq$700 & 05--11-04-2017 / 57848–57854 & $(1.15\pm0.19)\times 10^{-11}$  & $156.5\pm25.7$& \ref{sec:gamma-fermi}. \\
\emph{Fermi}-LAT & 10\, - \,100 GeV & $7.65\times10^{24}$  & $\simeq$500 & 05--11-04-2017 / 57848–57854 & $(6.50\pm2.95)\times 10^{-11}$  & $49.7\pm22.6$& \ref{sec:gamma-fermi}. \\
\emph{Fermi}-LAT & 100\, - \,1000 GeV & $7.65\times10^{25}$  & $\simeq$350 & 05--11-04-2017 / 57848–57854 & $<1.82\times 10^{-13}$  & $<139.5$ & \ref{sec:gamma-fermi}. \\
H.E.S.S. & 0.105 - 0.153\,TeV & $3.07\times10^{25}$ & $\simeq\!360$ & 24--25-03-2017 / 57836-57837 & $<2.17\times 10^{-14}$  & $<6.67$ & \ref{sec:gamma-hess}. \\
H.E.S.S. & 0.153 - 0.27\,TeV & $4.90\times10^{25}$ & $\simeq\!360$ & 24--25-03-2017 / 57836-57837 & $<3.54 \times 10^{-15}$  & $<1.74$ & \ref{sec:gamma-hess}. \\
H.E.S.S. & 0.27 - 0.57\,TeV & $9.45\times10^{25}$ & $\simeq\!360$ & 24--25-03-2017 / 57836-57837 & $<8.32 \times 10^{-16}$  & $<0.79$ & \ref{sec:gamma-hess}. \\
H.E.S.S. & 0.57 - 1.0\,TeV & $1.83\times10^{26}$ & $\simeq\!360$ & 24--25-03-2017 / 57836-57837 & $<7.93 \times 10^{-17}$  & $<0.14$ & \ref{sec:gamma-hess}. \\
MAGIC & 85\,GeV & $2.06\times10^{25}$ & $\simeq\!330$ & 01--02-04-2017 / 57849–57850 & $<5.28 \times 10^{-13}$  & $<108.7$ & \ref{sec:gamma-magic}. \\
MAGIC & 115\,GeV & $2.79\times10^{25}$ & $\simeq\!330$ & 01--02-04-2017 / 57849–57850 & $<3.08\times 10^{-14}$  & $<8.58$ & \ref{sec:gamma-magic}. \\
MAGIC & 157\,GeV & $3.79\times10^{25}$ & $\simeq\!330$ & 01--02-04-2017 / 57849–57850 & $<2.23 \times 10^{-14}$  & $<8.47$ & \ref{sec:gamma-magic}. \\
MAGIC & 213\,GeV & $5.14\times10^{25}$ & $\simeq\!330$ & 01--02-04-2017 / 57849–57850 & $<1.65 \times 10^{-14}$  & $<8.47$ & \ref{sec:gamma-magic}. \\
MAGIC & 288\,GeV & $6.96\times10^{25}$ & $\simeq\!330$ & 01--02-04-2017 / 57849–57850 & $<8.41 \times 10^{-15}$  & $<5.85$ & \ref{sec:gamma-magic}. \\ 
\end{longtable}
\tablefoot{
    \tablefoottext{1}{Spatial scale of the source emission (FWHM of the instrumental beam or PSF). For non-EHT radio observations, if the core is not resolved we use the flux peak obtained from the beam size as an upper limit on the core emission. For elliptical beams, we list the average of the axes as a representative angular scale. At higher energies the source is not spatially resolved; therefore, we quote the angular scale corresponding to the instrument’s angular resolution (PSF) in the relative energy band.}\\
    \tablefoottext{2}{The source presents a photon flux of $\Phi=(5.01\pm0.57)\times10^{-3}\,\mathrm{cm}^2\,\mathrm{s}^{-1}$ and a photon index of $\Gamma=1.46\pm0.19$, see Sect. \ref{sec:xray-swiftxrt}.}\\
    }
\normalsize
\clearpage 
\twocolumn

\end{appendix}

\end{document}